\newcommand{\va}{\boldsymbol{a}}

\newcommand{\vl}{\boldsymbol{l}}

\newcommand{\vt}{\boldsymbol{t}}
\newcommand{\vx}{\boldsymbol{x}}
\newcommand{\vy}{\boldsymbol{y}}

\newcommand{\vK}{\boldsymbol{K}}
\newcommand{\vI}{\boldsymbol{I}}
\newcommand{\vP}{\boldsymbol{P}}
\newcommand{\vR}{\boldsymbol{R}}

\newcommand{\modified}[1]{\textcolor{black}{#1}}

\documentclass[10pt,journal,compsoc]{IEEEtran}
\usepackage{makeidx}  
\usepackage{url}
\usepackage{latexsym}
\usepackage{marginnote}
\usepackage{lipsum}
\usepackage{hyperref}
\hypersetup{  
    colorlinks=true,
    linkcolor=cyan,
    filecolor=cyan,  
    citecolor=cyan,   
    urlcolor=cyan,  
}
\usepackage[table]{xcolor}
\usepackage{subfigure}
\usepackage{csquotes}
\usepackage{multirow}

\ifCLASSOPTIONcompsoc
  \usepackage[compress]{cite}
\else
  \usepackage{cite}
\fi

\usepackage[pdftex]{graphicx}

\usepackage{amsmath}
\usepackage{amssymb}

\begin{document}


\title{Learning Perspective Deformation in X-Ray Transmission Imaging}

\author{Yixing~Huang, Andreas~Maier,~\IEEEmembership{Senior Member,~IEEE}, Fuxin~Fan, Bj\"orn~Kreher, Xiaolin~Huang,~\IEEEmembership{Senior Member,~IEEE}, Rainer~Fietkau, Christoph~Bert$^\ast$, Florian~Putz$^\ast$
\IEEEcompsocitemizethanks{\IEEEcompsocthanksitem Y.~Huang, ~R.~Fietkau, C.~Bert and F.~Putz are with Department of Radiation Oncology, University Hospital Erlangen, Friedrich-Alexander-Universit\"at Erlangen-N\"urnberg, 91054 Erlangen, Germany. They are also with Comprehensive Cancer Center Erlangen-EMN (CCC ER-EMN), Erlangen, Germany.\\
E-mail: yixing.yh.huang@fau.de
\IEEEcompsocthanksitem A.~Maier \modified{and F.~Fan} are with Pattern Recognition Lab, Friedrich-Alexander-Universit\"at Erlangen-N\"urnberg, 91058 Erlangen, Germany.
\IEEEcompsocthanksitem \modified{B.~Kreher is with Siemens Healthcare GmbH, 91301 Forchheim, Germany.}
\IEEEcompsocthanksitem \modified{X.~Huang is with the Department of Automation, Shanghai Jiao Tong University, Shanghai, 200240, China}
}
\thanks{$\ast$: contribute equally.}
\thanks{This version was last edited on 21.Oct.2022.}
}


\IEEEtitleabstractindextext{%
\begin{abstract}
  \modified{In cone-beam X-ray transmission imaging, perspective deformation causes difficulty in direct, accurate geometric assessments of anatomical structures. In this work, the perspective deformation correction problem is formulated and addressed in a framework using two complementary ($180^\circ$) views. The complementary view setting provides a practical way to identify perspectively deformed structures by assessing the deviation between the two views. In addition, it provides bounding information and reduces uncertainty for learning perspective deformation. Two representative networks Pix2pixGAN and TransU-Net for correcting perspective deformation are investigated.
Experiments on numerical bead phantom data demonstrate the advantage of complementary views over orthogonal views or a single view. 
They show that Pix2pixGAN as a fully convolutional network achieves better performance in polar space than Cartesian space, while TransU-Net as a transformer-based hybrid network achieves comparable performance in Cartesian space to polar space. Further study demonstrates that the trained model has certain tolerance to geometric inaccuracy such as source-to-isocenter distances, rotation angles, detector principal point shifts and respiratory motion within calibration accuracy. This indicates that one model trained from one cone-beam computed tomography (CBCT) scanner is applicable to other scanners of the same type.
The experiments on the chest and head data demonstrate that our method has the potential for accurate cardiothoracic ratio measurement and cephalometric imaging.
The efficacy of the proposed framework on real cadaver CBCT projection data and its robustness in the presence of bulky metal implants and surgical screws indicate the promising aspects of future real applications.}
  
\end{abstract}

}

\maketitle

\IEEEdisplaynontitleabstractindextext

\IEEEpeerreviewmaketitle

\section{Introduction}
\IEEEPARstart{C}{one-beam} X-ray imaging along with a flat-panel detector is widely used for disease diagnosis, treatment planning, and intervention guiding. It is used for direct two-dimensional (2D) imaging such as fluoroscopic angiography \cite{ghoshhajra2017real} and chest radiographs \cite{hata2021dynamic} as well as three-dimensional (3D) volume reconstruction in cone-beam computed tomography (CBCT) \cite{bosetti2020cone,van2020tumor}. In cone-beam X-ray imaging, because of the divergence of X-rays, imaged structures with different depths have different magnification factors on the X-ray detector. As a consequence, acquired images suffer from geometric distortions, which is called perspective deformation. Perspective deformation causes difficulty in direct, accurate geometric assessments of structures of interest (SOI) in many practical applications, e.g., anatomical landmark detection \cite{kumar2007comparison,montufar2018hybrid}, fluoroscopic image stitching \cite{fotouhi2021reconstruction}, fiducial marker registration \cite{george2011robust,li2008comparison,hoegele2011stochastic}, and dual-modality image fusion \cite{syben2020known}. Therefore, orthogonal projections of SOI are preferred over perspective projections in many applications. 

\begin{figure}
\centering
\begin{minipage}[b]{0.4\linewidth}
\subfigure[Cephalometer]{
\includegraphics[width=\linewidth]{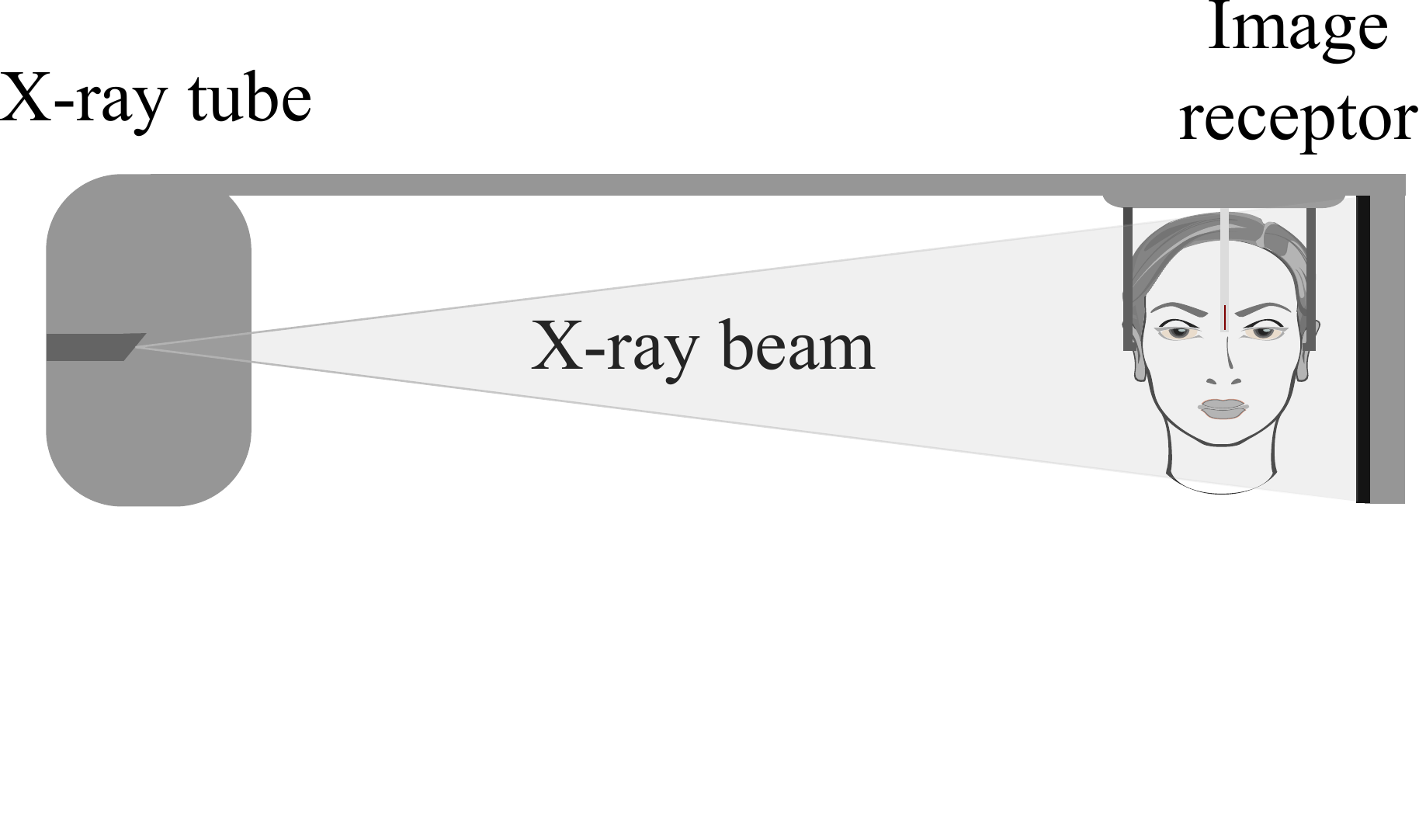}
\label{subfig:cephalometer}
}
\end{minipage}
\hspace{5pt}
\begin{minipage}[b]{0.4\linewidth}
\subfigure[Chest X-ray]{
\includegraphics[width=\linewidth]{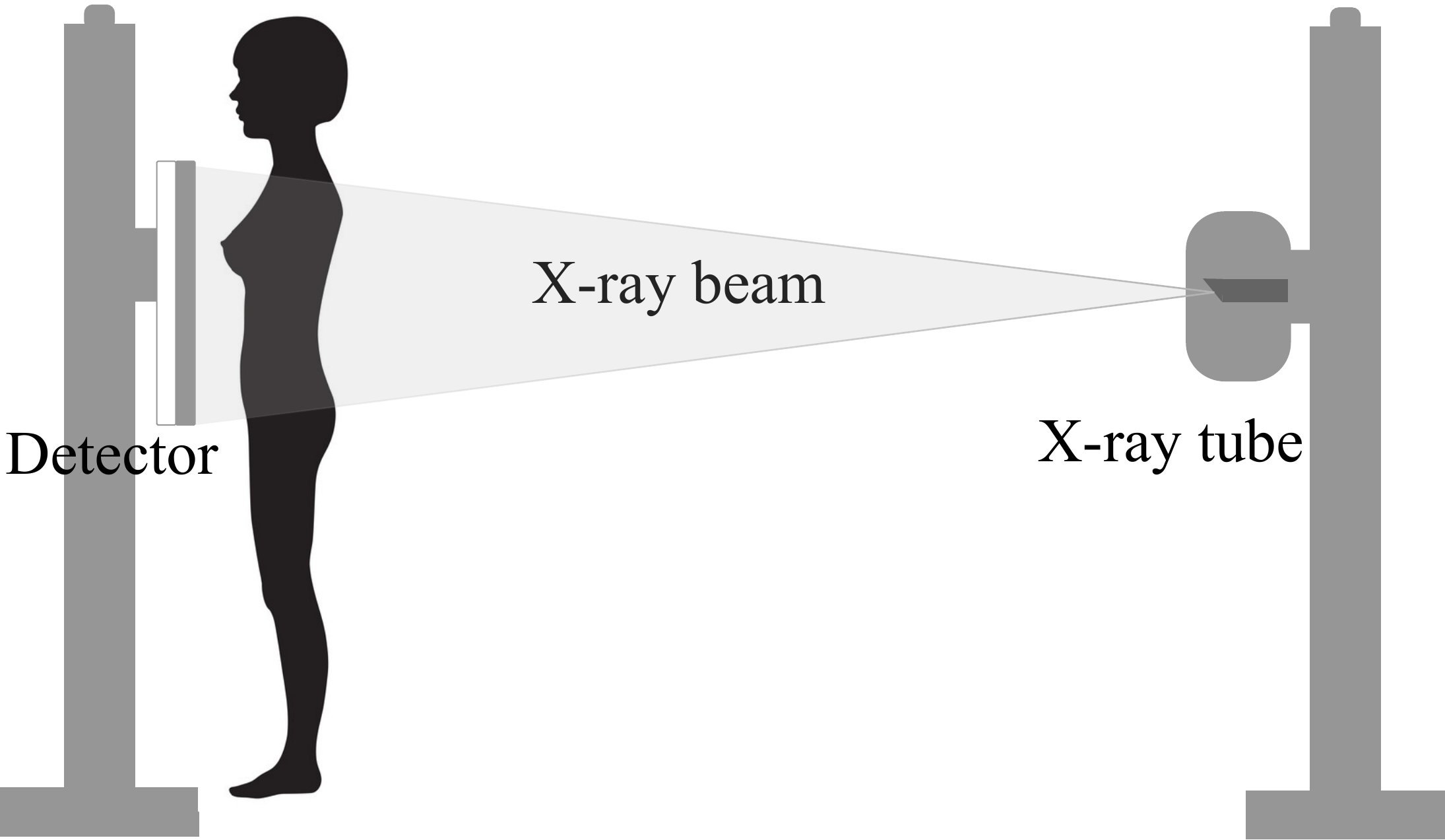}
\label{subfig:chestXRay}
}
\end{minipage}
\caption{\modified{Illustrations of cephalometer (a) and chest X-ray (b) systems. A cephalometer typically has a source-to-isocenter distance of 152.4\,cm \cite{kumar2007comparison} and a chest X-ray system has a source-to-detector/patient distance close to 2\,m \cite{hata2021dynamic}, while regular CBCT systems have a source-to-isocenter distance of around 65\,cm.}}
\label{Fig:CephalAndChestSystems}
\end{figure}

To reduce perspective deformation in practice, specialized cone-beam X-ray devices are designed for certain applications. For example, cephalometers \cite{broadbent1931new} and chest X-ray systems \cite{hata2021dynamic} are designed specially for cephalometric analysis and chest imaging respectively, both of which have a relatively large source-to-detector distance and a short object-to-detector distance, as displayed in Fig.\,\ref{Fig:CephalAndChestSystems}. However, in regular cone-beam X-ray imaging systems, e.g., C-arm CBCT systems, such a large source-to-detector distance is not available. In such systems, perspective deformation remains an issue for applications like chest X-ray imaging and cephalometric analysis.

 Alternative to specialized devices, to deal with perspective deformation, digitally reconstructed radiographs (DRRs) generated from an intermediate 3D volume are commonly used. For example, in dental CBCT, DRRs using orthogonal projection are used as synthetic cephalograms \cite{kumar2007comparison}, which provide more cephalometric landmark accuracy than those with perspective projection. However, obtaining such 3D CBCT volumes brings additional dose exposure to patients, since hundreds of projections are acquired for 3D reconstruction. Another potential application is in hybrid magnetic resonance imaging (MRI) and X-ray imaging \cite{fahrig2001truly,stimpel2019projection}. 
 Obtaining 3D MRI volumes to generate DRRs is time-consuming, taking around 30\,min for each scan. More importantly, pre-acquired 3D MRI volumes cannot provide accurate information on daily organ changes \cite{wu2008line}. Therefore, a solution for fast 2D MRI/X-ray hybrid imaging in one exam without patient repositioning \cite{fahrig2001truly,fahrig2001truly2} is desired for future potential applications in interventional surgery and radiation therapy \cite{raaymakers2008feasibility}.
According to Fourier sampling theorems, 2D parallel-beam MRI images are much more natural and efficient to acquire than 2D cone-beam MRI images. Hence, the registration between 2D cone-beam X-ray images and 2D parallel-beam MRI images is necessary, which requires the conversion between perspective and orthogonal projections \cite{syben2020known}.
Therefore, learning perspective deformation, which directly converts 2D perspective (cone-beam) projections to 2D orthogonal (parallel-beam) projections, has important value in many applications.


Perspective deformation is also a common problem in optical imaging \cite{valente2015perspective,lutz2019deep,mantel2020method,del2020blind}. But optical images are reflection (surface) images, and such perspective deformation is typically caused by the distortion of camera lens. Therefore, perspective deformation in X-ray transmission imaging has a substantial difference from that in optical imaging. As a result, perspective distortion correction algorithms for optical imaging cannot be applied.


Learning perspective deformation in X-ray transmission imaging is a novel and important topic for CBCT and deep learning, which has great clinical impacts. Our work is the first one to formulate and address this problem in a framework using two views with the proposed framework being verified by experiments on multiple datasets. Specifically, the contributions of this work mainly lie in the following aspects: 

 a) Formulate the perspective deformation learning problem and provide a framework to investigate it;
 
 b) Propose a simple complementary view setting, which provides bounding information and hence reduces uncertainty as well as required receptive field size for learning perspective deformation in X-ray transmission imaging;
 
 c) Provide a practical way to identify which structures suffer from perspective deformation by assessing deviations between the two complementary views, which can be easily observed in an RGB stack of complementary views; 
 
 d) Demonstrate the efficacy of the proposed framework with representative networks, e.g., pix2pixGAN as a fully convolutional network (FCN) representative and TransU-Net as a transformer representative; 
 
 e) Investigate the influence of spatial coordinate systems (Cartesian and polar space in particular) and geometric inaccuracy (rotation angles, source-to-isocenter distances, detector principal point shifts and respiratory motion); 
 
 f) Evaluate the efficacy of the proposed framework on multiple datasets, from simulated general spherical bead phantom data to DRRs calculated from real patients' CT datasets (chest and head data) as well as acquired real CBCT projection data (knee data), which indicates the promising aspects of future real applications.


\section{Problem and Learning Framework}
\subsection{Perspective Deformation in CBCT}
\label{subsect:PDinCBCT}

\begin{figure}
\centering
\begin{minipage}[t]{0.45\linewidth}
\subfigure[]{
\includegraphics[width=\linewidth]{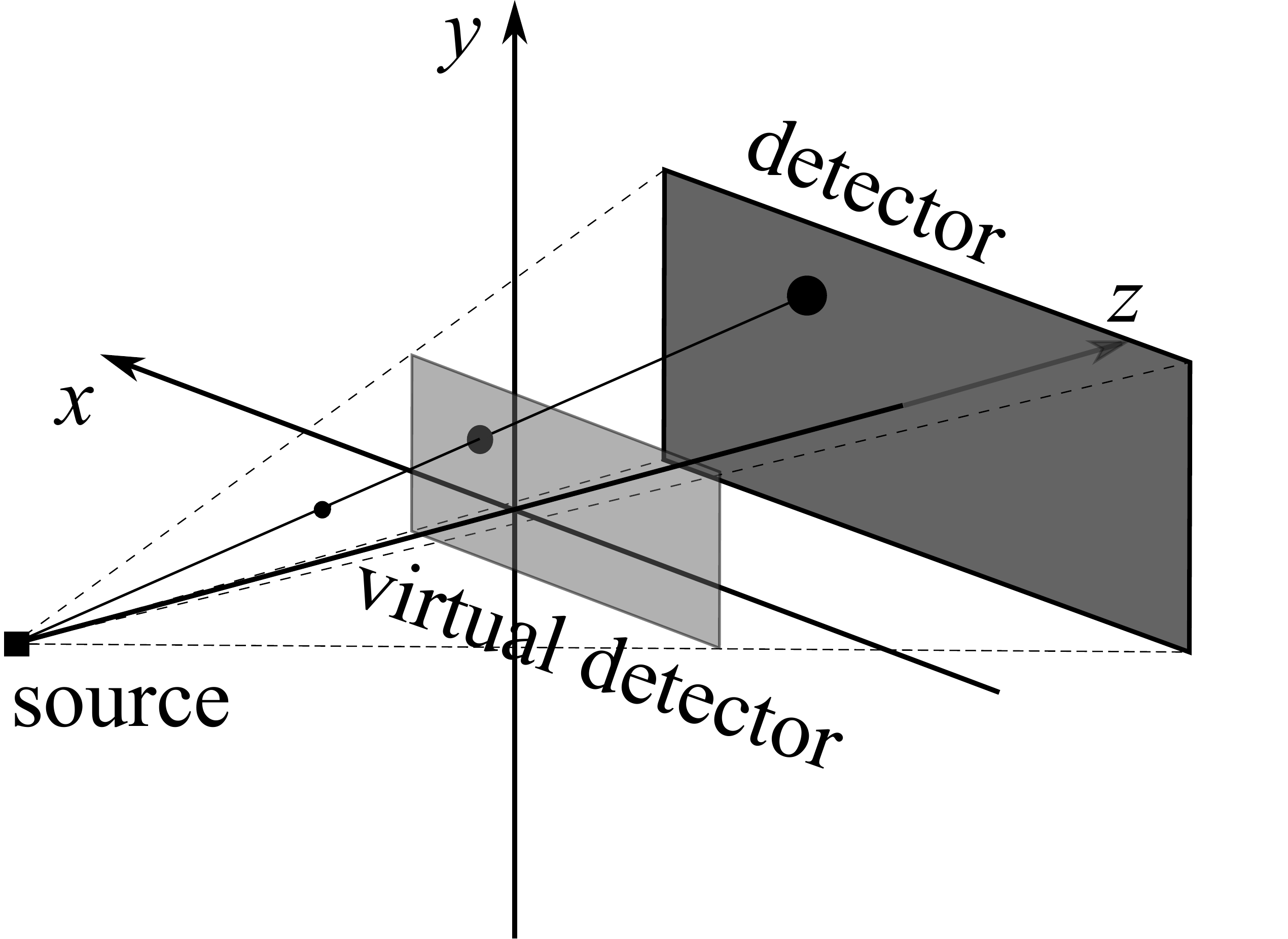}
\label{subfig:virtualDetector}
}
\end{minipage}
\begin{minipage}[t]{0.45\linewidth}
\subfigure[Single view]{
\includegraphics[width=\linewidth]{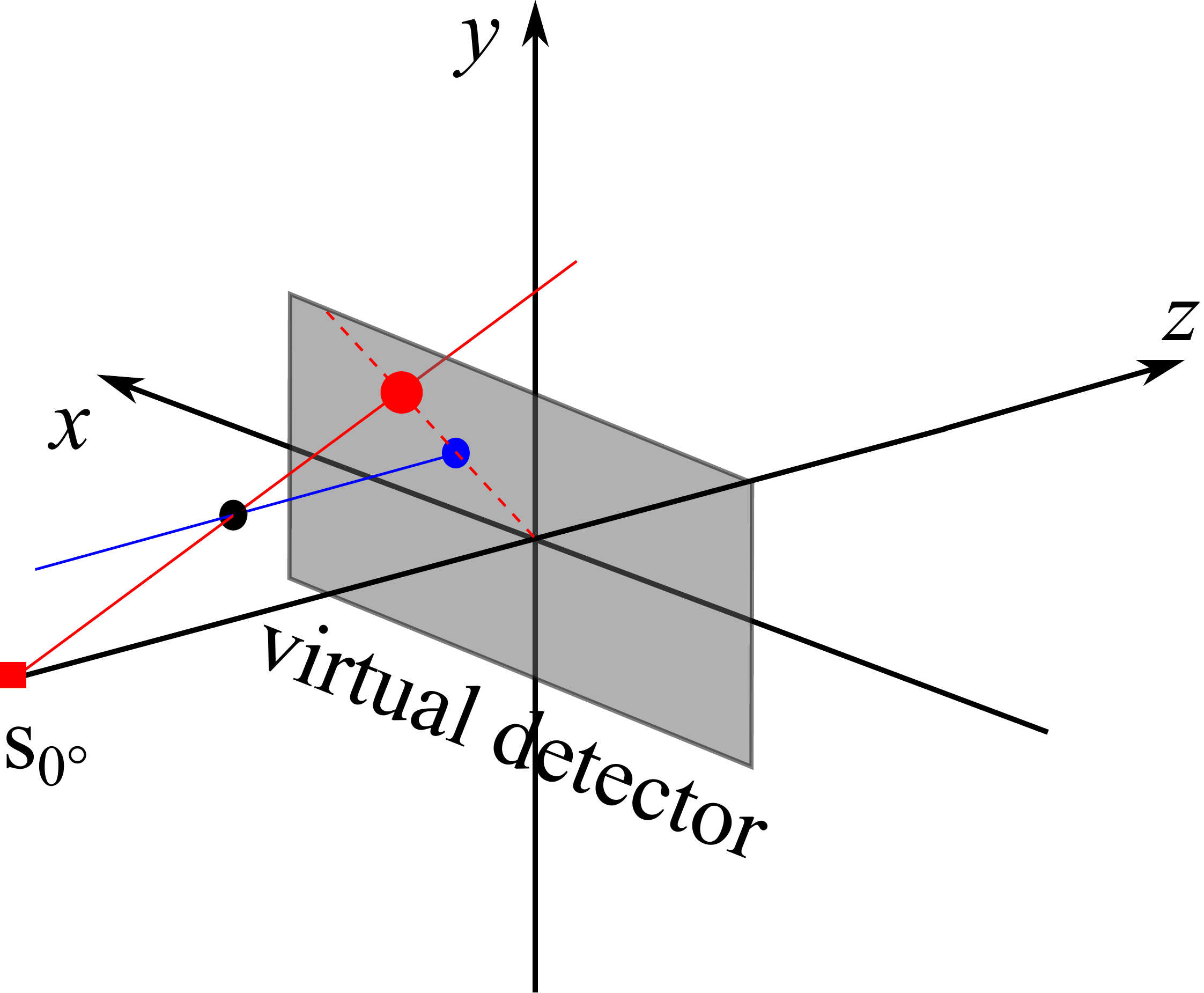}
\label{subfig:singleView}
}
\end{minipage}

\begin{minipage}[t]{0.45\linewidth}
\subfigure[Dual orthogonal views]{
\includegraphics[width=\linewidth]{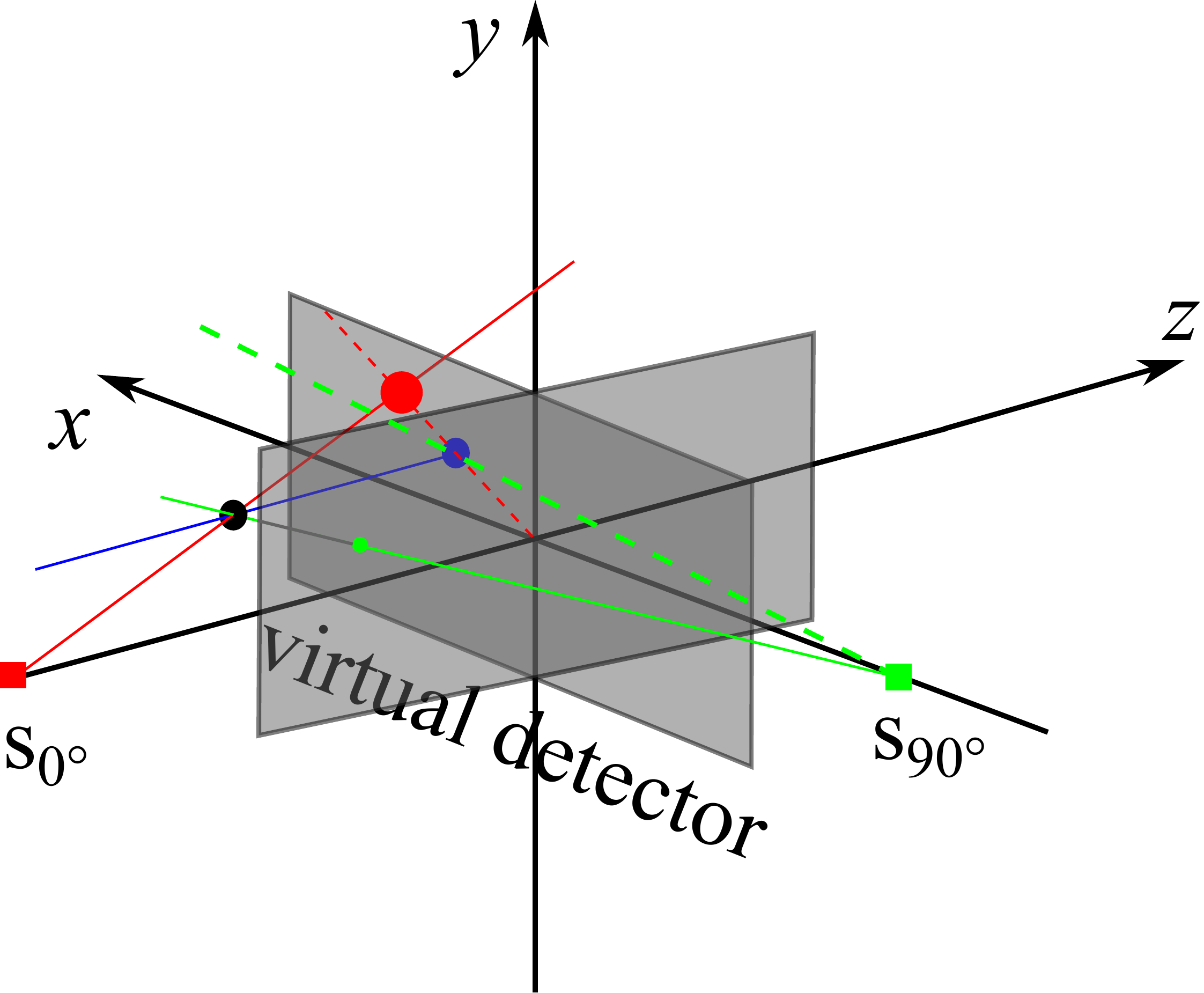}
\label{subfig:orthoViewBPOP}
}
\end{minipage}
\begin{minipage}[t]{0.45\linewidth}
\subfigure[Dual complementary views]{
\includegraphics[width=\linewidth]{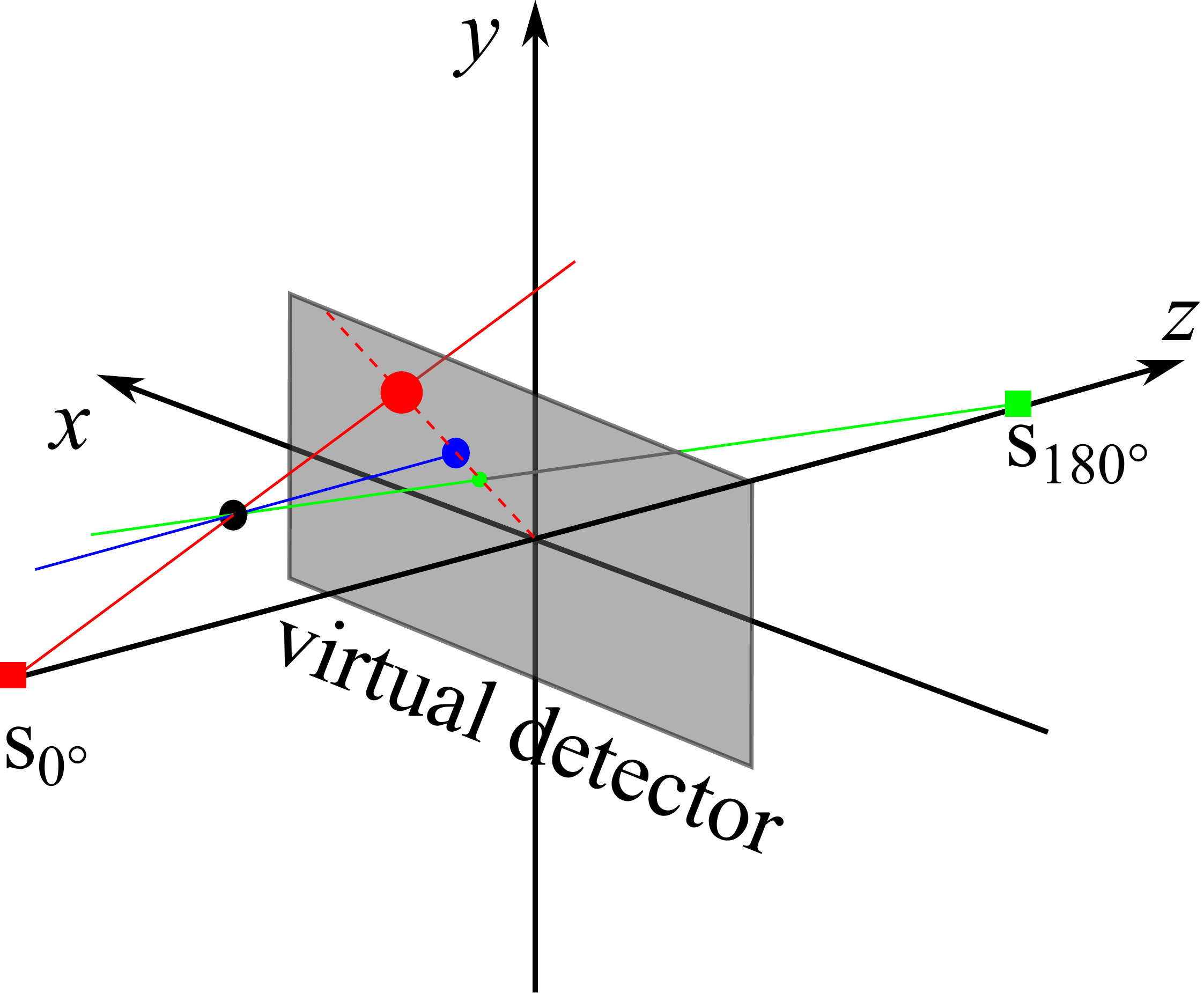}
\label{subfig:complementaryViews}
}
\end{minipage}
\caption{\modified{A CBCT system with a virtual detector at the isocenter (a) and perspective deformation learning with different views: (b) a single view, (c) two orthogonal views, and (d) two complementary views.} The red point is the perspective projection of the black point from the $0^\circ$ view, and the blue point is the orthogonal projection of the black point. The green point is the perspective projection of the black point from the $90^\circ$ view in (b), while it is from the $180^\circ$ view in (c). }
\label{Fig:viewIllustrations}
\end{figure}

Perspective deformation can be described by perspective projection matrices in general CBCT systems. A perspective projection matrix $\vP$ with a size of $3 \times 4$ can be decomposed into an intrinsic parameter matrix $\vK^{3\times 3}$, a rotation matrix $\vR^{3\times 3}$ and a translation vector $\vt^{3\times 1}$,
\begin{equation}
\vP = \vK [\vR | \vt].
\end{equation}
For a CBCT system with a source-to-detector distance $D_{\text{sd}}$, a source-to-isocenter distance $D_{\text{si}}$, a flat-panel detector pixel spacing $s_u$ and $s_v$, and its principle point coordinates $p_u$ and $p_v$ (where the principle ray hits at the detector), the intrinsic parameter matrix $\vK$ is defined as follows \cite{hartley2003multiple},
\begin{equation}
\vK = \begin{bmatrix}
D_{\text{sd}}/s_u & 0 & p_u\\
0 & D_{\text{sd}}/s_v & p_v\\
0 & 0 & 1
\end{bmatrix},
\end{equation}
where the presence of $s_u$ and $s_v$ converts physical coordinates to detector pixel coordinates.
The rotation matrix $\vR$ is determined by the camera orientation, i.e., the rotation angles along the three axes $\theta_x$, $\theta_y$ and $\theta_z$. The translation vector $\vt$ is the location of the world origin in camera coordinates. With an initial state displayed in Fig.\,\ref{subfig:virtualDetector}, $\vt$ is $[0, 0, D_{\text{si}}]^\top$ and $\vR$ is the identity matrix $\vI^{3\times3}$.

In a CBCT system, structures acquired in the flat-panel detector can be rebinned to a virtual detector located at the isocenter (Fig.\,\ref{subfig:virtualDetector}) to reduce the magnification factor caused by the ratio of $D_{\text{sd}}$ to $D_{\text{si}}$. The new intrinsic parameter matrix $\vK'$ and projection matrix $\hat{\vP}$ for the virtual detector is
\begin{equation}
\vK' = \begin{bmatrix}
D_{\text{si}}/s'_u & 0 & p_u\\
0 & D_{\text{si}}/s'_v & p_v\\
0 & 0 & 1
\end{bmatrix},
\label{eqn:KvirtualDetector}
\end{equation}
and,
\begin{equation}
\hat{\vP} = \vK' [\vR | \vt].
\end{equation}
When the virtual detector pixel spacing $s'_u = s_u \cdot D_{\text{si}}/D_{\text{sd}}$ and $s'_v = s_v \cdot D_{\text{si}}/D_{\text{sd}}$, the intrinsic parameter matrix $\vK'$ is the same as $\vK$ and hence the projection matrix $\hat{\vP}$ remains the same as well. 
In the following, without the loss of generality an isotropic virtual detector pixel spacing $s$ is used, i.e., $s'_u=s'_v=s$.

The orthogonal projection with the same virtual detector has the following projection matrix \cite{hartley2003multiple},
\begin{equation}
\bar{\vP} = \begin{bmatrix}
1/s & 0 & p_u\\
0 & 1/s & p_v\\
0 & 0 & 1
\end{bmatrix}
\begin{bmatrix}
\tilde{\vR} & \boldsymbol{0}\\
\boldsymbol{0}^\top & 1 \\
\end{bmatrix},
\label{eqn:parallelProjMatrix}
\end{equation}
where $\tilde{\vR}^{2\times 3}$ is formed by the first two rows of $\vR$.

After rebinning into the virtual detector, structures in the mid-sagittal plane (where the virtual detector is located) have no magnification. However, structures in other sagittal planes are either magnified or shrunk depending on their depths. For a point $\va=(x, y, z, 1)$ in homogeneous coordinates, its corresponding points in perspective projection and orthogonal projection are denoted by $\hat{\va}$ and $\bar{\va}$, respectively,
\begin{equation}
\begin{array}{l}
\hat{\va} = \hat{\vP} \va,\\
 \bar{\va} = \bar{\vP} \va.
\end{array}
\end{equation}
Then perspective deformation (PD) is calculated as,
\begin{equation}
d_\text{PD} = ||\hat{\va} -  \bar{\va}||.
\end{equation}

\begin{figure}[t]
\centering
\begin{minipage}[t]{0.31\linewidth}
\subfigure[Cartesian]{
\includegraphics[width=\linewidth]{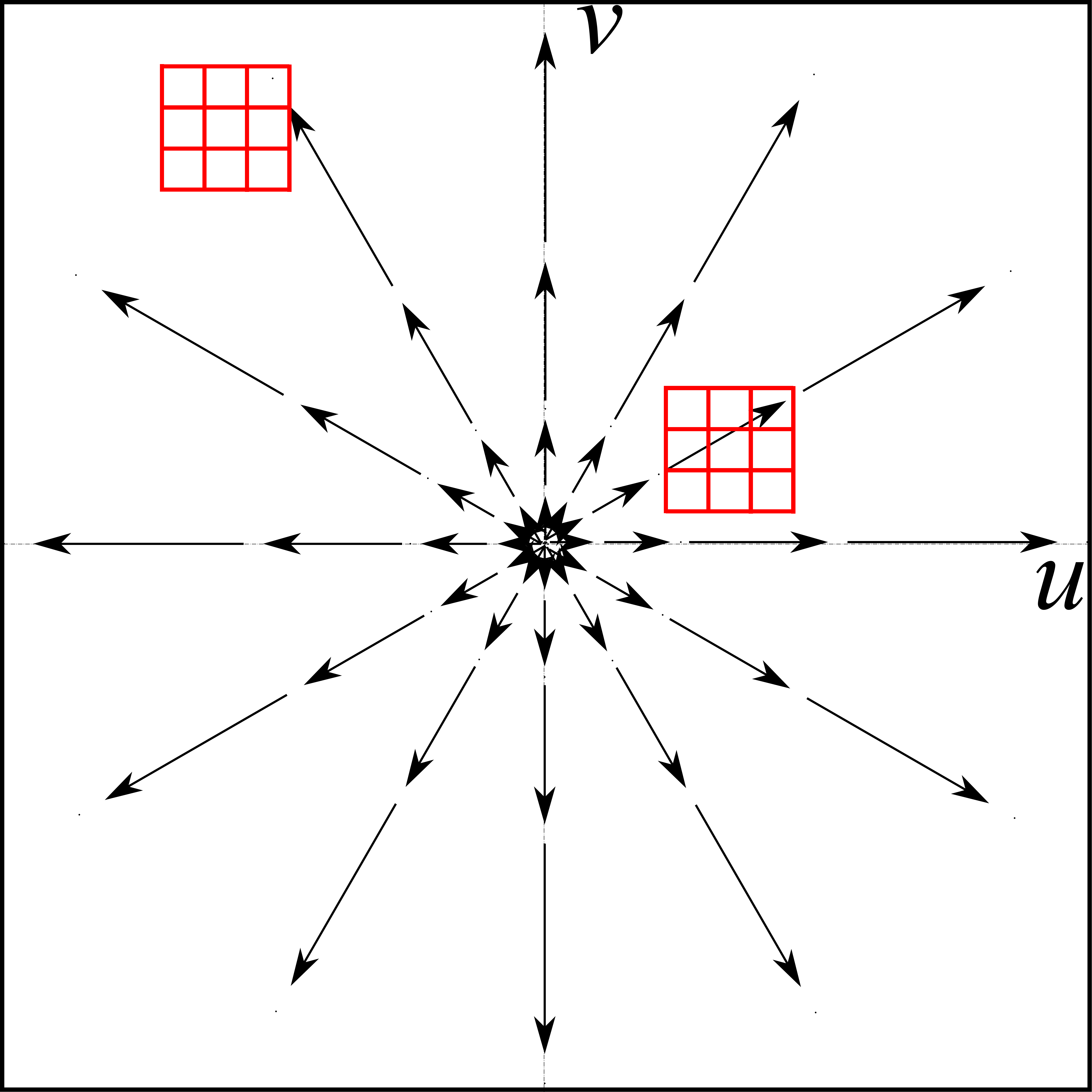}
\label{subfig:cartisianPerspectiveDeformation}
}
\end{minipage}
\hspace{5pt}
\begin{minipage}[t]{0.31\linewidth}
\subfigure[Polar]{
\includegraphics[width=\linewidth]{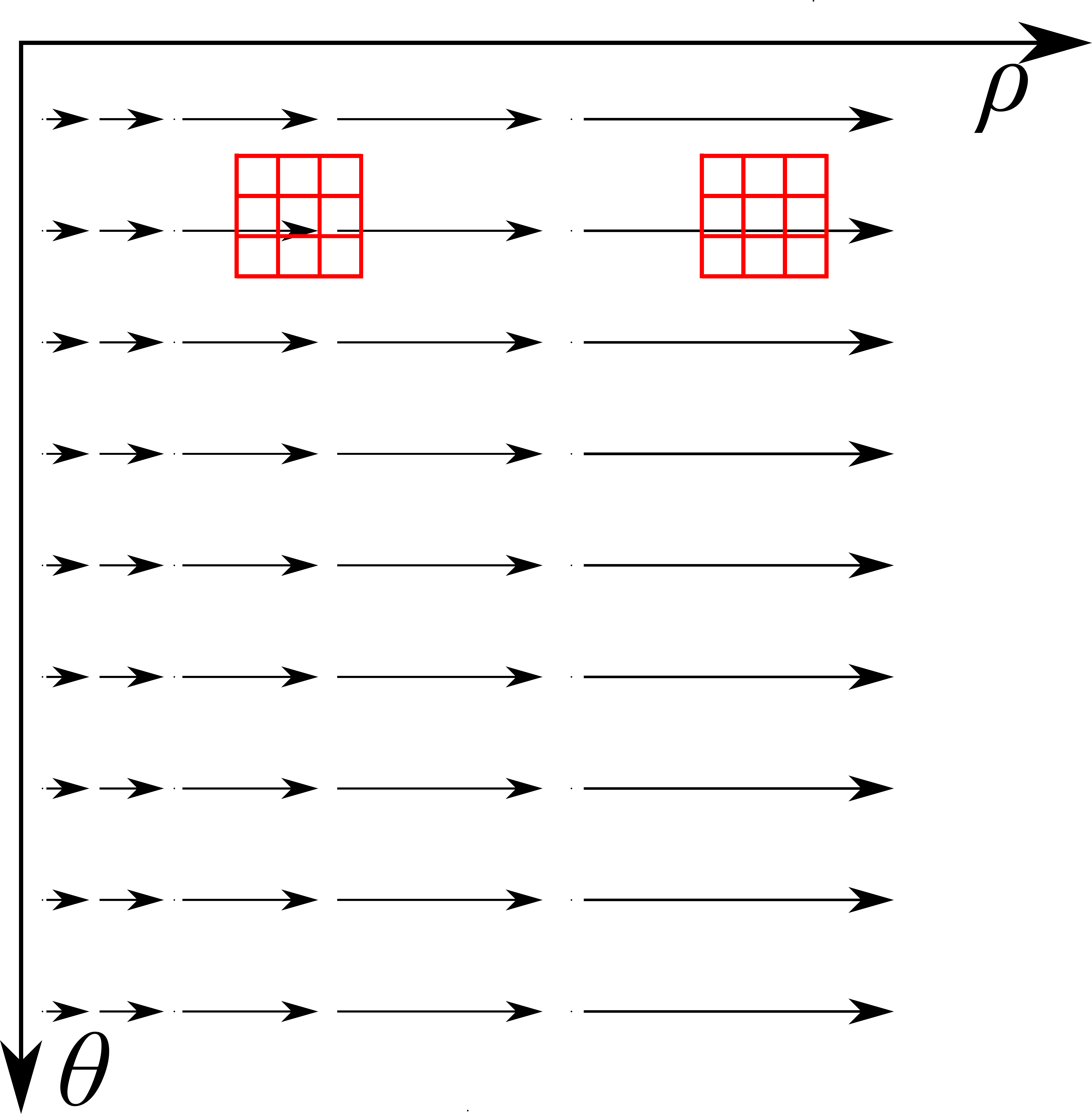}
\label{subfig:polarPerspectiveDeformation}
}
\end{minipage}
\caption{Properties of perspective deformation in Cartesian and polar coordinate systems, given a fixed depth ($m>1$ in the illustration as an example). The orientation and the length of an arrow reflect the direction and the magnitude of perspective deformation at the corresponding position. The red 3\,$\times$\,3 grids represent 3\,$\times$\,3 convolutional kernels. }
\label{Fig:spaceDifference}
\end{figure}

When the CBCT system has the orientation and position as displayed in Fig.\,\ref{subfig:virtualDetector},
\begin{equation}
\begin{array}{l}
\hat{\va} = 
\begin{bmatrix}
x \cdot D_\text{si}/s + p_u\cdot ( D_\text{si} + z)\\
y \cdot D_\text{si}/s + p_v\cdot ( D_\text{si} + z)\\
 z + D_\text{si} 
\end{bmatrix}
=
\begin{bmatrix}
\frac{D_\text{si}}{D_\text{si} + z} \frac{x}{s} + p_u\\
\frac{D_\text{si}}{D_\text{si} + z} \frac{y}{s} + p_v\\
 1\\ 
\end{bmatrix},\\
 \bar{\va} = 
 \begin{bmatrix}
 \frac{x}{s} + p_u\\
\frac{y}{s} + p_v\\
 1\\ 
\end{bmatrix}.
\label{eqn:0viewProjections}
\end{array}
\end{equation}
Given the principal point on the detector $\boldsymbol{O}_{\text{det}} = [p_u, p_v, 1]^\top$, it is clear that $\hat{\va}$, $\bar{\va}$ and $\boldsymbol{O}_{\text{det}}$ are collinear, which indicates that perspective deformation occurs along radial directions as described in Fig.\,\ref{subfig:cartisianPerspectiveDeformation}. And the amount of perspective deformation is,
\begin{equation}
d_\text{PD} = |m-1|\sqrt{x^2 + y^2}/s= |m-1|\rho_0,
\label{eqn:magnification}
\end{equation}
where $m = D_{\text{si}}/( D_{\text{si}} + z)$ is the magnification factor and $\rho_0$ is the distance of $\bar{\va}$ to the principle point $\boldsymbol{O}_{\text{det}}$. It tells us that the further a point is away from the principal ray, the larger its perspective deformation magnitude is, given a fixed depth $z$. This position dependency is reflected by the different lengths and orientations of arrows in Fig.\,\ref{subfig:cartisianPerspectiveDeformation}. 

\subsection{Key Issues}
\label{subsect:KeyIssues}
 To learn perspective deformation, two key issues need to be addressed: view design and generative model choice.

 \textit{View design:} As perspective projection matrices are underdetermined, learning perspective deformation from a single view leads to nonunique solutions. Therefore, an auxiliary view is necessary. In practice, an orthogonal view \cite{melhem2016eos,chow2020robust,ying2019x2ct} is commonly used to provide depth information. Such depth information can be estimated by human experts with their domain knowledge and experience. Instead, two complementary ($180^\circ$) views look very similar to humans and hence are rarely used in practice. In this work, the following two questions will be addressed: \textit{Is an additional auxiliary view beneficial as well for a neural network to learn perspective deformation? Is an orthogonal view better than a complementary view as well for a neural network to learn perspective deformation?}
 
\textit{Generative model choice:} Converting perspective projection images into orthogonal projection images is an image-to-image generation task. Hence, various state-of-the-art generative models can be applied, for example, FCNs \cite{isola2017image} and transformer-based networks \cite{dosovitskiy2020image,chen2021transunet}. As described in Subsection \ref{subsect:PDinCBCT}, perspective deformation is position dependent. \textit{Are FCNs and transformers capable of learning such position dependent features?} Since radial deformation in Cartesian space becomes translational in polar space as displayed in Fig.\,\ref{subfig:polarPerspectiveDeformation}, \textit{is it better for FCNs or transformers to learn perspective deformation in polar space than Cartesian space?} Learning perspective deformation typically requires considering long-range dependencies. \textit{Are FCNs with limited perceptive fields able to learn it? Will transformers perform better than FCNs in this task?} There is always a gap between research experimental data and real-world data. \textit{Are the models sensitive to geometric system errors? Will the models generalize well in real data?} The above questions will be explored in this work.

\subsection{Framework for Learning Perspective Deformation}
\label{subsect:framework}
\begin{figure}[t]
\centering
\includegraphics[width=\linewidth]{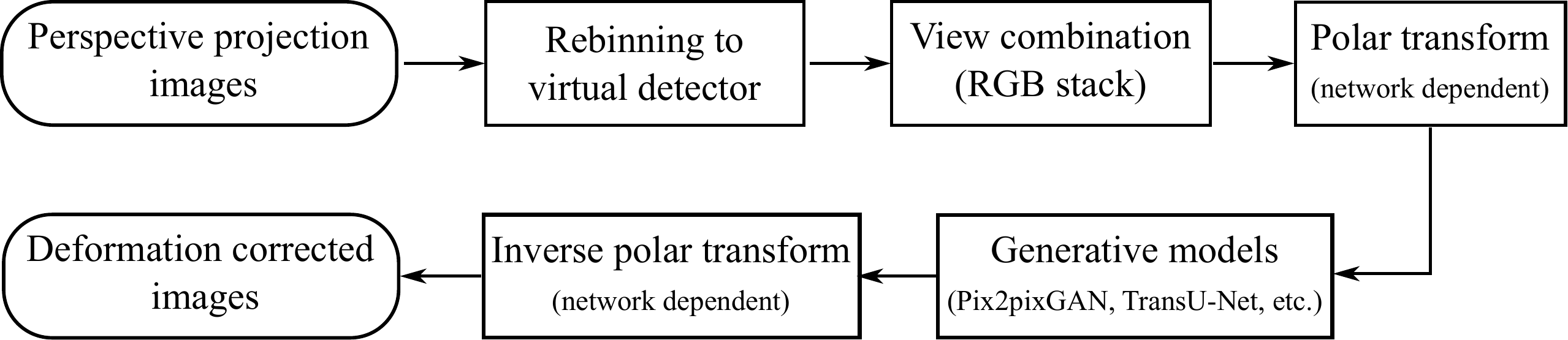}
\caption{\modified{The pipeline for learning perspective deformation.}}
\label{Fig:pipeline}
\end{figure}
To address the key issues described above, a framework for learning perspective deformation is proposed in Fig.\,\ref{Fig:pipeline}, which is explained in detail in the following.

\section{View Design}
\subsection{Geometric Modelling}

\subsubsection{Perspective deformation learning via a single view}

\modified{Perspective projection matrices are underdetermined. As a result, the same perspective projection point $\hat{\va}$ corresponds to nonunique points in the imaged object, which are located at its back-projection line, i.e., 
\begin{equation}
\vl = \hat{\vP}^{\dagger} \hat{\va},
\end{equation}
where $\dagger$ stands for pseudo inverse.}
For example, in Fig.\,\ref{subfig:singleView}, the black point can be located anywhere along the red solid ray, if no depth information is available. Therefore, all the points along the red dashed line, including the blue point, are potential orthogonal projection candidates of the red point. \modified{Here the red dashed line is the orthogonal projection of its back-projection (OPBP) on the virtual detector,
\begin{equation}
\vl_{\text{OPBP}} = \bar{\vP}\hat{\vP}^{\dagger} \hat{\va}.
\end{equation}  
}
As a consequence, it is very challenging to learn perspective deformation from one single view. Therefore, a second view is necessary. 

\subsubsection{Perspective deformation learning via dual orthogonal views}

In practice, biplanar X-ray systems \cite{melhem2016eos,chow2020robust,ying2019x2ct} are widely used in interventional surgery for depth estimation. In a biplanar system, an additional orthogonal view is utilized, as illustrated in Fig.\,\ref{subfig:orthoViewBPOP}. 
Given the updated rotation matrix $\vR$ with $90^\circ$ rotation along the Y-axis, $\vR = [0, 0, -1;0, 1, 0;1, 0,0]$,
the perspective projection matrix $\hat{\vP}_{90^\circ}$ for the orthogonal view is obtained. Correspondingly, the perspective projection of $\va$ is,
\begin{equation}
\hat{\va}_{90^\circ} = \hat{\vP}_{90^\circ}\va = 
\begin{bmatrix}
\frac{D_\text{si}}{D_\text{si} + x} \frac{-z}{s} + p_u\\
\frac{D_\text{si}}{D_\text{si} + x} \frac{y}{s} + p_v\\
 1\\ 
\end{bmatrix}.
\end{equation}

Hence, the point-to-point distance from two orthogonal views in the same detector pixel coordinates can be calculated as the following,
\begin{equation}
d_{90^\circ} = ||\hat{\va}_{90^\circ} - \hat{\va}||.
\label{eq:d90}
\end{equation}

Ideally, two views of the same point of interest can determine its 3D position, which is the intersection point of the two corresponding rays. For example, the black point in Fig.\,\ref{subfig:orthoViewBPOP} is the intersection point of the red and green solid lines. 
In the $0^\circ$-view virtual detector, ideally the orthogonal projection of a point can be determined as well by the intersection point of two OPBP lines. For example, in Fig.\,\ref{subfig:orthoViewBPOP} the target blue point is the intersection point of the red and green dashed lines, where the green dashed line is defined as,
\begin{equation}
\vl_{\text{OPBP},90^\circ} = \bar{\vP}\hat{\vP}^{\dagger}_{90^\circ} \hat{\va}_{90^\circ}.
\end{equation} 
Alternatively, the target blue point is determined by moving the red point along the radial direction to the green dashed line.
Such ideal cases are based on the assumption that the point-to-point correspondence from two views is determined. However, for different points $\va$ in the imaged object, $d_{90^\circ}$ has a large variance as it depends not only on the depth $z$ in the $0^\circ$ view but also the depth $x$ in the $90^\circ$ view. As a consequence, such correspondence is very challenging to determine.

\begin{figure}[t]
\centering

\begin{minipage}[t]{0.31\linewidth}
\subfigure[3D bead phantom]{
\includegraphics[width=\linewidth]{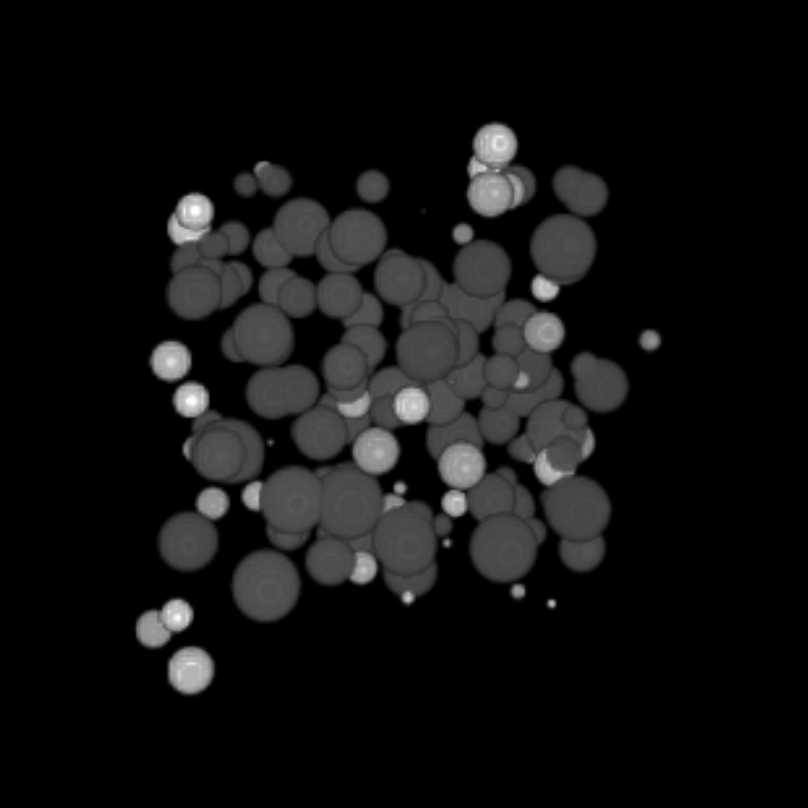}
\label{subfig:3DBeadPhantom}
}
\end{minipage}
\begin{minipage}[t]{0.31\linewidth}
\subfigure[Perspective $0^\circ$]{
\includegraphics[width=\linewidth]{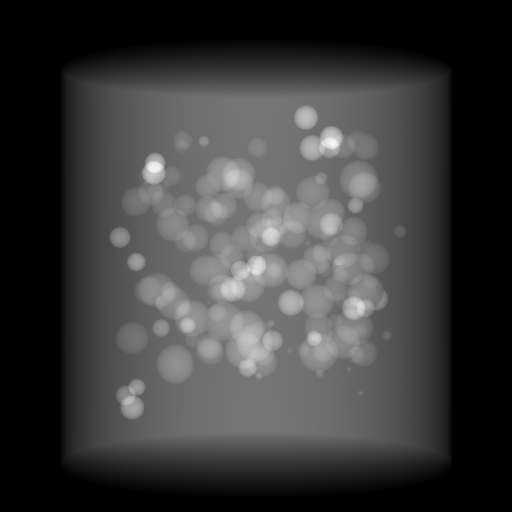}
\label{subfig:perspective0degExample}
}
\end{minipage}
\begin{minipage}[t]{0.31\linewidth}
\subfigure[Perspective $90^\circ$]{
\includegraphics[width=\linewidth]{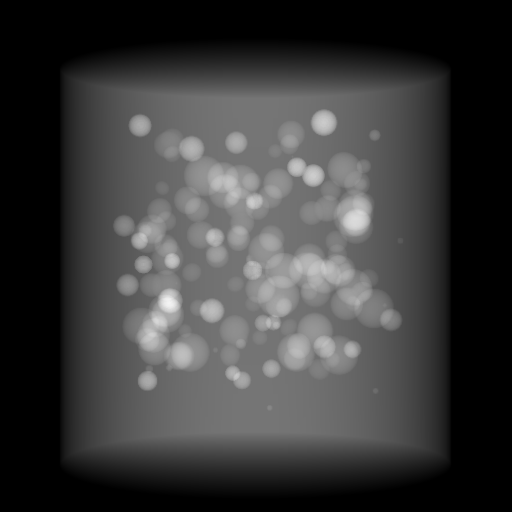}
\label{subfig:perspective90degExample}
}
\end{minipage}
\begin{minipage}[t]{0.31\linewidth}
\subfigure[$0^\circ$ and $90^\circ$]{
\includegraphics[width=\linewidth]{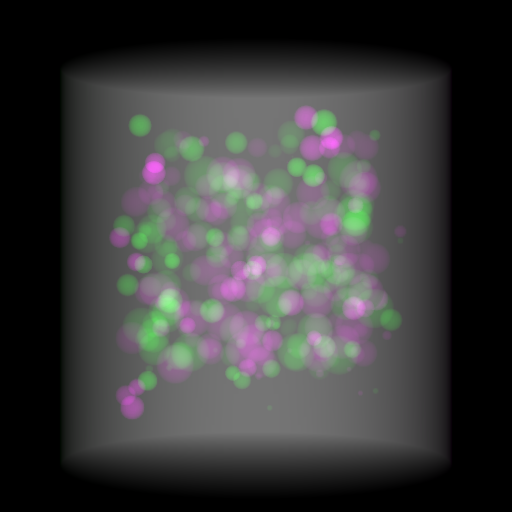}
\label{subfig:perspectiveRGB0and90degExample}
}
\end{minipage}
\begin{minipage}[t]{0.31\linewidth}
\subfigure[$0^\circ$ and $90^\circ$ OPBP]{
\includegraphics[width=\linewidth]{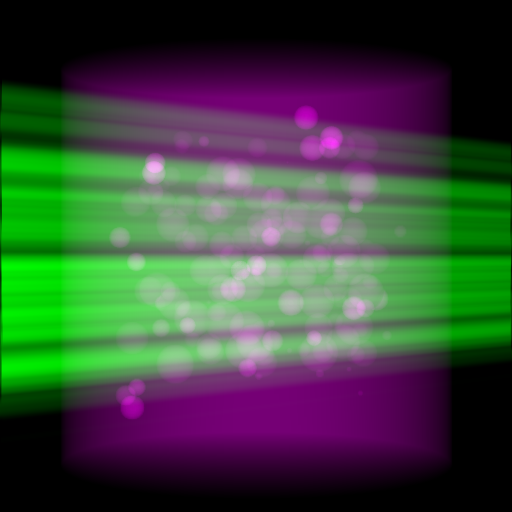}
\label{subfig:OPBPData1}
}
\end{minipage}
\begin{minipage}[t]{0.31\linewidth}
\ 
\end{minipage}
\caption{Perspective projection images from two orthogonal views for a 3D bead phantom as an example: (a) a 3D bead phantom where the background cylinder is omitted for better visualization of the beads; (b) the perspective projection of the 3D bead phantom from the $0^\circ$ view; (c) the perspective projection of the 3D bead phantom from the $90^\circ$ view; (d) the RGB stack of the $0^\circ$ and $90^\circ$ perspective projection images; (e) the RGB stack of the $0^\circ$ and $90^\circ$ OPBP images. The intensity range [0, 11] in projection images is converted to [0, 255] for visualization.}
\label{Fig:orthogonalBeadPhantomExmaples}
\end{figure}

As an illustrative example, the two perspective projection images from two orthogonal views for a 3D bead phantom are displayed in Fig.\,\ref{Fig:orthogonalBeadPhantomExmaples}. 
To better compare the perspective projection images from the two orthogonal views, an RGB image is formed in Fig.\,\ref{subfig:perspectiveRGB0and90degExample}, where the red and blue channels use images from the $0^\circ$ perspective view, while the green channel uses images from the $90^\circ$ perspective view. Here we fill up three channels so that pixels, which have similar intensity from both views, appear grey such as the background cylinder area. In the formed RGB image, the magenta beads from the $0^\circ$ view and the green beads from $90^\circ$ view are located in different positions. 
\modified{Similarly, an RGB image consisting of the $0^\circ$ perspective projection image (red and blue channels) and the $90^\circ$ OPBP image (green channel) is displayed in Fig.\,\ref{subfig:OPBPData1}.
In Fig.\,\ref{subfig:perspectiveRGB0and90degExample} the bead-to-bead (or point-to-point) correspondence between two views is not straightforward because of the large number of beads. It is the same for the bead-to-stripe (point-to-line) correspondence in Fig.\,\ref{subfig:OPBPData1}. Therefore, it is challenging for neural networks to learn perspective deformation from two orthogonal views.}

\subsubsection{Perspective deformation learning via dual complementary views}
\modified{In contrast to practice}, an additional complementary ($180^\circ$) view (Fig.\,\ref{subfig:complementaryViews}) is proposed in this work for learning perspective deformation. 
Given the updated rotation matrix $\vR$ with $180^\circ$ rotation along the Y-axis, $\vR = [-1, 0, 0; 0, 1, 0; 0, 0, -1]$,
the perspective projection matrix $\hat{\vP}_{180^\circ}$ for the complementary view is obtained. Correspondingly, the perspective projection of $\va$ is,
\begin{equation}
\hat{\va}_{180^\circ} = \hat{\vP}_{180^\circ}\va = 
\begin{bmatrix}
\frac{D_\text{si}}{D_\text{si} -z} \frac{-x}{s} + p_u\\
\frac{D_\text{si}}{D_\text{si} -z} \frac{y}{s} + p_v\\
 1\\ 
\end{bmatrix}.
\end{equation}
Compared with $\hat{\va}$ in Eqn.\,(\ref{eqn:0viewProjections}), only the signs of $x$ and $z$ are changed. Here $m'=D_\text{si}/(D_\text{si} -z)$ is the magnification factor from the $180^\circ$ view. 
 In Fig.\,\ref{subfig:complementaryViews}, it is easily observed that the OPBP line of $\hat{\va}_{180^\circ}$ is identical to that of $\va$,
\begin{equation}
\vl_{\text{OPBP},180^\circ} = \bar{\vP}\hat{\vP}^{\dagger}_{180^\circ} \hat{\va}_{180^\circ} = \vl_{\text{OPBP}},
\end{equation}
 as the blue, red  and green points are collinearly located in $\vl_{\text{OPBP}}$.

By horizontally flipping the perspective projection of the complementary view with respect to the principal point, the $180^\circ$ view perspective projection image is switched to $0^\circ$ view detector pixel coordinates, i.e.,
\begin{equation}
\hat{\va}'_{180^\circ} = 
\begin{bmatrix}
\frac{D_\text{si}}{D_\text{si} -z} \frac{x}{s} + p_u\\
\frac{D_\text{si}}{D_\text{si} -z} \frac{y}{s} + p_v\\
 1\\ 
\end{bmatrix}.
\end{equation}
The point-to-point distance from two complementary views can be calculated as the following,
\begin{equation}
d_{180^\circ} = ||\hat{\va}'_{180^\circ} - \hat{\va}||=|m'-m|\rho_0.
\label{eqn:d180}
\end{equation}

A complementary view in parallel-beam X-ray imaging is fully redundant after flipping. However, in cone-beam X-ray imaging, because of the cone-angle, \modified{the complementary view provides additional beneficial information about the position of a target point. More specifically, with the constraint $1/m + 1/m' = 2$, one of $m$ and $m'$ is $\geq$ 1 and the other is $\leq$ 1. Because of this, $\bar{\va}$ is located between $\hat{\va}$ and $\hat{\va}'_{180^\circ}$. Hence,} a complementary view straightforwardly provides an interval where the orthogonal projection of the target point should be located. For example, in Fig.\,\ref{subfig:complementaryViews} the blue point is located between the red point and the green point. \modified{Therefore, the relative position of $\bar{\va}$ in the interval of $\hat{\va}$ and $\hat{\va}'_{180^\circ}$ can be determined by the following ratio $\alpha$,
\begin{equation}
\alpha = d_\text{PD}/d_{180^\circ}.
\label{eqn:alphaDefinition}
\end{equation}}

\begin{figure}[t]
\centering

\begin{minipage}[t]{0.31\linewidth}
\subfigure[Orthogonal $0^\circ$]{
\includegraphics[width=\linewidth]{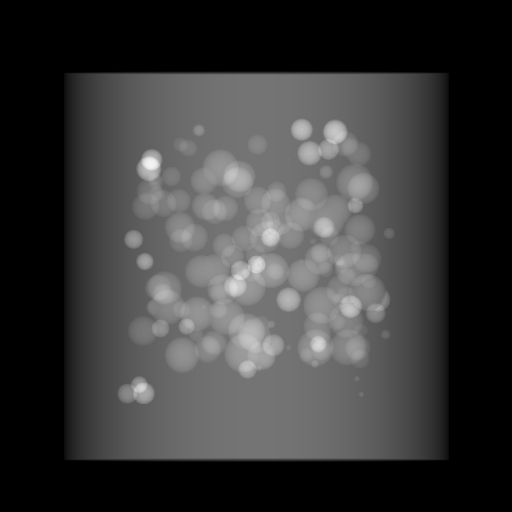}
\label{subfig:orthogonalExample}
}
\end{minipage}
\begin{minipage}[t]{0.31\linewidth}
\subfigure[Perspective $0^\circ$]{
\includegraphics[width=\linewidth]{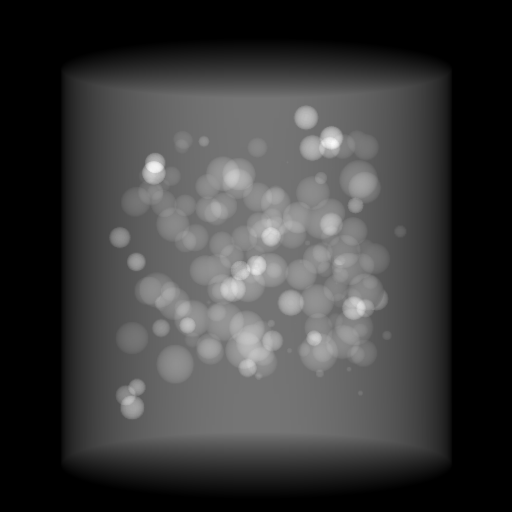}
\label{subfig:perspective0degExample}
}
\end{minipage}
\begin{minipage}[t]{0.31\linewidth}
\subfigure[\modified{Perspective deformation ((b)-(a))}]{
\includegraphics[width=\linewidth]{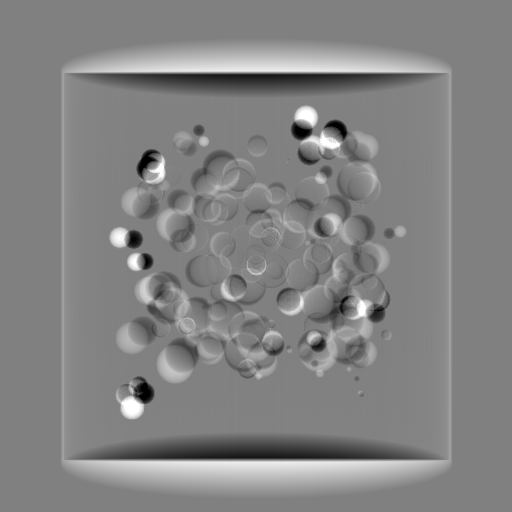}
\label{subfig:diffPersSubOrthoExample}
}
\end{minipage}
\begin{minipage}[t]{0.31\linewidth}
\subfigure[Perspective $180^\circ$]{
\includegraphics[width=\linewidth]{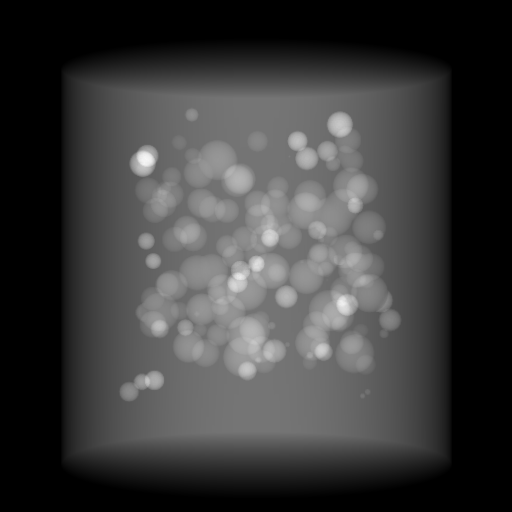}
\label{subfig:perspective180degExample}
}
\end{minipage}
\begin{minipage}[t]{0.31\linewidth}
\subfigure[Difference of Perspective $0^\circ$  and $180^\circ$]{
\includegraphics[width=\linewidth]{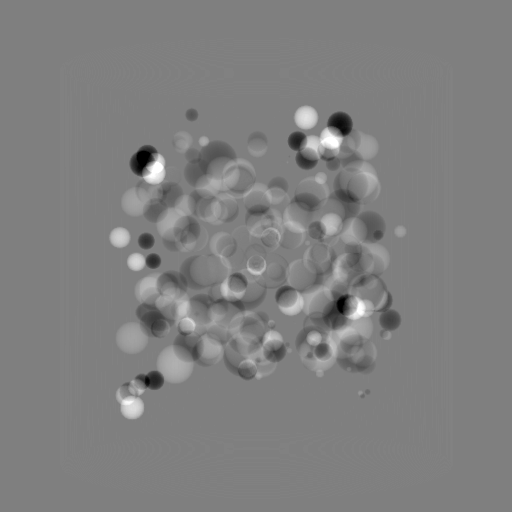}
\label{subfig:diffPers0Sub180Example}
}
\end{minipage}
\begin{minipage}[t]{0.31\linewidth}
\subfigure[RGB $0^\circ$ and $180^\circ$]{
\includegraphics[width=\linewidth]{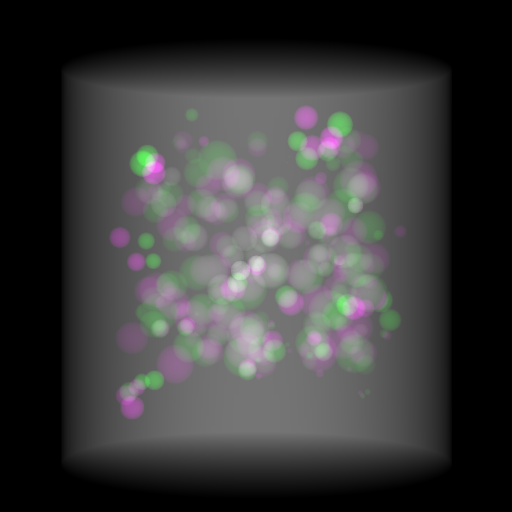}
\label{subfig:perspectiveRGB0and180degExample}
}
\end{minipage}

\caption{An example for perspective deformation learning from dual complementary views: (a) the orthogonal projection of the 3D bead phantom from the $0^\circ$ view; (b) the perspective projection from the $0^\circ$ view; \modified{(c) the difference image between (b) and (a);} (d) the perspective projection from the $180^\circ$ view; (e) the difference image between (d) and (b); (f) the RGB stack of the $0^\circ$ and $180^\circ$ perspective projection images, where the magenta and green areas correspond to the positive (bright) and negative (dark) areas in (e). The intensity range [0, 11] is converted to [0, 255] for visualization.}
\label{Fig:BeadPhantom2View}
\end{figure}

\modified{Note that when and only when $z=0$, which is $m'=m=1$, $d_{180^\circ}$ and $d_\text{PD}$ both equal to 0. This provides a practical way to observe which structures suffer from perspective deformation by assessing deviations between the two complementary views.}

\modified{The orthogonal and perspective projection images of the same 3D bead phantom in a complementary view setting are displayed in Fig.\,\ref{Fig:BeadPhantom2View}.}
Fig.\,\ref{subfig:diffPersSubOrthoExample} shows the perspective deformation, which is the difference between the perspective projection (Fig.\,\ref{subfig:perspective0degExample}) and the orthogonal projection (Fig.\,\ref{subfig:orthogonalExample}) from the $0^\circ$ view. \modified{Fig.\,\ref{subfig:diffPersSubOrthoExample} clearly demonstrates that the magnitude of perspective deformation increases from the center outwards radially.}
The $180^\circ$ perspective projection image is displayed in Fig.\,\ref{subfig:perspective180degExample} and its difference with respect to the $0^\circ$ perspective projection is displayed in Fig.\,\ref{subfig:diffPers0Sub180Example}. Fig.\,\ref{subfig:diffPers0Sub180Example} is similar to Fig.\,\ref{subfig:diffPersSubOrthoExample} in bead areas, \modified{which illustrates that the deviation between two complementary views has strong correlation to perspective deformation}. To integrate such dual-view information, like Fig.\,\ref{subfig:perspectiveRGB0and90degExample}, we convert the perspective projections images from the $0^\circ$ and $180^\circ$ views to a 3-channel RGB image in Fig.\,\ref{subfig:perspectiveRGB0and180degExample}. The red and blue channels use images from the $0^\circ$ view, while the green channel uses images from the $180^\circ$ view. 
 In the RGB images, the color reveals the intensity difference between the $0^\circ$ and $180^\circ$ perspective projection images. Grey areas contain close intensity values from both views. Instead, magenta and green areas indicate larger intensity values from the $0^\circ$ and $180^\circ$ views respectively, where perspective deformation correction is necessary. They correspond to the positive (bright) and negative (dark) areas in the difference image in Fig.\,\ref{subfig:diffPers0Sub180Example}.  \modified{In Fig.\,\ref{subfig:perspectiveRGB0and180degExample}, the magenta beads and their corresponding green beads are located close to each other, which allows a network with limited receptive field size to capture bead-to-bead (point-to-point) dependency.}

\subsection{\modified{Uncertainty Analysis}}

\begin{figure}
\centering
\begin{minipage}[t]{0.485\linewidth}
\subfigure[]{
\includegraphics[width=\linewidth]{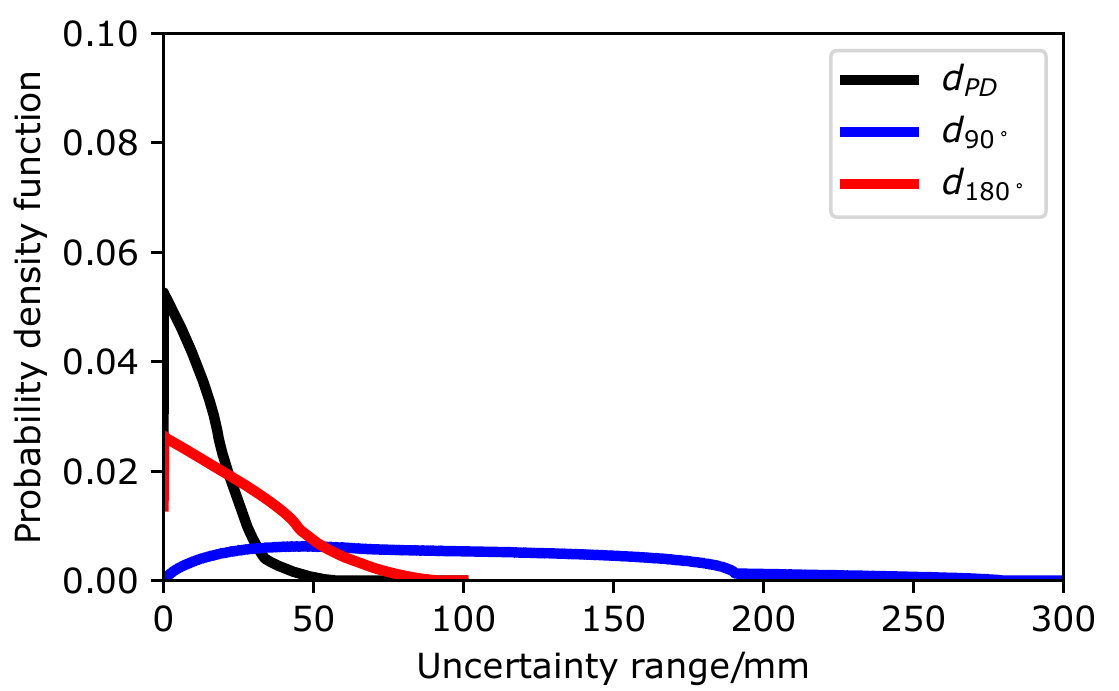}
\label{subfig:absoluteDistanceComplementary}
}
\end{minipage}
\begin{minipage}[t]{0.485\linewidth}
\subfigure[]{
\includegraphics[width=\linewidth]{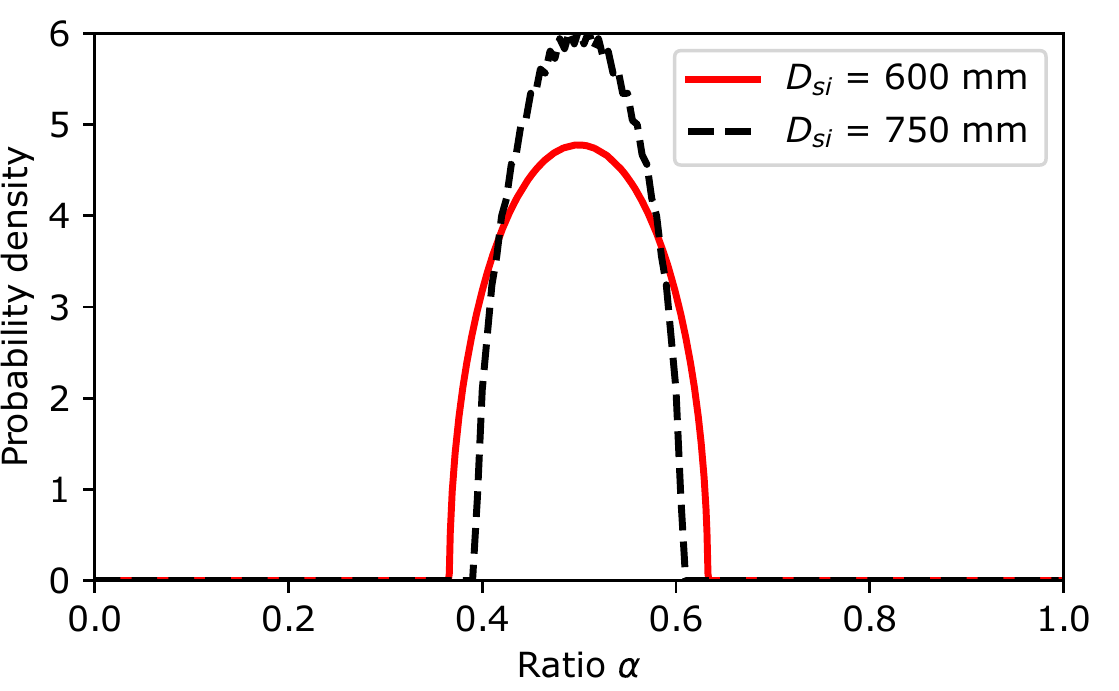}
\label{subfig:ratioComplementary}
}
\end{minipage}
\caption{\modified{Numerical uncertainty analysis for different view settings: (a) The distributions of point-to-point distances when $D_\text{si} = 600$\,mm; (b) the distribution of ratio $\alpha$ in a complementary view setting for $D_\text{si} = 600$\,mm and $D_\text{si} = 750$\,mm.}}
\label{Fig:uncertaintyAnalysis}
\end{figure}
The distributions of point-to-point distances $d_\text{PD}$, $d_{90^\circ}$, and $d_{180^\circ}$ for an exemplary cylinder object (diameter 320\,mm, height 320\,mm) in a CBCT system with $D_\text{si} = 600\,$mm are numerically analyzed in Fig.\,\ref{subfig:absoluteDistanceComplementary}. The distances are multiplied by the detector pixel spacing and displayed in physical units mm. Fig.\,\ref{subfig:absoluteDistanceComplementary} shows that the perspective deformation $d_\text{PD}$ is mainly distributed in a range of [0, 55]\,mm, which indicates the minimum size of receptive fields for neural networks to learn perspective deformation. For orthogonal views, $d_{90^\circ}$ is distributed in a wide range of [0, 280]\,mm, which requires long-distance dependency and imposes large uncertainty for neural networks to learn perspective deformation. In striking contrast, $d_{180^\circ}$ is mainly distributed in a range of [0, 90]\,mm, within the double distance of $d_\text{PD}$. Moreover, the distribution of $\alpha$ is displayed in Fig.\,\ref{subfig:ratioComplementary}. When $D_\text{si} = 600\,$mm, $\alpha$ is distributed in a narrow range of [0.364, 0.636]. When $D_\text{si}$ increases to 750\,mm, $\alpha$ is distributed in a narrower range of [0.395, 0.605]. The distributions of $\alpha$ indicates that $\bar{\va}$ is located near the middle point of $\hat{\va}$ and $\hat{\va}'_{180^\circ}$, which drastically reduces uncertainty. Due to the relatively short-distance dependency and low uncertainty, learning perspective deformation with complementary views is more effective for neural networks than orthogonal views or a single view.

\section{Generative Model Choice}
\subsection{Neural Networks}

Various state-of-the-art as well as newly emerging generative neural networks can be plugged into our framework in Fig.\,\ref{Fig:pipeline} to learn perspective deformation. As described in Subsection \ref{subsect:PDinCBCT}, perspective deformation is position dependent. Such position dependency can be learned by transformer-based networks, e.g., the vision transformer (ViT), where a position embedding is typically used to extract position-dependent features. FCNs are effective in learning translational features, but are suboptimal, or even incapable of learning rotational features in general. Therefore, for FCNs, learning perspective deformation in polar space is expected to be advantageous over learning directly in Cartesian space, since radial deformation in Cartesian space becomes translational in polar space as displayed in Fig.\,\ref{subfig:polarPerspectiveDeformation}.

In this work, two representative neural networks are evaluated as demonstration examples for the proposed framework: Pix2pixGAN \cite{isola2017image} as a representative of FCNs and TransU-Net \cite{chen2021transunet} as a representative of transformers.

 Pix2pixGAN uses a standard U-Net as its generator and a 5-layer CNN as its discriminator. It is trained with the adversarial loss and the $\ell_1$ loss together with the perceptual loss using the pretrained VGG-16 model.
 
For tasks that require long-range dependencies, FCNs exhibit limited performance due to the intrinsic locality of convolution operations. Transformers are capable of learning long-range relationships and transferring to downstream tasks under large-scale pre-training. Vision transformer (ViT) \cite{dosovitskiy2020image}, the pioneer network from language to vision, and its variants \cite{chen2021transunet,liu2021swin,chu2021twins} have demonstrated success in various computer vision tasks. However, ViT-based transformers have the limitation of low resolution in computer vision tasks which require dense inference \cite{chen2021transunet,liu2021swin}, e.g., semantic segmentation and image synthesis. This is because transformers treat image patches as 1D token sequences and exclusively focus on modeling global contexts, which result in low-resolution features \cite{chen2021transunet}. Such transformers are sufficient for coarse-scale image classification and segmentation tasks, but not for image synthesis (a regression task) in this work. TransU-Net is a hybrid CNN-Transformer network built upon ViT, which leverages both high-resolution spatial information from CNN features and global context with long-range dependencies from transformers \cite{chen2021transunet}. In this work, the encoder of TransU-Net uses ResNet50 + ViT-B/16 configuration. The decoder is the expansive path of a regular U-Net.

\subsection{Proof-of-Concept Experiment}
\label{subsect:ProofBead}
\begin{figure}[t]
\centering
\begin{minipage}[t]{0.32\linewidth}
\subfigure[Reference, Cartesian]{
\includegraphics[width=\linewidth]{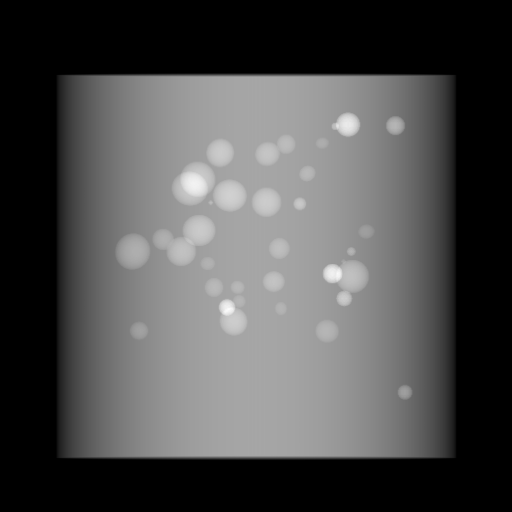}
\label{subfig:target22}
}
\end{minipage}
\begin{minipage}[t]{0.32\linewidth}
\subfigure[Reference, polar]{
\includegraphics[width=\linewidth]{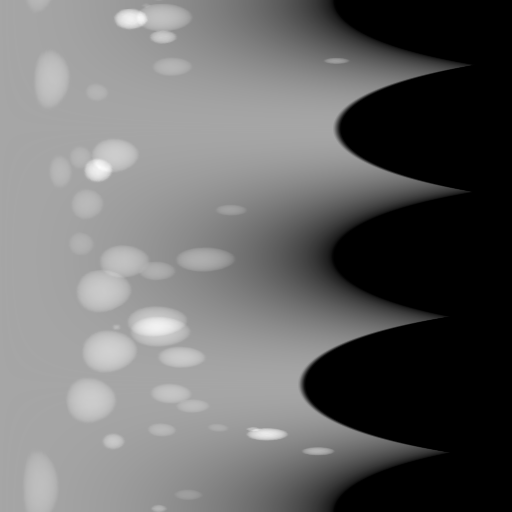}
\label{subfig:TargetPolar}
}
\end{minipage}
\begin{minipage}[t]{0.32\linewidth}
\subfigure[$0^\circ$ and $180^\circ$ perspective projection, polar]{
\includegraphics[width=\linewidth]{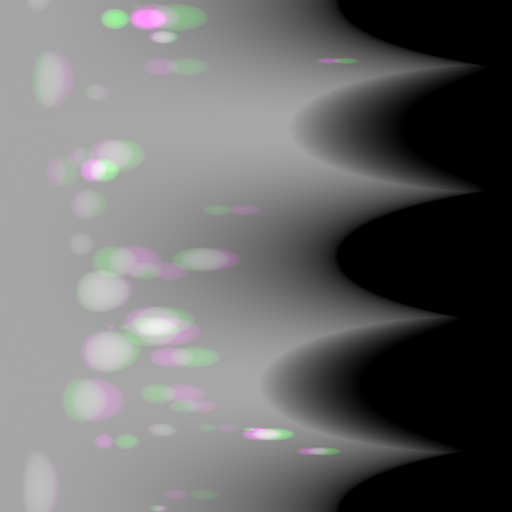}
\label{subfig:dual80InputPolar}
}
\end{minipage}
\caption{\modified{The reference image in Cartesian and polar coordinates and its perspective projection images in polar coordinates in the complementary view.} }
\label{Fig:DataIInputs}
\end{figure}
\begin{table*}
\caption{Quantitative evaluation of different methods on the bead phantom data.}
\label{Tab:OverallComparisonBead}
\centering
\begin{tabular}{|l|l|l|c|c|c|c|c|c|}
\hline
Network &Metric & Space & $0^\circ$ & $0^\circ \& 90^\circ$ naive & $0^\circ \& 90^\circ$ OPBP & $0^\circ \& 180^\circ$ & $0^\circ \& 180^\circ +$ & $0^\circ, 90^\circ \& 180^\circ$ \\
\hline
\multirow{4}{*}{Pix2pixGAN} &\multirow{2}{*}{RMSE} & Cartesian & 8.45 &8.44 &8.09 & 5.31 &3.28 & 5.11\\
& & Polar & 4.68 &4.20 &3.87 & \modified{\textbf{1.40$\pm$0.29}} & 1.76 & 1.29\\
\cline{2-9}
& \multirow{2}{*}{SSIM} & Cartesian & 0.8794 &0.8582 &0.8443  & 0.9378 &0.9614 & 0.8835\\
& & Polar & 0.9667 &0.9717 & 0.9745 & \modified{\textbf{0.9935$\pm$0.0022}} &0.9906 & 0.9948\\
\hline
\multirow{4}{*}{TransU-Net} &\multirow{2}{*}{RMSE} & Cartesian & 4.78 &4.46 &4.20 & 2.95 &1.63 & 2.27\\
& & Polar & 4.42 &4.10 &3.82 & \textbf{1.73} & \textbf{1.33} & 1.73\\
\cline{2-9}
& \multirow{2}{*}{SSIM} & Cartesian & 0.9722 &0.9749 &0.9767 & 0.9943 &0.9961 & 0.9949\\
& & Polar & 0.9747 & 0.9759 &0.9775  & \textbf{0.9959} &\textbf{0.9960} & 0.9958\\
\hline

\end{tabular}
\end{table*}

\begin{figure}[t]
\centering

\begin{minipage}[t]{0.06\linewidth}
\ 
\end{minipage}
\begin{minipage}[b]{0.3\linewidth}
\centering
\footnotesize{Single view (0$^\circ$)}
\end{minipage}
\begin{minipage}[b]{0.3\linewidth}
\centering
\footnotesize{Dual orthogonal views}
\end{minipage}
\begin{minipage}[b]{0.3\linewidth}
\centering
\footnotesize{Dual complementary views}
\end{minipage}

\begin{minipage}[t]{0.06\linewidth}
\centering
{\rotatebox{90}{\footnotesize{Cartesian}}}
\end{minipage}
\begin{minipage}[b]{0.3\linewidth}
\subfigure[10.32, 0.8768]{
\includegraphics[width=\linewidth]{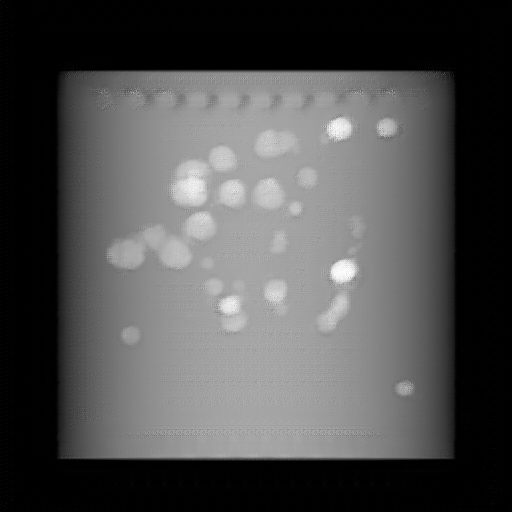}
\label{subfig:singularOutput}
}
\end{minipage}
\begin{minipage}[b]{0.3\linewidth}
\subfigure[6.41, 0.8555]{
\includegraphics[width=\linewidth]{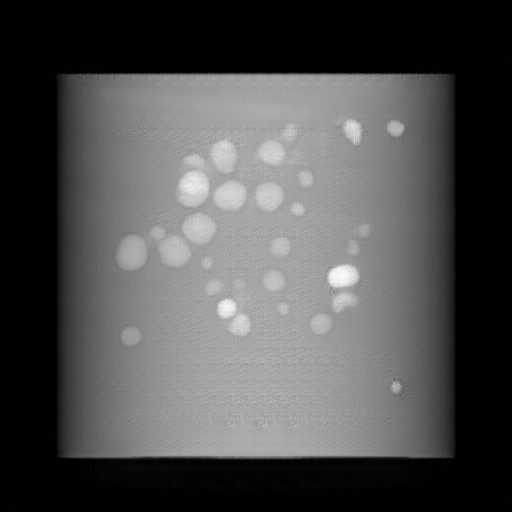}
\label{subfig:dual90OutputOPBP}
}
\end{minipage}
\begin{minipage}[b]{0.3\linewidth}
\subfigure[5.85, 0.9204]{
\includegraphics[width=\linewidth]{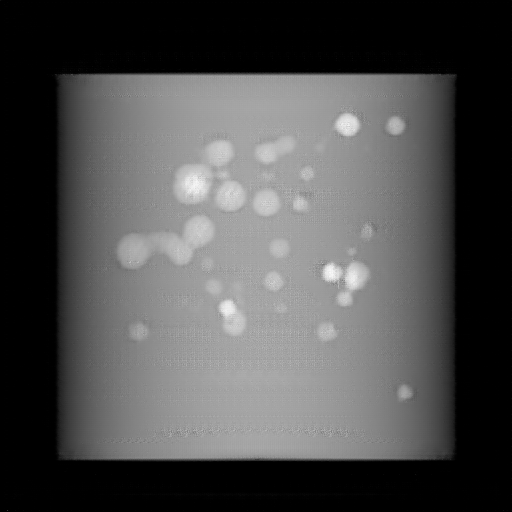}
\label{subfig:dual180Output}
}
\end{minipage}

\begin{minipage}[t]{0.06\linewidth}
\centering
{\rotatebox{90}{\footnotesize{Polar}}}
\end{minipage}
\begin{minipage}[b]{0.3\linewidth}
\subfigure[5.41, 0.9664]{
\includegraphics[width=\linewidth]{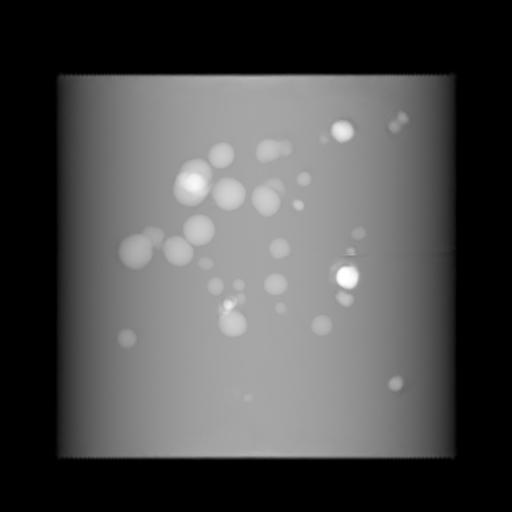}
\label{subfig:singularPolarOutput}
}
\end{minipage}
\begin{minipage}[b]{0.3\linewidth}
\subfigure[3.92, 0.9774]{
\includegraphics[width=\linewidth]{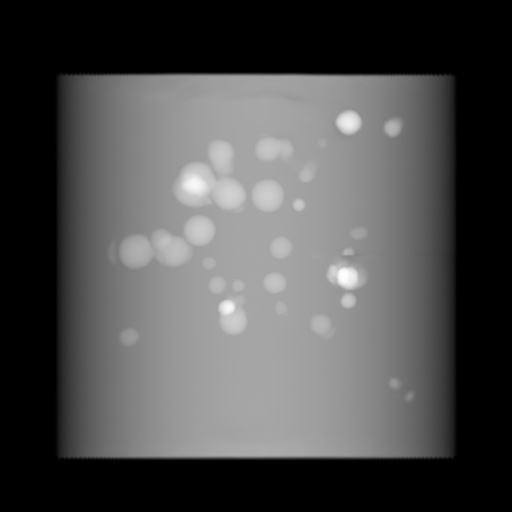}
\label{subfig:dual90PolarOutputOPBP}
}
\end{minipage}
\begin{minipage}[b]{0.3\linewidth}
\subfigure[1.18, 0.9946]{
\includegraphics[width=\linewidth]{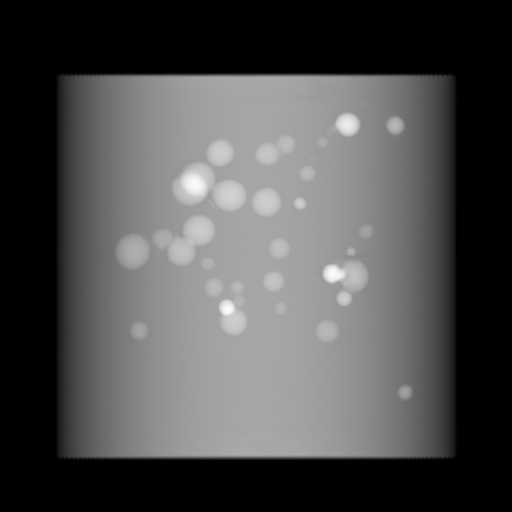}
\label{subfig:dual180PolarOutput}
}
\end{minipage}

\begin{minipage}[t]{0.06\linewidth}
\centering
\rotatebox{90}{\footnotesize{Cartesian error}}
\end{minipage}
\begin{minipage}[b]{0.3\linewidth}
\subfigure[10.32, 0.8768]{
\includegraphics[width=\linewidth]{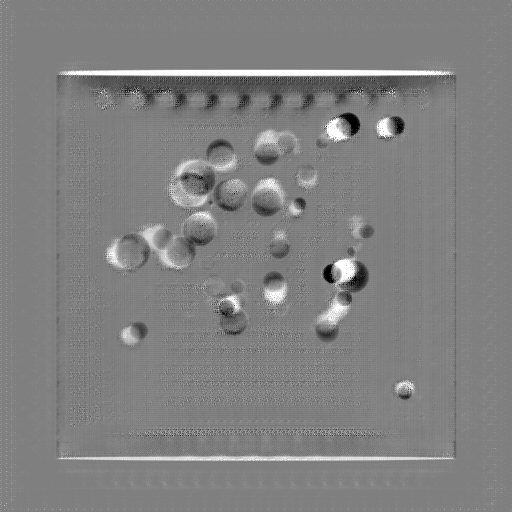}
\label{subfig:singularOutputDiff}
}
\end{minipage}
\begin{minipage}[b]{0.3\linewidth}
\subfigure[6.41, 0.8555]{
\includegraphics[width=\linewidth]{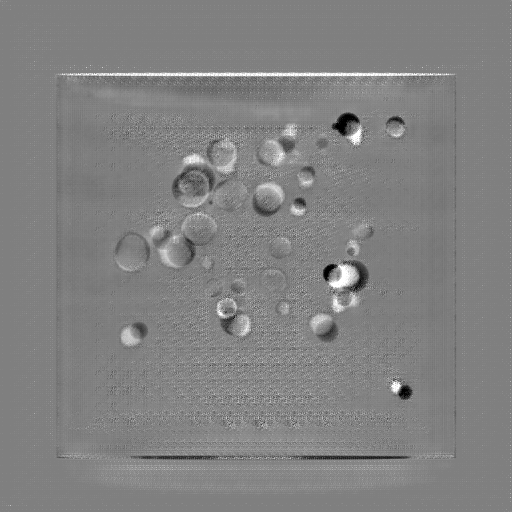}
\label{subfig:dual90OutputDiff}
}
\end{minipage}
\begin{minipage}[b]{0.3\linewidth}
\subfigure[5.85, 0.9204]{
\includegraphics[width=\linewidth]{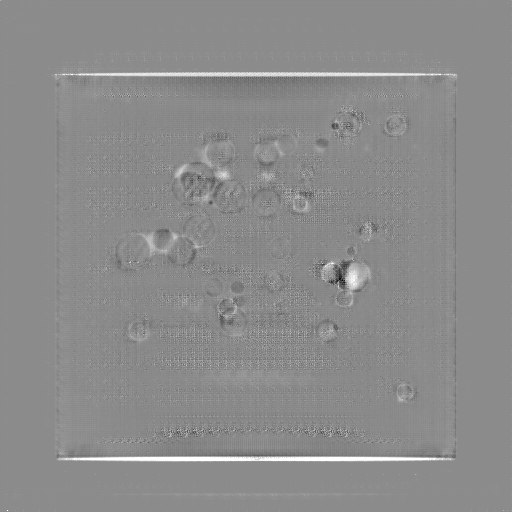}
\label{subfig:dual180OutputDiff}
}
\end{minipage}

\begin{minipage}[t]{0.06\linewidth}
\centering
\rotatebox{90}{\footnotesize{Polar error}}
\end{minipage}
\begin{minipage}[b]{0.3\linewidth}
\subfigure[5.41, 0.9664]{
\includegraphics[width=\linewidth]{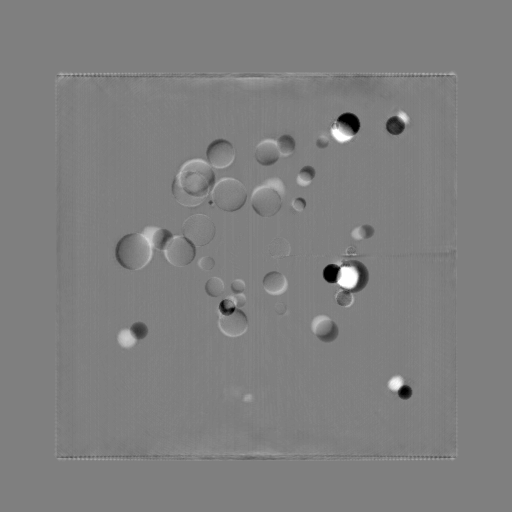}
\label{subfig:singularPolarOutputDiff}
}
\end{minipage}
\begin{minipage}[b]{0.3\linewidth}
\subfigure[3.92, 0.9774]{
\includegraphics[width=\linewidth]{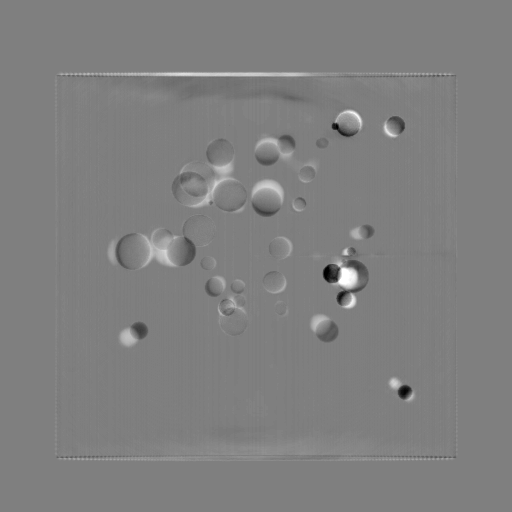}
\label{subfig:dual90PolarOutputDiff}
}
\end{minipage}
\begin{minipage}[b]{0.3\linewidth}
\subfigure[1.18, 0.9946]{
\includegraphics[width=\linewidth]{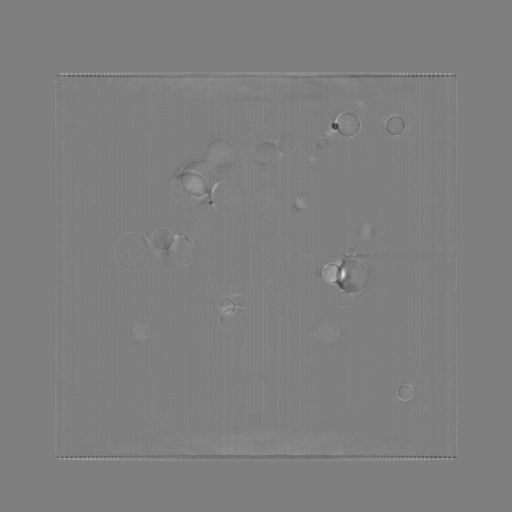}
\label{subfig:dual18PolarOutputDiff}
}
\end{minipage}
\caption{\modified{Predictions of Pix2pixGAN in different spaces with different views. Error images are displayed in a window of [-50, 50].}}
\label{Fig:DataIPredictions}
\end{figure}

The above described methods for learning perspective deformation are investigated on simulated bead phantom data as a proof-of-concept experiment.  One reference image in Cartesian and polar coordinates is displayed in Fig.\,\ref{Fig:DataIInputs}, while its corresponding Pix2pixGAN predictions with different configurations are displayed in Fig.\,\ref{Fig:DataIPredictions}. Fig.\,\ref{subfig:dual80InputPolar} demonstrates the appearance of perspective projection images from complementary views in the polar coordinates, where the network only needs to move a bead horizontally to a position between its corresponding magenta and green pair. To save space, other input images are not displayed. In Fig.\,\ref{Fig:DataIPredictions}, the prediction results are displayed in (a)-(f), while their error images are displayed in (g)-(l). Fig.\,\ref{subfig:singularOutput} and Fig.\,\ref{subfig:singularOutputDiff} use the single $0^\circ$ view. In Fig.\,\ref{subfig:singularOutput}, many beads are distorted, losing their circular shapes. They are blurry as well. Although the background cylinder is restored to have a rectangular shape, aliasing is observed. In Fig.\,\ref{subfig:singularPolarOutput}, the beads have better shapes and less blur than those in Fig.\,\ref{subfig:singularOutput}. However, artifacts occur near certain beads and some tiny beads are still blurry.
The error images in \modified{Fig.\,\ref{subfig:singularOutputDiff} and Fig.\,\ref{subfig:singularPolarOutputDiff}} also show that learning perspective deformation in polar coordinates is better than that in Cartesian coordinates. 
The results of dual orthogonal views \modified{using $0^\circ$ and $90^\circ$ OPBP input images (like Fig.\,\ref{subfig:OPBPData1})} are displayed in \modified{Fig.\,\ref{subfig:dual90OutputOPBP} and Fig.\,\ref{subfig:dual90PolarOutputOPBP}}. Compared with Fig.\,\ref{subfig:singularOutput}, Fig.\,\ref{subfig:dual90OutputOPBP} has less aliasing and less shape distortion. However, Fig.\,\ref{subfig:dual90OutputDiff} indicates that the error in Fig.\,\ref{subfig:dual90OutputOPBP} is still large, which has a root-mean-square error (RMSE) value of 6.41. 
The results of dual complementary views are displayed in \modified{Fig.\,\ref{subfig:dual180Output} and Fig.\,\ref{subfig:dual180PolarOutput}}. In Fig.\,\ref{subfig:dual180Output}, aliasing still remains. In Fig.\,\ref{subfig:dual180PolarOutput}, all the beads have decent appearance. Among all the error images, Fig.\,\ref{subfig:dual18PolarOutputDiff} has the least error.

For overall image quality comparison, the mean RMSE and structure similarity index measurement (SSIM) values for the prediction results with different configurations are displayed in Tab.\,\ref{Tab:OverallComparisonBead}. In addition to the results in Fig.\,\ref{Fig:DataIPredictions}, three more results are included for comparison: a) combining $0^\circ$ and $90^\circ$ perspective projection images naively like Fig.\,\ref{subfig:perspectiveRGB0and90degExample} as the input of the neural network, denoted by \enquote{$0^\circ \& 90^\circ$ naive} in Tab.\,\ref{Tab:OverallComparisonBead}; b) Using the difference image between $0^\circ$ and $180^\circ$ perspective projection images like Fig.\,\ref{subfig:diffPers0Sub180Example} as the third channel of the RGB image instead of using the $0^\circ$ perspective projection again, denoted by \enquote{$0^\circ \& 180^\circ +$} in Tab.\,\ref{Tab:OverallComparisonBead}; c) The direct combination of $0^\circ$, $90^\circ$, and $180^\circ$ perspective projection images as the three channels of an RGB input, denoted by \enquote{$0^\circ, 90^\circ \& 180^\circ$}. The experiments of \enquote{$0^\circ \& 180^\circ$} are repeated ten times to avoid the influence of random weight initialization. Regarding image space, Tab.\,\ref{Tab:OverallComparisonBead} demonstrates that learning perspective deformation in polar coordinates can drastically improve image quality \modified{for Pix2pixGAN} compared with that in Cartesian coordinates.

Regarding acquisition views, with the naive combination of orthogonal views, the RMSE values have no considerable improvement over those with a single view. With OPBP, the RMSE is 0.35 less, which is not a large improvement, although it demonstrates that an orthogonal view with OPBP is beneficial for learning perspective deformation. Regarding using the difference image as the third channel (comparison between \enquote{$0^\circ \& 180^\circ$} and \enquote{$0^\circ \& 180^\circ +$}), it improves image quality in Cartesian coordinates. However, in polar coordinates, it has no large influence on the image quality. In addition, naively combining three views (\enquote{$0^\circ,\,90^\circ \& 180^\circ +$}) has no considerable improvement compared with \enquote{$0^\circ \& 180^\circ$}. All in all, Tab.\,\ref{Tab:OverallComparisonBead} recommends using two complementary views in polar coordinates for \modified{Pix2pixGAN to learn} perspective deformation.

To save space, the TransU-Net predictions of the same image in Fig.\,\ref{Fig:DataIPredictions} are displayed in Fig.\,6 in the supplementary material. The quantitative evaluation of TransU-Net using different spatial coordinate systems with different views are displayed in Tab.\,\ref{Tab:OverallComparisonBead} as well. Regarding spatial coordinate systems, learning perspective deformation has comparable performance in the Cartesian space and the polar space for TransU-Net, although learning in the polar space has slightly better performance. Regarding acquisition views, consistent with Pix2pixGAN, using complementary views in general is better than using orthogonal views or a single view for TransU-Net. Regarding the choice of \enquote{$0^\circ \& 180^\circ +$} or \enquote{$0^\circ \& 180^\circ$} for view combination, further discussion can be found in the supplementary material.

\section{Experiments on patient data}
\label{Sect:ExpPatientData}
To evaluate the generalizability of the proposed complementary view setting, further experiments on projection data simulated from patients' CT datasets (including chest and head CT data) as well as real CBCT projection data are performed. The real data contains twenty cadaver knee scans from a Siemens Healthcare CBCT scanner. The reference images for real data are orthogonal projection images of iterative reweighted total variation reconstruction \cite{huang2018scale} volumes from measured CBCT projection data. More details on the datasets are described in the supplementary material.
\subsection{Chest Data}

\begin{figure}
\centering

\begin{minipage}[t]{0.06\linewidth}
\centering
{\rotatebox{90}{\footnotesize{\ }}}
\end{minipage}
\begin{minipage}[b]{0.3\linewidth}
\subfigure[Reference]{
\includegraphics[width=\linewidth]{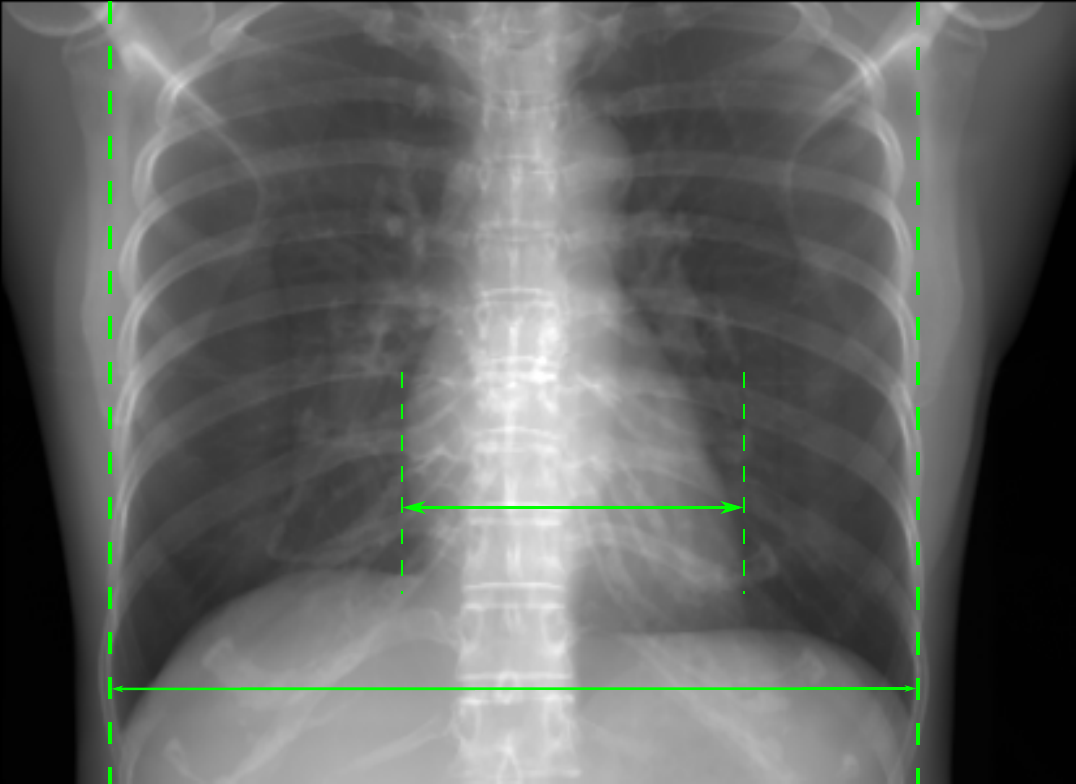}
\label{subfig:torsoTarget}
}
\end{minipage}
\begin{minipage}[b]{0.3\linewidth}
\subfigure[$0^\circ$ perspective]{
\includegraphics[width=\linewidth]{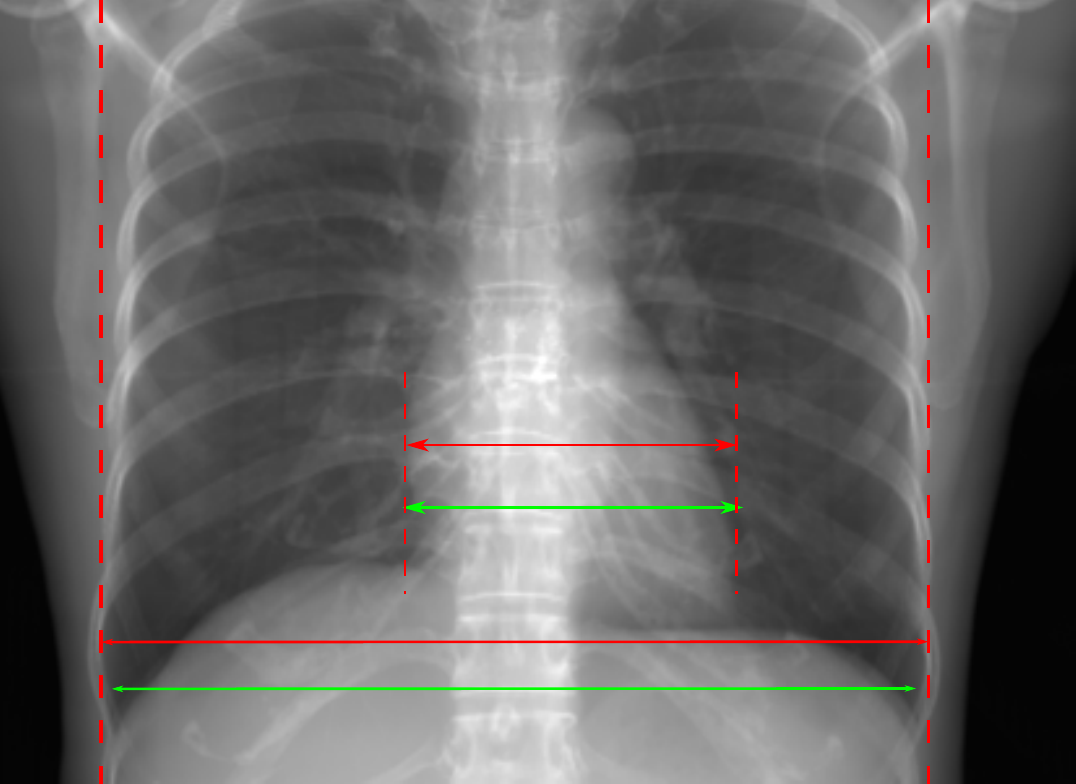}
\label{subfig:torsoPerspective}
}
\end{minipage}
\begin{minipage}[b]{0.3\linewidth}
\subfigure[(b)-(a)]{
\includegraphics[width=\linewidth]{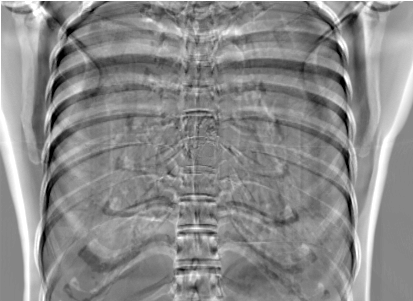}
\label{subfig:torsoPerspectiveDiff}
}
\end{minipage}

\begin{minipage}[t]{0.06\linewidth}
\centering
{\rotatebox{90}{\footnotesize{\ }}}
\end{minipage}
\begin{minipage}[b]{0.3\linewidth}
\centering
\scriptsize{0.4237, $-$, $-$}
\vspace{5pt}
\end{minipage}
\begin{minipage}[b]{0.3\linewidth}
\centering
\scriptsize{0.4002, 17.01, 0.6532}
\vspace{5pt}
\end{minipage}
\begin{minipage}[b]{0.3\linewidth}
\centering
\ 
\vspace{5pt}
\end{minipage}

\begin{minipage}[t]{0.06\linewidth}
\centering
{\rotatebox{90}{\footnotesize{Pix2pixGAN}}}
\end{minipage}
\begin{minipage}[b]{0.3\linewidth}
\subfigure[$0^\circ$ Cartesian]{
\includegraphics[width=\linewidth]{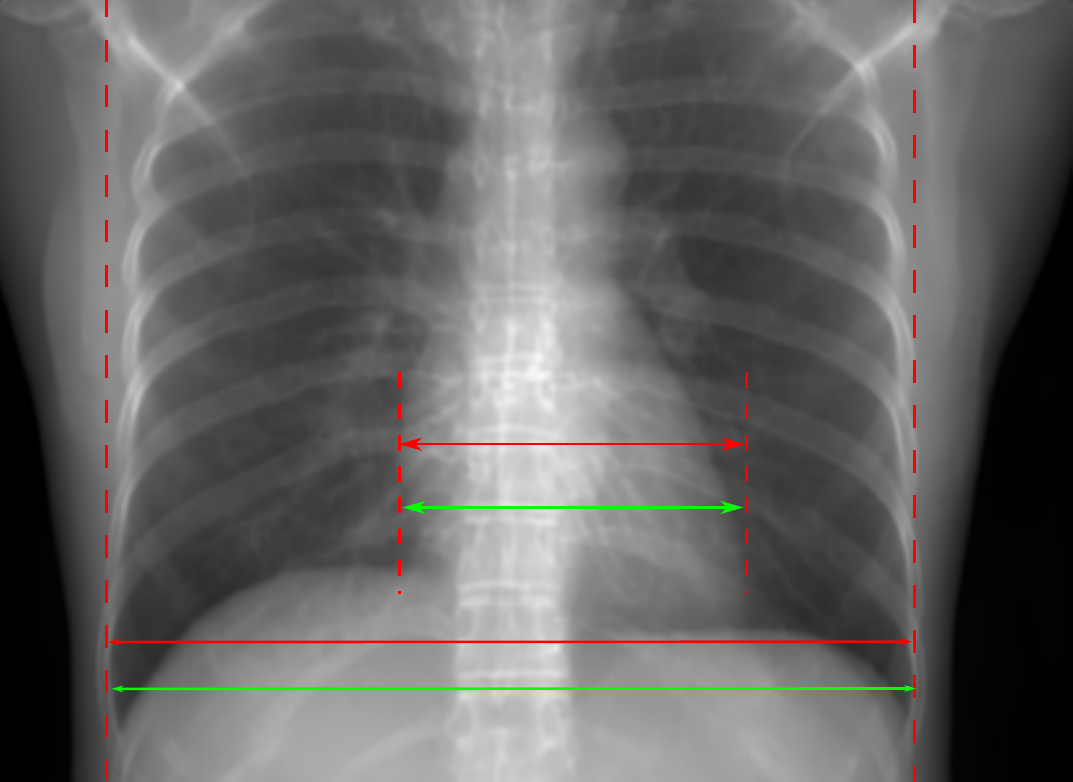}
\label{subfig:torsoSingleOutput}
}
\end{minipage}
\begin{minipage}[b]{0.3\linewidth}
\subfigure[$0^\circ \& 180^\circ$ Cartesian]{
\includegraphics[width=\linewidth]{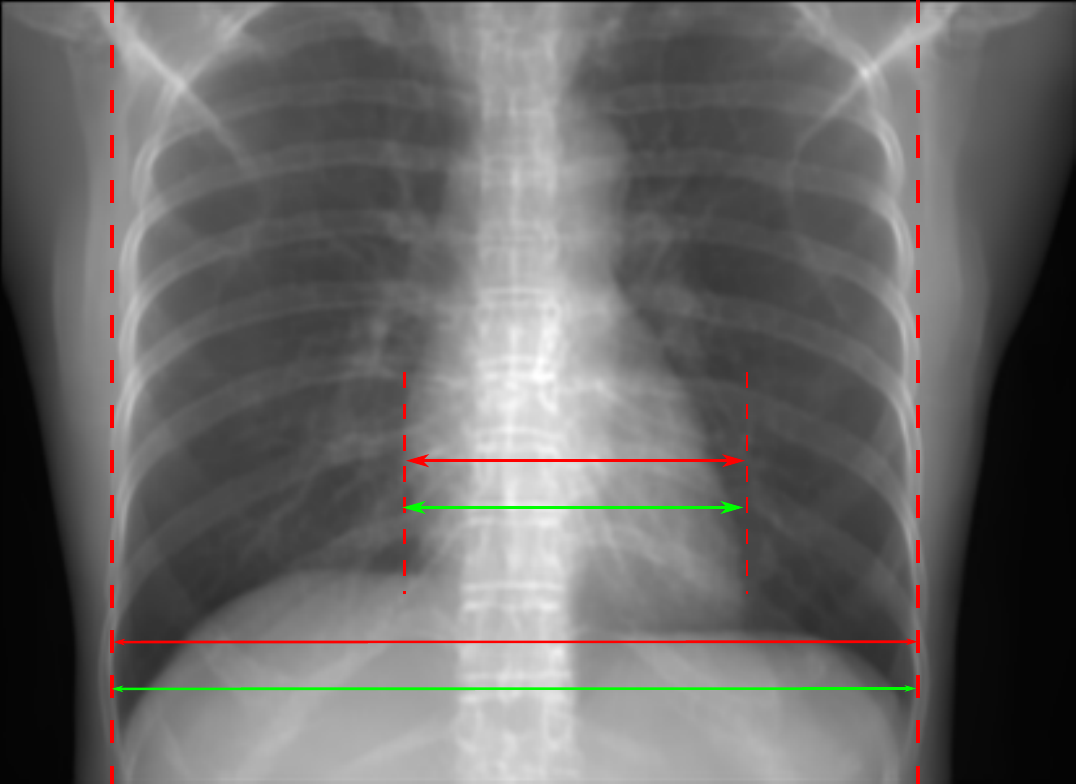}
\label{subfig:torsoDualOutput}
}
\end{minipage}
\begin{minipage}[b]{0.3\linewidth}
\subfigure[$0^\circ \& 180^\circ$ polar]{
\includegraphics[width=\linewidth]{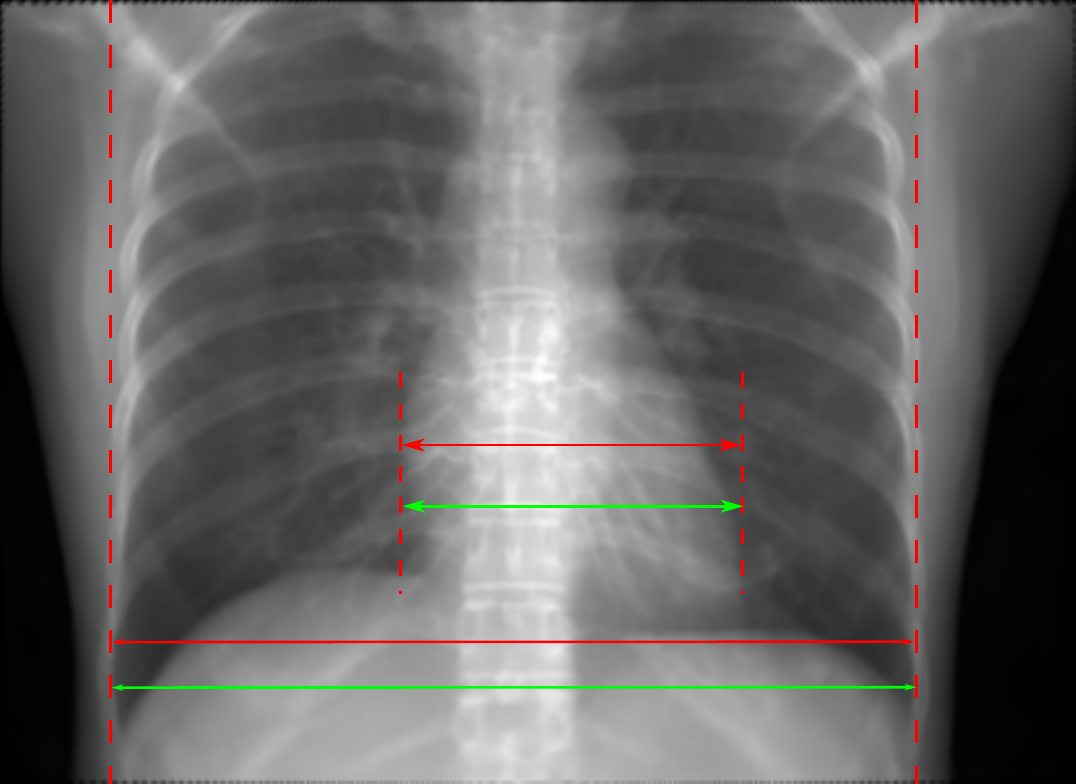}
\label{subfig:torsoDualPolarOutput}
}
\end{minipage}

\begin{minipage}[t]{0.06\linewidth}
\centering
{\rotatebox{90}{\footnotesize{\ }}}
\end{minipage}
\begin{minipage}[b]{0.3\linewidth}
\centering
\scriptsize{0.4303, 7.08, 0.8535}
\vspace{5pt}
\end{minipage}
\begin{minipage}[b]{0.3\linewidth}
\centering
\scriptsize{0.4214, 5.37, 0.9098}
\vspace{5pt}
\end{minipage}
\begin{minipage}[b]{0.3\linewidth}
\centering
\scriptsize{0.4240, 3.83, 0.9536}
\vspace{5pt}
\end{minipage}

\begin{minipage}[t]{0.06\linewidth}
\centering
{\rotatebox{90}{\footnotesize{Pix2pixGAN}}}
\end{minipage}
\begin{minipage}[b]{0.3\linewidth}
\subfigure[(d)-(a)]{
\includegraphics[width=\linewidth]{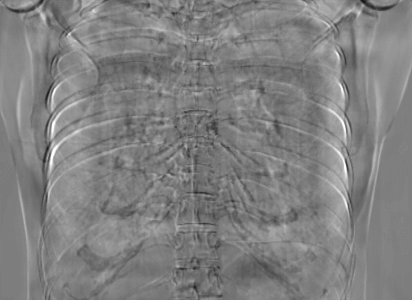}
\label{subfig:torsoSingleOutputDiff}
}
\end{minipage}
\begin{minipage}[b]{0.3\linewidth}
\subfigure[(e)-(a)]{
\includegraphics[width=\linewidth]{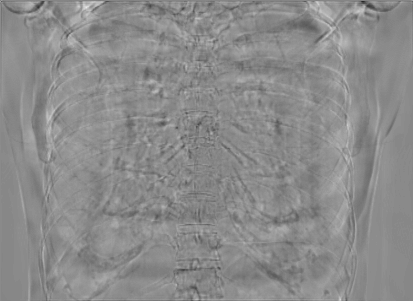}
\label{subfig:torsoDualOutputDiff}
}
\end{minipage}
\begin{minipage}[b]{0.3\linewidth}
\subfigure[(f)-(a)]{
\includegraphics[width=\linewidth]{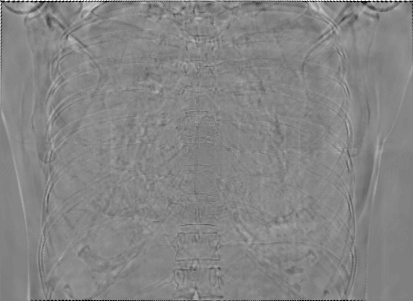}
\label{subfig:torsoDualPolarOutputDiff}
}
\end{minipage}

\begin{minipage}[t]{0.06\linewidth}
\centering
{\rotatebox{90}{\footnotesize{\modified{TransU-Net}}}}
\end{minipage}
\begin{minipage}[b]{0.3\linewidth}
\subfigure[$0^\circ$ Cartesian]{
\includegraphics[width=\linewidth]{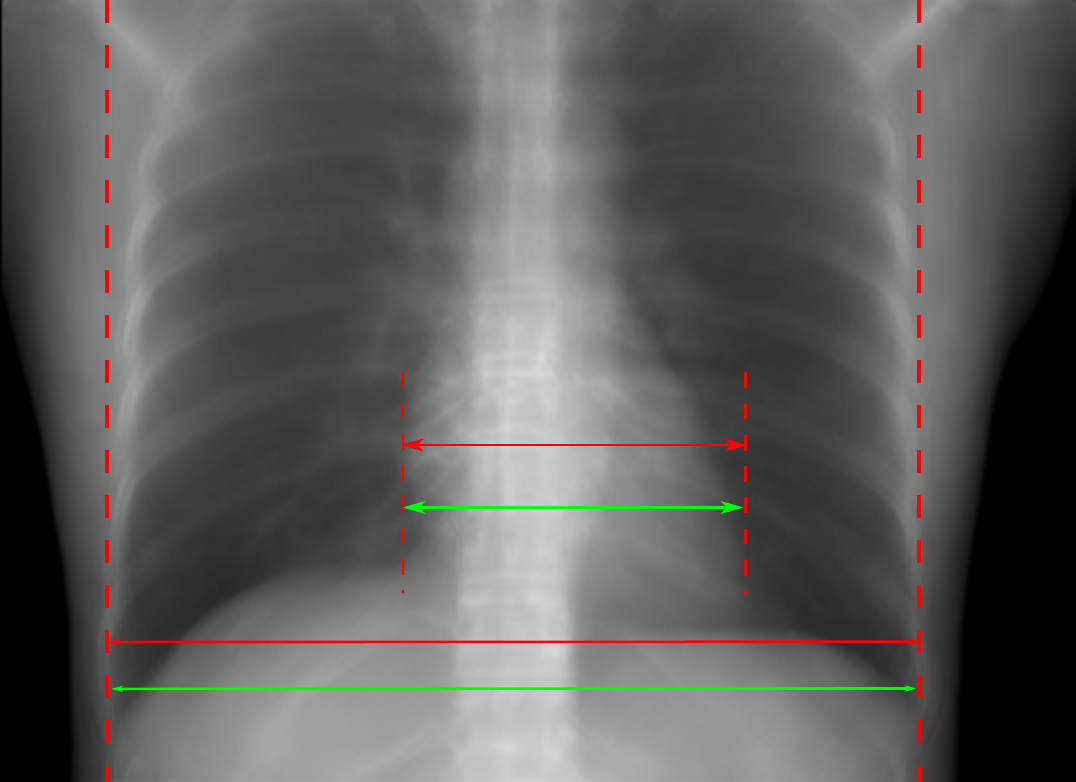}
\label{subfig:torsoSingleOutputTransUNet}
}
\end{minipage}
\begin{minipage}[b]{0.3\linewidth}
\subfigure[$0^\circ \& 180^\circ$ Cartesian]{
\includegraphics[width=\linewidth]{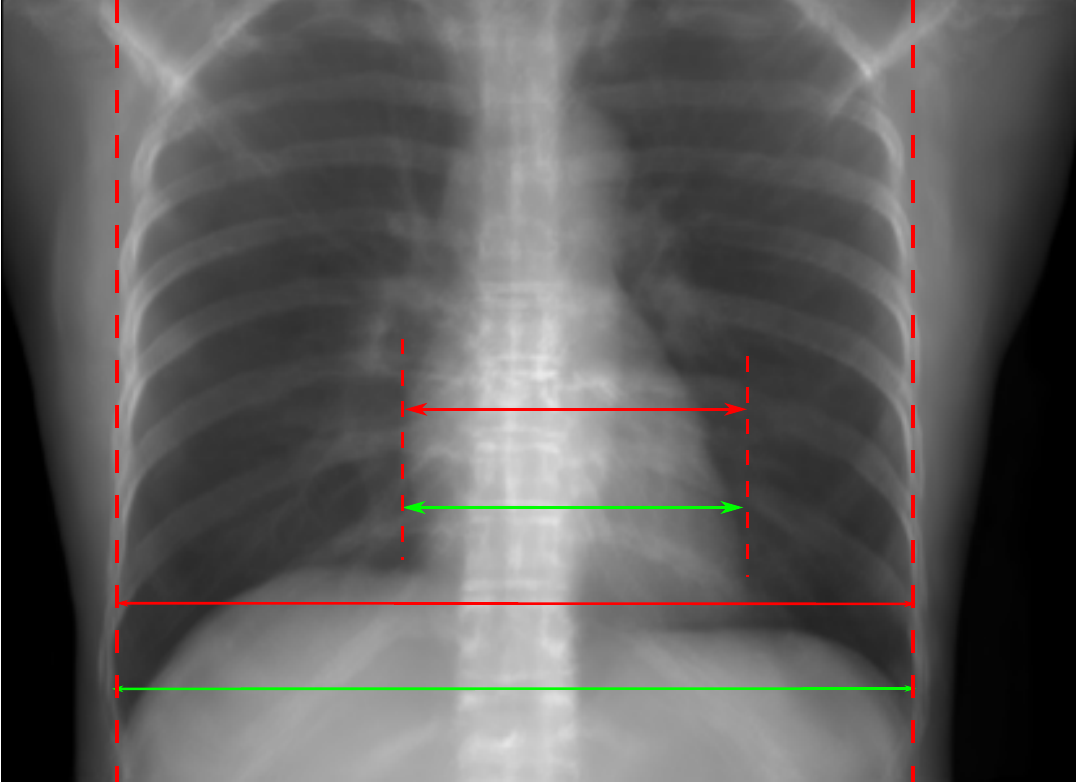}
\label{subfig:torsoDualOutputTransUNet}
}
\end{minipage}
\begin{minipage}[b]{0.3\linewidth}
\subfigure[$0^\circ \& 180^\circ$ polar]{
\includegraphics[width=\linewidth]{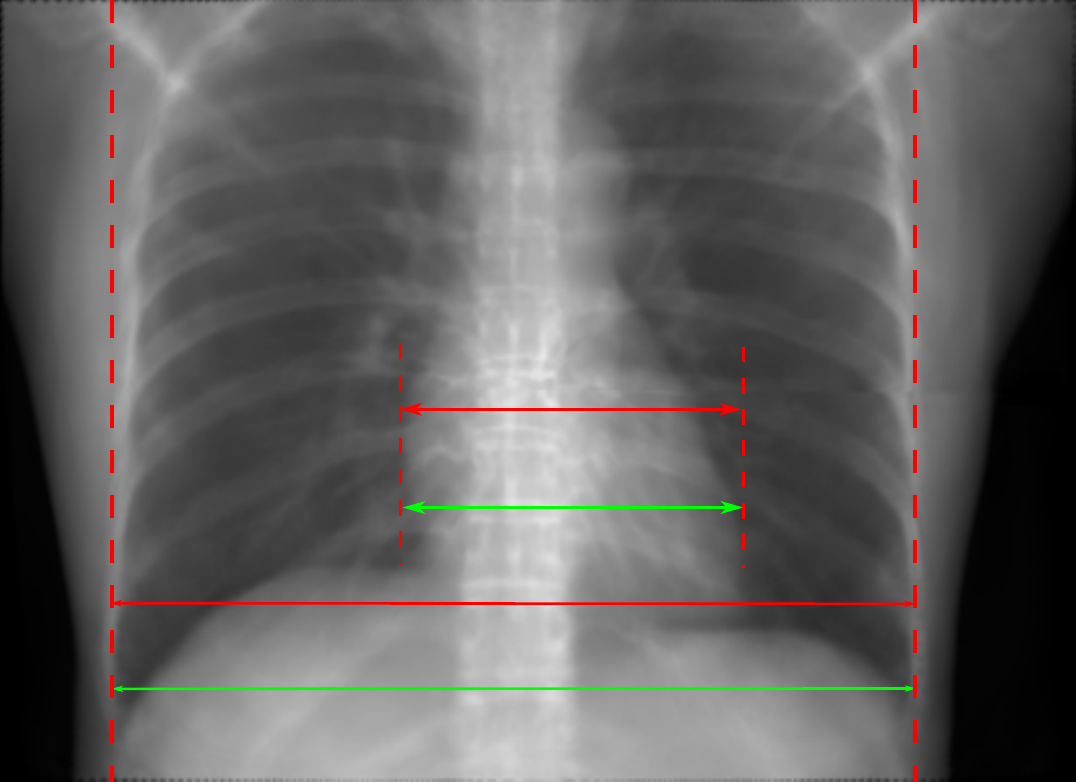}
\label{subfig:torsoDualPolarOutputTransUNet}
}
\end{minipage}

\begin{minipage}[t]{0.06\linewidth}
\centering
{\rotatebox{90}{\footnotesize{\ }}}
\end{minipage}
\begin{minipage}[b]{0.3\linewidth}
\centering
\scriptsize{0.4272, 10.92, 0.8222}
\vspace{5pt}
\end{minipage}
\begin{minipage}[b]{0.3\linewidth}
\centering
\scriptsize{0.4281, 9.37, 0.8424}
\vspace{5pt}
\end{minipage}
\begin{minipage}[b]{0.3\linewidth}
\centering
\scriptsize{0.4248, 8.12, 0.8859}
\vspace{5pt}
\end{minipage}

\begin{minipage}[t]{0.06\linewidth}
\centering
{\rotatebox{90}{\footnotesize{\modified{TransU-Net}}}}
\end{minipage}
\begin{minipage}[b]{0.3\linewidth}
\subfigure[(j)-(a)]{
\includegraphics[width=\linewidth]{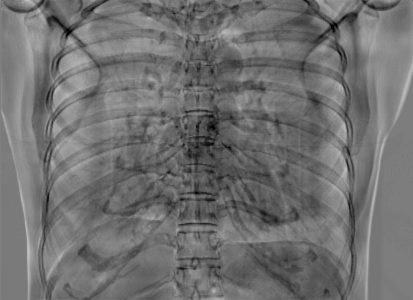}
\label{subfig:torsoSingleOutputDiffTransUNet}
}
\end{minipage}
\begin{minipage}[b]{0.3\linewidth}
\subfigure[(k)-(a)]{
\includegraphics[width=\linewidth]{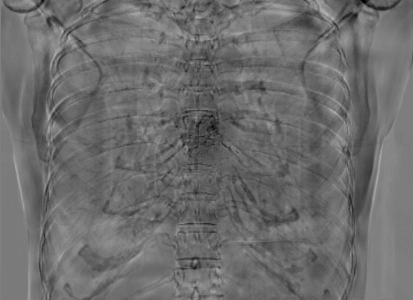}
\label{subfig:torsoDualOutputDiffTransUNet}
}
\end{minipage}
\begin{minipage}[b]{0.3\linewidth}
\subfigure[(l)-(a)]{
\includegraphics[width=\linewidth]{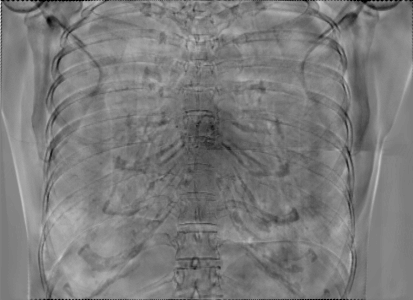}
\label{subfig:torsoDualPolarOutputDiffTransUNet}
}
\end{minipage}
\caption{Perspective deformation learning in one exemplary patient case for chest X-ray imaging. The maximal horizontal cardiac diameter and the maximal horizontal thoracic diameter in (b) and (d)-(f) are indicated by the horizontal red lines, while those in the reference image (a) are green lines. The cardiothoracic ratio, RMSE, and SSIM for each image is displayed in its corresponding subcaption. }
\label{Fig:TorsoDataResults}
\end{figure}

\begin{table}
\centering
\caption{Quantitative evaluation of different methods on chest data.}
\label{Tab：Torso}
\begin{tabular}{|l|l|c|c|c|c|}
\hline
Method & Metric  & $0^\circ$ input &$0^\circ$ & $0^\circ\&180^\circ$   & $0^\circ\&180^\circ$ \\
&  &perspective & Cart. & Cart. & polar  \\
\hline
Pix2pix &RMSE & 18.68 & 11.88 & 7.90 & 4.98\\
\cline{2-6}
GAN& SSIM & 0.6401 & 0.8103 & 0.8944 & 0.9493\\
\hline
Trans &RMSE & 18.68 & 17.21 & 12.12 & 13.06\\
\cline{2-6}
U-Net& SSIM & 0.6401 & 0.7744 & 0.8950 & 0.8899\\
\hline
\end{tabular}
\end{table}

The results of one patient in chest X-ray imaging are displayed in Fig.\,\ref{Fig:TorsoDataResults}, where the cardiothoracic ratio is assessed as an exemplary clinical application \cite{truszkiewicz2021radiological}. In the reference image (Fig.\,\ref{subfig:torsoTarget}), the maximal horizontal cardiac diameter (MHCD) and the maximal horizontal thoracic diameter (MHTD) are indicated by two green horizontal lines. Its cardiothoracic ratio is 0.4237. In the $0^\circ$ perspective projection image (Fig.\,\ref{subfig:torsoPerspective}), all the anatomical structures can be visualized with fine resolution. However, due to perspective deformation, anatomical structures, e.g. the ribs and the spine, are deformed. The deformations are visualized better in the difference image Fig.\,\ref{subfig:torsoPerspectiveDiff}. Compared with the ribs and the spine, the heart has less deformation as its location is closer to the isocenter. In Fig.\,\ref{subfig:torsoPerspective}, the MHCD and the MHTD are indicated by two red horizontal lines, while the green lines are those of the reference image. While the MHCD has changed little from 10.47\,cm to 10.16\,cm, the MHTD has changed considerably from 24.71\,cm to 25.40\,cm. As a consequence, the cardiothoracic ratio becomes 0.4002, which is below the normal range of 0.42\,-\,0.50 \cite{truszkiewicz2021radiological}. The result of learning perspective deformation from $0^\circ$ single view is displayed in Fig.\,\ref{subfig:torsoSingleOutput}, where the MHCD and the MHTD are 10.63\,cm and 24.71\,cm, respectively. The MHTD of Fig.\,\ref{subfig:torsoSingleOutput} is the same as that of the reference image. This is also reflected by the difference image Fig.\,\ref{subfig:torsoSingleOutputDiff}, where the lower ribs have small errors. However, the upper ribs as well as the spine still have considerable errors. The results of perspective deformation learning from $0^\circ \& 180^\circ$ views in Cartesian and polar coordinates are displayed in Fig.\,\ref{subfig:torsoDualOutput} and Fig.\,\ref{subfig:torsoDualPolarOutput}, respectively. The measured MHCDs and MHTDs in these two images are very close to the reference ones. Hence, their cardiothoracic ratios, 0.4214 and 0.4240 respectively, are close to the reference ratio as well. In the difference images (Fig.\,\ref{subfig:torsoDualOutputDiff} and Fig.\,\ref{subfig:torsoDualPolarOutputDiff}), the errors of ribs and spine decrease as their boundaries are no longer apparently visible. Nevertheless, Fig.\,\ref{subfig:torsoDualPolarOutputDiff} has less error than Fig.\,\ref{subfig:torsoDualOutputDiff}, achieving the smallest RMSE value of 3.83. The quantitative evaluation of all the 162 testing datasets is displayed in Tab.\,\ref{Tab：Torso}, where learning perspective deformation from two complementary views in polar coordinates achieves the least RMSE 4.98 and highest SSIM 0.9517, demonstrating the superiority of learning perspective deformation from two complementary views in polar coordinates.


 
The TransU-Net results are displayed in Figs.\,\ref{Fig:TorsoDataResults}(j)-(l). Compared with their corresponding Pix2pixGAN results, the TransU-Net prediction images are more blurry, although the same perceptual loss is used. The error images in Figs.\,\ref{Fig:TorsoDataResults}(m)-(o) indicate that TransU-Net reduces perspective deformation better with complementary views than a single view. The quantitative evaluation in Tab.\,\ref{Tab：Torso} shows that TransU-Net cannot effectively reduce perspective deformation with a single view. With complementary views in both Cartesian and polar coordinate systems, TransU-Net achieves comparable performance, which is still considerably worse than that of Pix2pixGAN. The inferior performance of TransU-Net to Pix2pixGAN on the chest data is potentially caused by the repetitive nature of the segmental rib anatomy, which leads TransU-Net to be ineffective in extracting position-dependent features. 
\subsection{Head Data}

\begin{figure}[t]
\centering

\begin{minipage}[t]{0.06\linewidth}
\centering
{\rotatebox{90}{\footnotesize{\ }}}
\end{minipage}
\begin{minipage}[b]{0.3\linewidth}
\subfigure[Reference]{
\includegraphics[width=\linewidth]{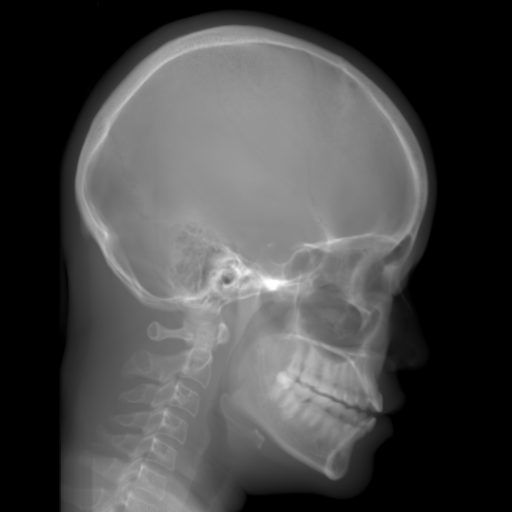}
\label{subfig:cephaloTarget}
}
\end{minipage}
\begin{minipage}[b]{0.3\linewidth}
\subfigure[$0^\circ$ perspective]{
\includegraphics[width=\linewidth]{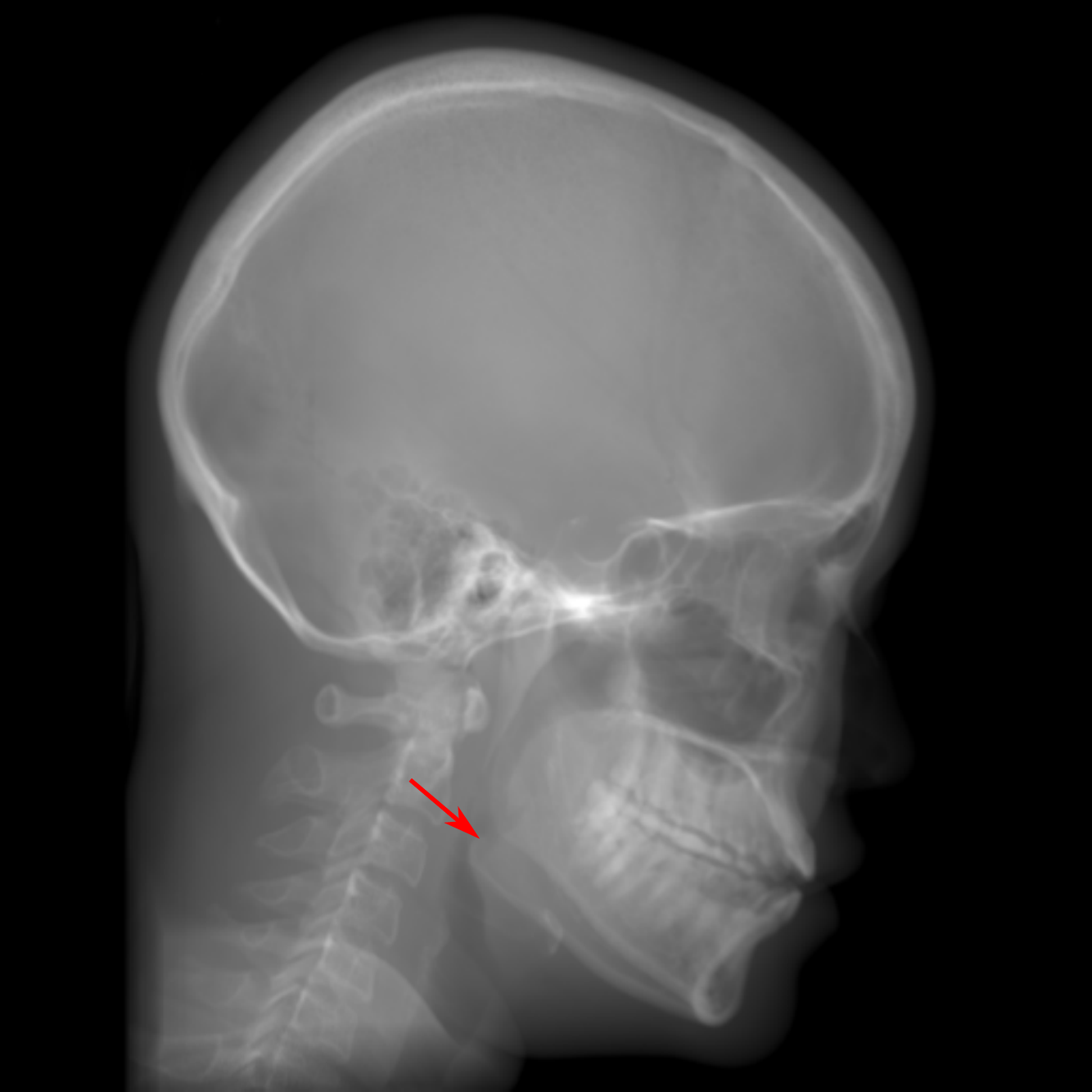}
\label{subfig:cephaloPerspective}
}
\end{minipage}
\begin{minipage}[b]{0.3\linewidth}
\subfigure[(b)-(a)]{
\includegraphics[width=\linewidth]{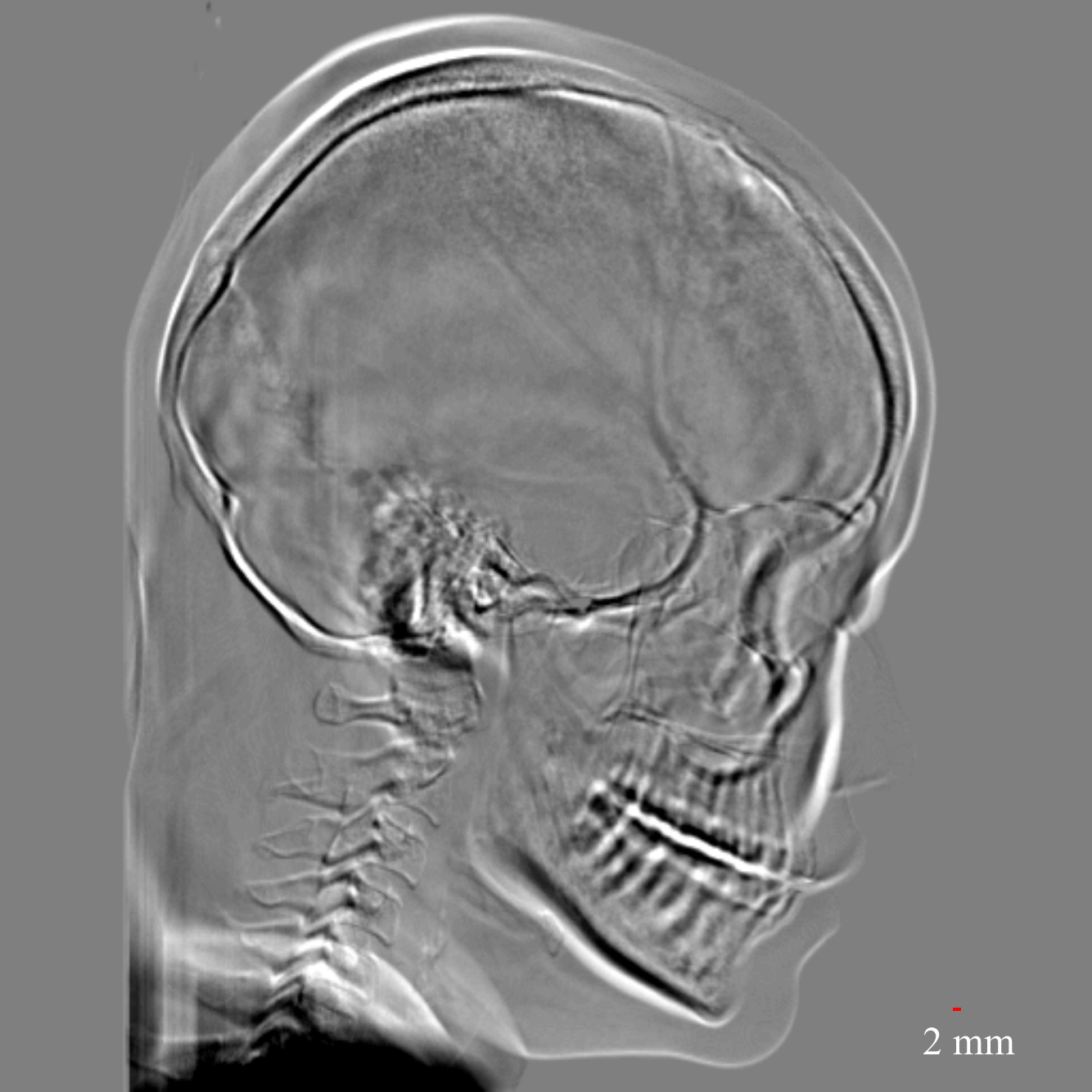}
\label{subfig:cephaloPerspectiveDiff}
}
\end{minipage}

\begin{minipage}[t]{0.06\linewidth}
\centering
{\rotatebox{90}{\footnotesize{\ }}}
\end{minipage}
\begin{minipage}[b]{0.3\linewidth}
\centering
\scriptsize{\  }
\vspace{5pt}
\end{minipage}
\begin{minipage}[b]{0.3\linewidth}
\centering
\scriptsize{7.80, 0.9093}
\vspace{5pt}
\end{minipage}
\begin{minipage}[b]{0.3\linewidth}
\centering
\scriptsize{\ }
\vspace{5pt}
\end{minipage}

\begin{minipage}[t]{0.06\linewidth}
\centering
{\rotatebox{90}{\footnotesize{Pix2pixGAN Prediction}}}
\end{minipage}
\begin{minipage}[b]{0.3\linewidth}
\subfigure[$0^\circ$ Cartesian]{
\includegraphics[width=\linewidth]{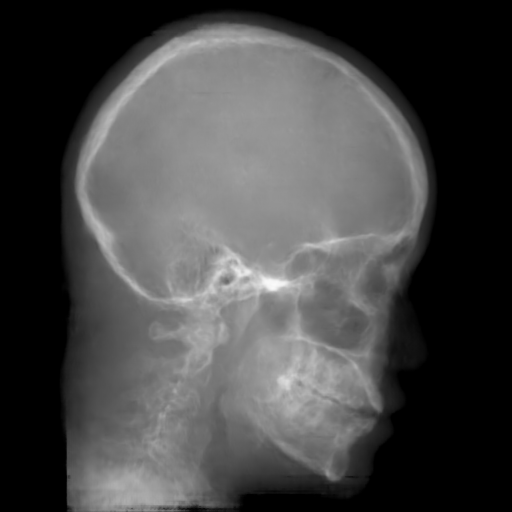}
\label{subfig:cephaloSingleOutput}
}
\end{minipage}
\begin{minipage}[b]{0.3\linewidth}
\subfigure[$0^\circ \& 180^\circ$ Cartesian]{
\includegraphics[width=\linewidth]{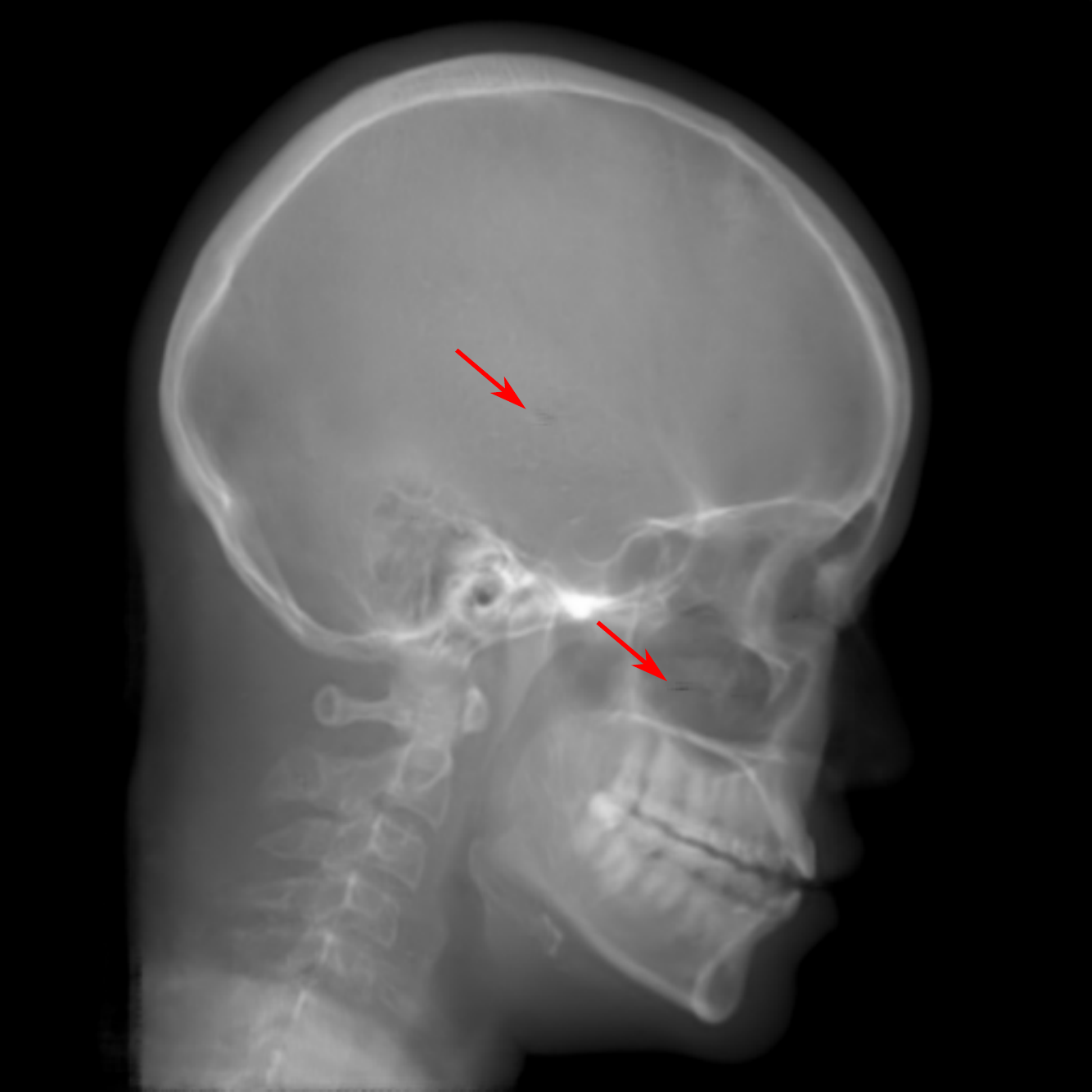}
\label{subfig:cephaloDualOutput}
}
\end{minipage}
\begin{minipage}[b]{0.3\linewidth}
\subfigure[$0^\circ \& 180^\circ$ polar]{
\includegraphics[width=\linewidth]{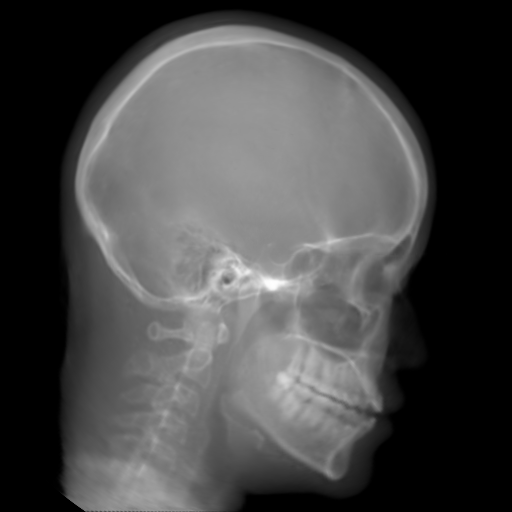}
\label{subfig:cephaloDualPolarOutput}
}
\end{minipage}

\begin{minipage}[t]{0.06\linewidth}
\centering
{\rotatebox{90}{\footnotesize{\ }}}
\end{minipage}
\begin{minipage}[b]{0.3\linewidth}
\centering
\scriptsize{6.06, 0.9249}
\vspace{5pt}
\end{minipage}
\begin{minipage}[b]{0.3\linewidth}
\centering
\scriptsize{2.99, 0.9820}
\vspace{5pt}
\end{minipage}
\begin{minipage}[b]{0.3\linewidth}
\centering
\scriptsize{2.07, 0.9842}
\vspace{5pt}
\end{minipage}

\begin{minipage}[t]{0.06\linewidth}
\centering
{\rotatebox{90}{\footnotesize{Pix2pixGAN difference}}}
\end{minipage}
\begin{minipage}[b]{0.3\linewidth}
\subfigure[(d)-(a)]{
\includegraphics[width=\linewidth]{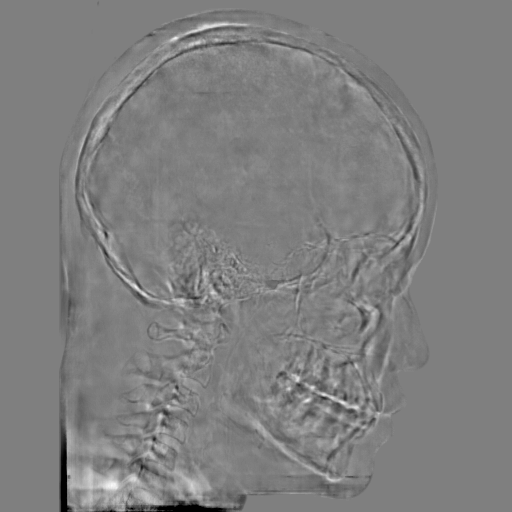}
\label{subfig:cephaloSingleOutputDiff}
}
\end{minipage}
\begin{minipage}[b]{0.3\linewidth}
\subfigure[(e)-(a)]{
\includegraphics[width=\linewidth]{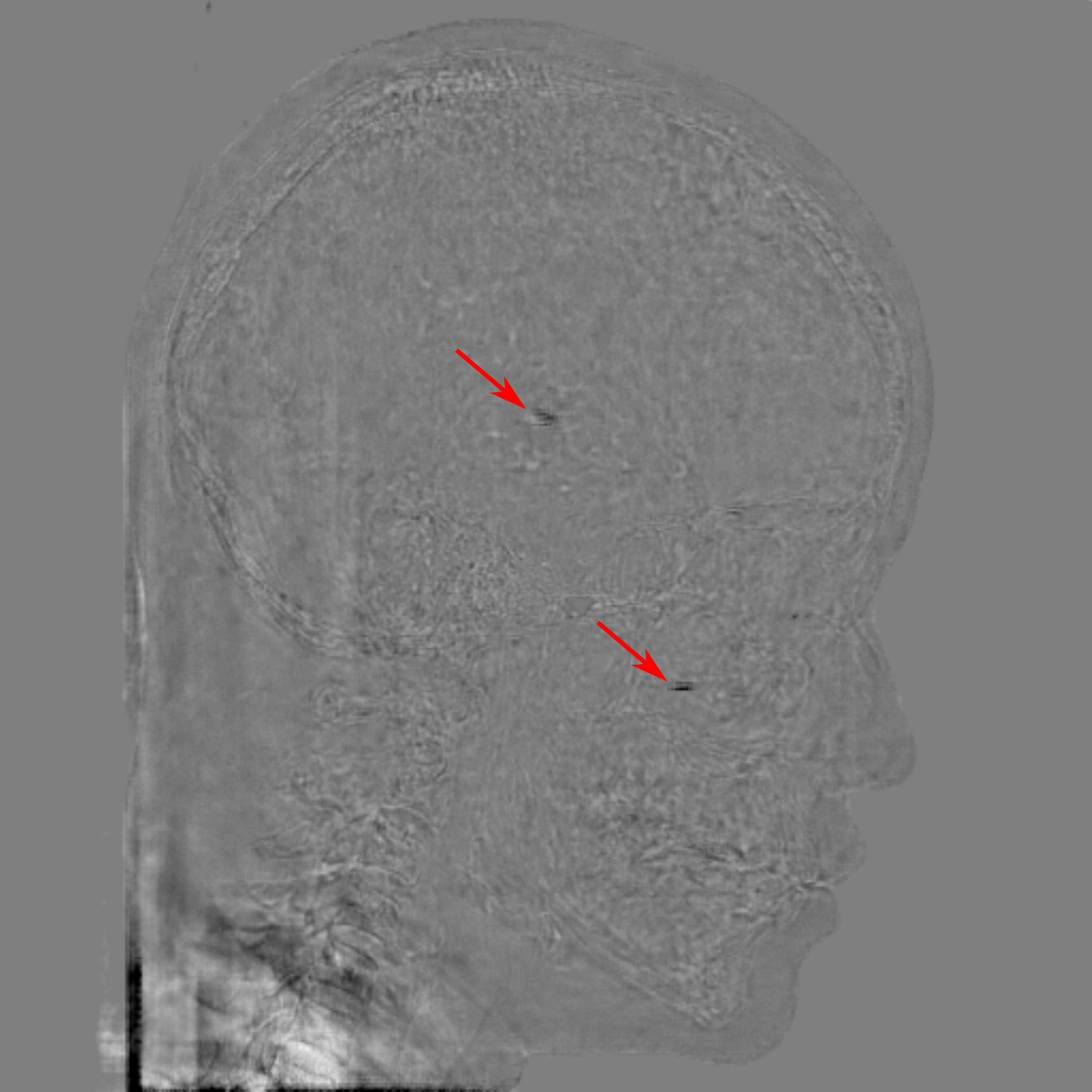}
\label{subfig:cephaloDualOutputDiff}
}
\end{minipage}
\begin{minipage}[b]{0.3\linewidth}
\subfigure[(f)-(a)]{
\includegraphics[width=\linewidth]{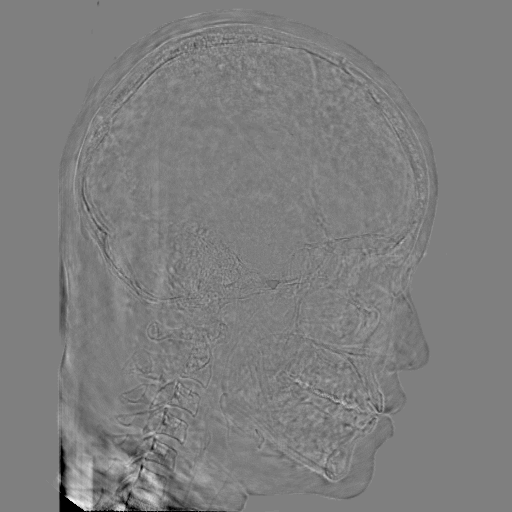}
\label{subfig:cephaloDualPolarOutputDiff}
}
\end{minipage}

\begin{minipage}[t]{0.06\linewidth}
\centering
{\rotatebox{90}{\footnotesize{\modified{TransU-Net Prediction}}}}
\end{minipage}
\begin{minipage}[b]{0.3\linewidth}
\subfigure[$0^\circ$ Cartesian]{
\includegraphics[width=\linewidth]{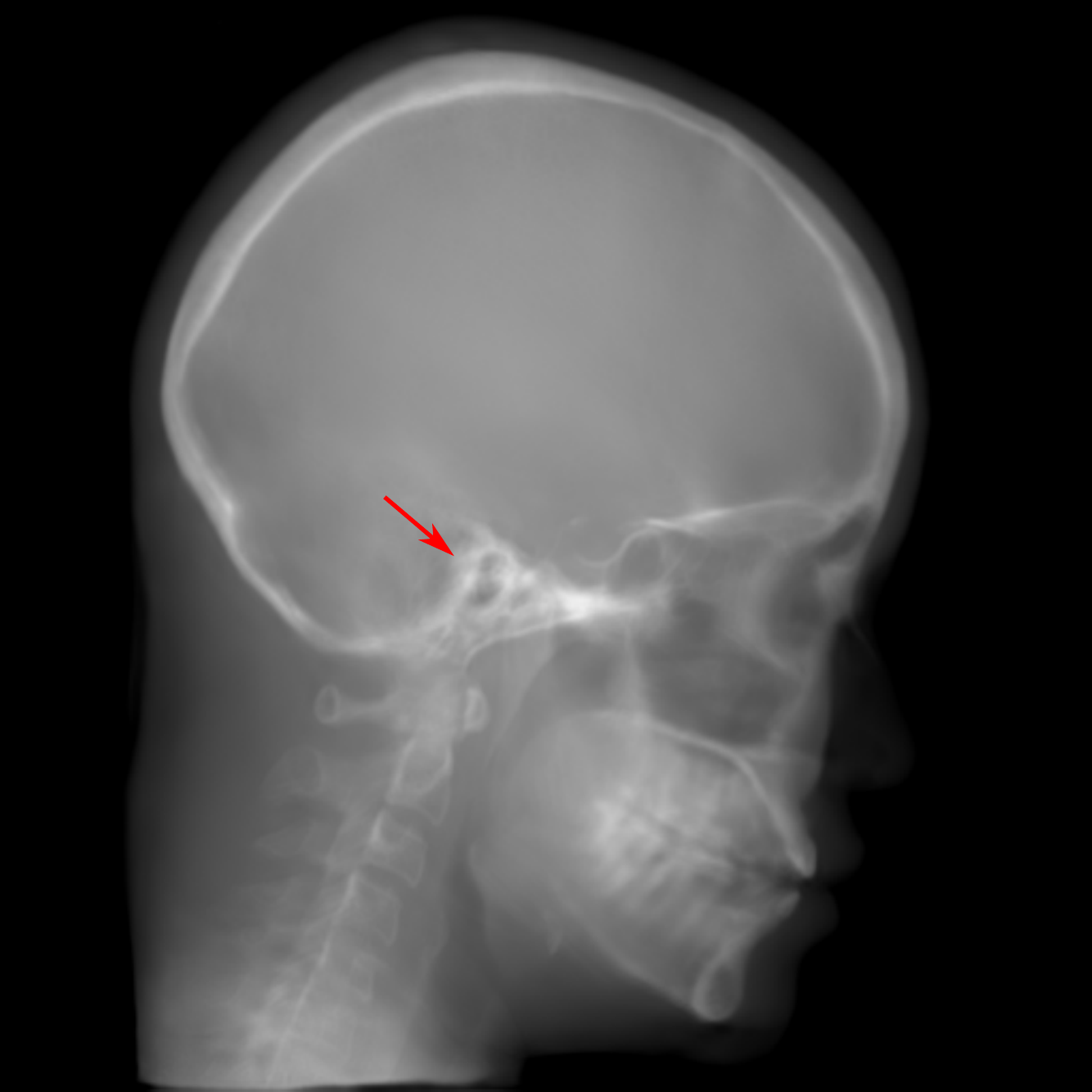}
\label{subfig:cephaloSingleOutputTransUNet}
}
\end{minipage}
\begin{minipage}[b]{0.3\linewidth}
\subfigure[$0^\circ \& 180^\circ$ Cartesian]{
\includegraphics[width=\linewidth]{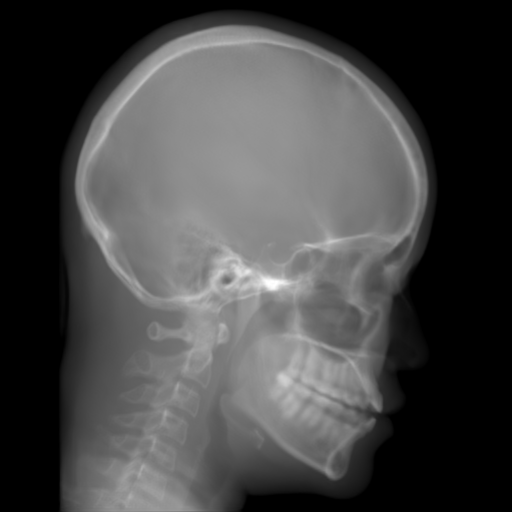}
\label{subfig:cephaloDualOutputTransUNet}
}
\end{minipage}
\begin{minipage}[b]{0.3\linewidth}
\subfigure[$0^\circ \& 180^\circ$ polar]{
\includegraphics[width=\linewidth]{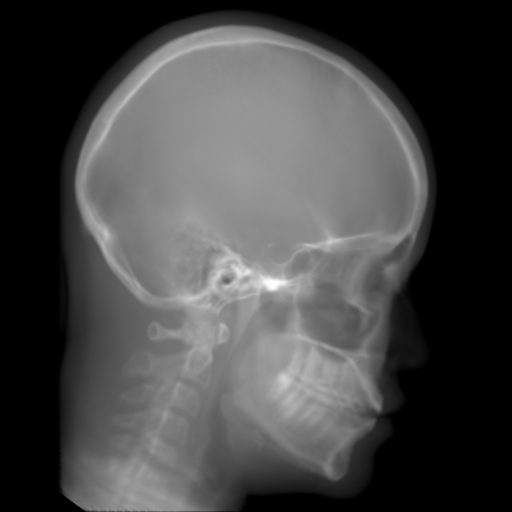}
\label{subfig:cephaloDualPolarOutputTransUNet}
}
\end{minipage}

\begin{minipage}[t]{0.06\linewidth}
\centering
{\rotatebox{90}{\footnotesize{\ }}}
\end{minipage}
\begin{minipage}[b]{0.3\linewidth}
\centering
\scriptsize{6.45, 0.9356}
\vspace{5pt}
\end{minipage}
\begin{minipage}[b]{0.3\linewidth}
\centering
\scriptsize{2.36, 0.9864}
\vspace{5pt}
\end{minipage}
\begin{minipage}[b]{0.3\linewidth}
\centering
\scriptsize{3.65, 0.9734}
\vspace{5pt}
\end{minipage}

\begin{minipage}[t]{0.06\linewidth}
\centering
{\rotatebox{90}{\footnotesize{\modified{TransU-Net Difference}}}}
\end{minipage}
\begin{minipage}[b]{0.3\linewidth}
\subfigure[(j)-(a)]{
\includegraphics[width=\linewidth]{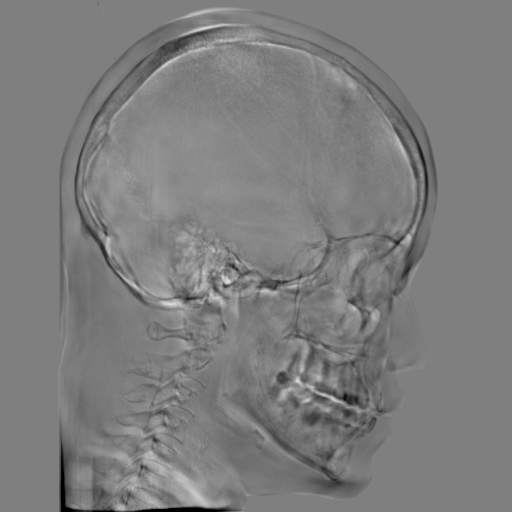}
\label{subfig:cephaloSingleOutputTransUNetDiff}
}
\end{minipage}
\begin{minipage}[b]{0.3\linewidth}
\subfigure[(k)-(a)]{
\includegraphics[width=\linewidth]{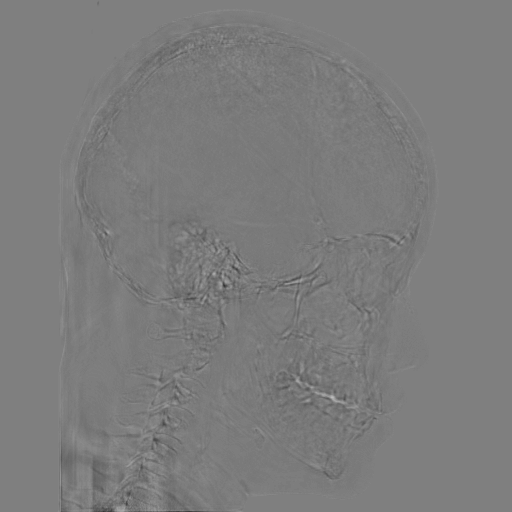}
\label{subfig:cephaloDualOutputTransUNetDiff}
}
\end{minipage}
\begin{minipage}[b]{0.3\linewidth}
\subfigure[(l)-(a)]{
\includegraphics[width=\linewidth]{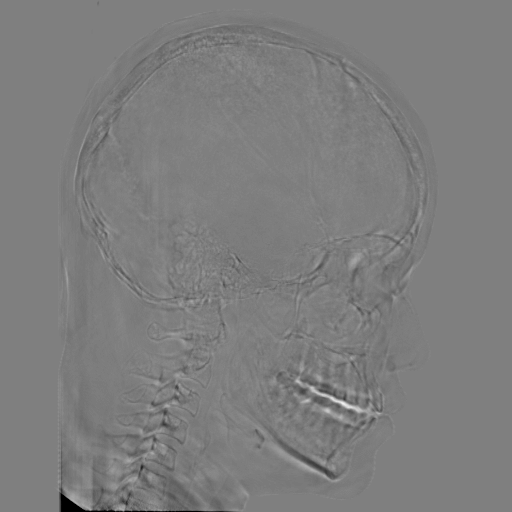}
\label{subfig:cephaloDualPolarOutputTransUNetDiff}
}
\end{minipage}
\caption{Perspective deformation learning in one exemplary patient case for cephalometric imaging. In (b), the left and right sides of the mandible do not overlap well, as indicated by the arrow. In (c), a scale bar of 2\,mm is displayed (zoom in for better visualization), as 2\,mm is the clinically acceptable precision for cephalometric landmark detection. In (e), (h) and \modified{(j)}, incorrect areas are marked by the red arrows.}
\label{Fig:HeadDataResults}
\end{figure}

\begin{table}
\centering
\caption{Quantitative evaluation of different methods on head data.}
\label{Tab：Head}
\begin{tabular}{|l|l|c|c|c|c|}
\hline
Method & Metric & $0^\circ$ input &$0^\circ$ & $0^\circ\&180^\circ$   & $0^\circ\&180^\circ$ \\
&  &perspective & Cart. & Cart. & polar  \\
\hline
Pix2pix & RMSE & 10.69 & 7.33 & 4.58 & 3.87\\
\cline{2-6}
GAN &SSIM & 0.8680 & 0.9053 & 0.9476 & 0.9625\\
\hline
Trans & RMSE & 10.69 & 8.13 & 3.36 & 3.22\\
\cline{2-6}
U-Net &SSIM & 0.8680 & 0.9257 & 0.9682 & 0.9719\\
\hline
\end{tabular}
\end{table}


The results of one exemplary patient for cephalometric imaging are displayed in Fig.\,\ref{Fig:HeadDataResults}. In the $0^\circ$ perspective projection image (Fig.\,\ref{subfig:cephaloPerspective}), because of perspective deformation, anatomical structures from the left and right sides do not overlap well, especially for the mandible as indicated by the red arrow in Fig.\,\ref{subfig:cephaloPerspective}. It causes inaccuracy in determining the cephalometric landmark of the gonion. The difference of Fig.\,\ref{subfig:cephaloPerspective} to the reference Fig.\,\ref{subfig:cephaloTarget} is displayed in Fig.\,\ref{subfig:cephaloPerspectiveDiff}. A scale bar of 2\,mm is displayed in Fig.\,\ref{subfig:cephaloPerspectiveDiff}, as 2\,mm is the clinically acceptable precision for cephalometric landmark detection. It is obvious that many anatomical structures in the $0^\circ$ perspective projection images have position shifts larger than 2\,mm. In the prediction image (Fig.\,\ref{subfig:cephaloSingleOutput}) using a single $0^\circ$ view in Cartesian coordinates, perspective deformation is reduced to some degree, as displayed in the difference image Fig.\,\ref{subfig:cephaloSingleOutputDiff}. For example, the mandible region has less error. However, Fig.\,\ref{subfig:cephaloSingleOutputDiff} also indicates that many bony structures still have deviations larger than 2\,mm. The results of learning from dual complementary views in Cartesian and polar coordinates are displayed in Fig.\,\ref{subfig:cephaloDualOutput} and Fig.\,\ref{subfig:cephaloDualPolarOutput}, respectively. Both images have little perspective deformation, as revealed by their difference images in Fig.\,\ref{subfig:cephaloDualOutputDiff} and Fig.\,\ref{subfig:cephaloDualPolarOutputDiff}. Nevertheless, in Fig.\,\ref{subfig:cephaloDualOutput}, two dark regions are indicated by the two arrows, which are better visualized in the difference image Fig.\,\ref{subfig:cephaloDualOutputDiff}.
\modified{The results of TransU-Net are displayed in Figs.\,\ref{Fig:HeadDataResults}(j)-(o). In Fig.\,\ref{subfig:cephaloSingleOutputTransUNetDiff}, the structures near the porion landmark are distorted, for example, the ear canal indicated by the arrow. Consistent with Pix2pixGAN, perspective deformation is largely reduced in the both TransU-Net prediction images using dual complementary views in Cartesian and polar coordinates. }

\subsection{Robustness to System Errors}

\begin{figure}
\centering
\begin{minipage}[t]{0.48\linewidth}
\subfigure[Rotation error]{
\includegraphics[width=\linewidth]{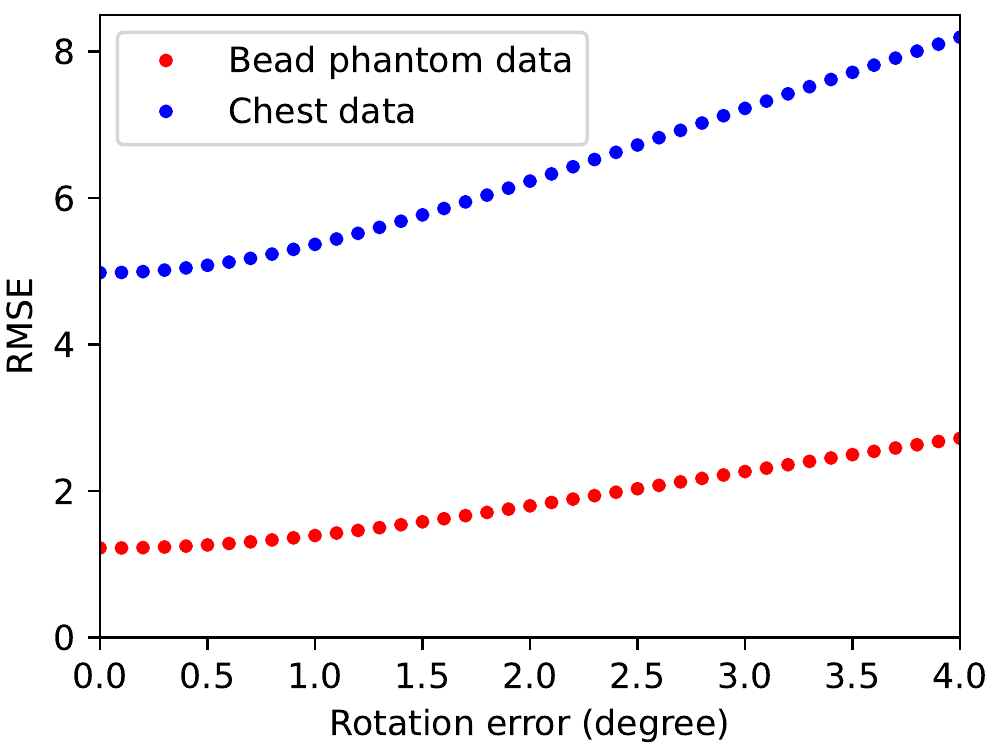}
\label{subfig:ErrorOverDeg}
}
\end{minipage}
\begin{minipage}[t]{0.48\linewidth}
\subfigure[$D_\text{si}$ error]{
\includegraphics[width=\linewidth]{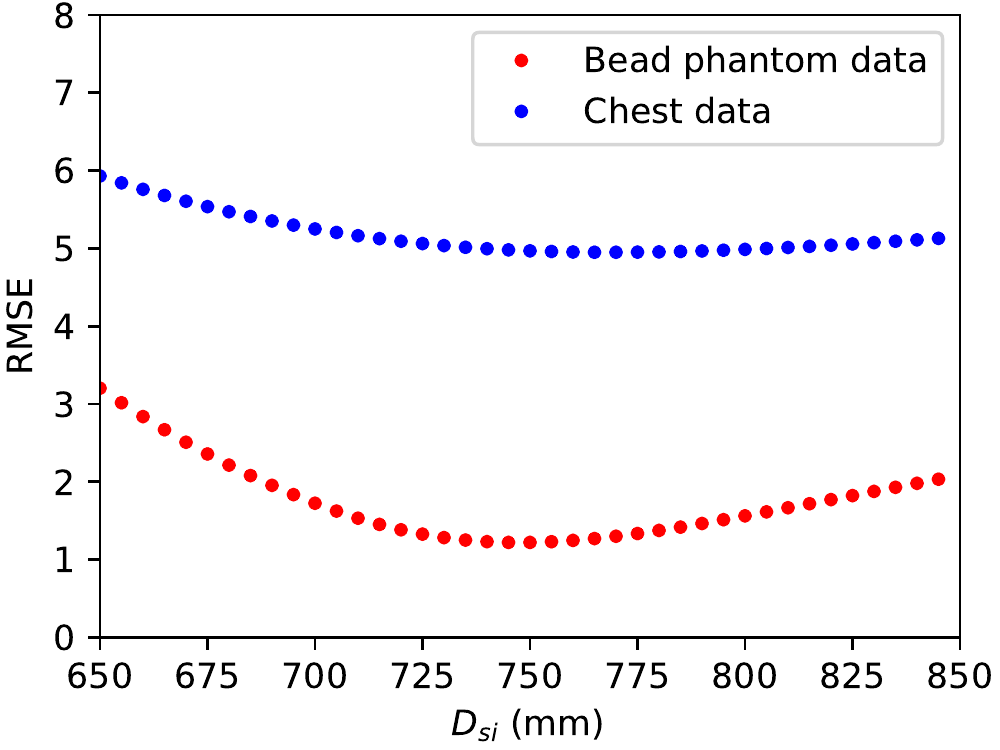}
\label{subfig:ErrorOverSID}
}
\end{minipage}

\begin{minipage}[t]{0.48\linewidth}
\subfigure[Principal point shift error]{
\includegraphics[width=\linewidth]{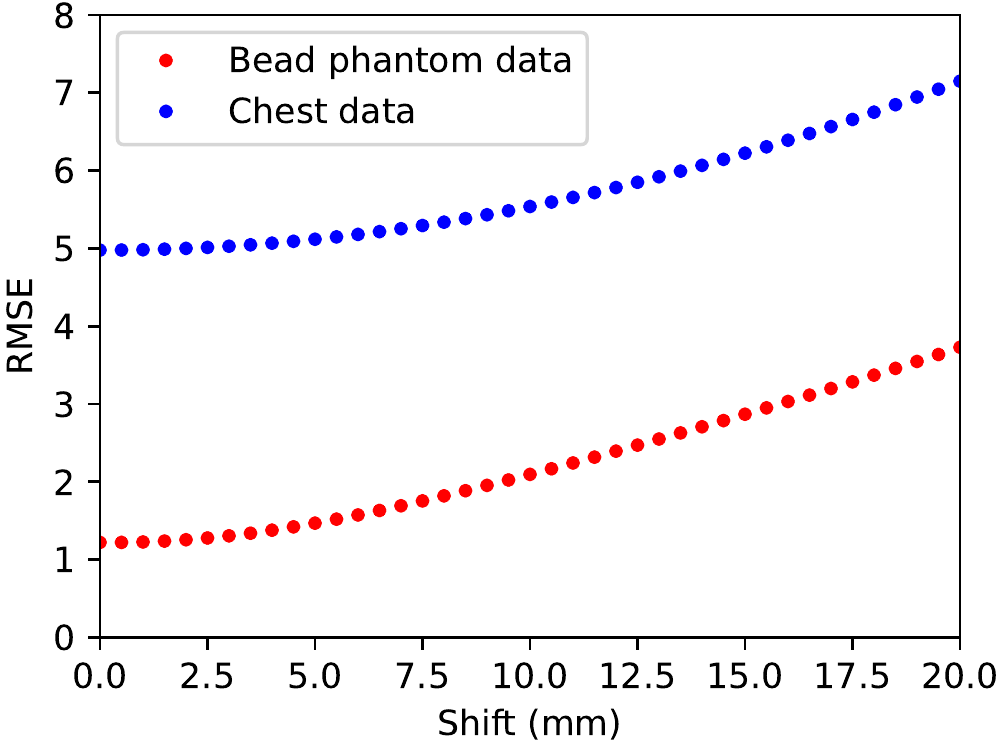}
\label{subfig:ErrorOverPP}
}
\end{minipage}
\begin{minipage}[t]{0.48\linewidth}
\subfigure[Respiratory phase error]{
\includegraphics[width=\linewidth]{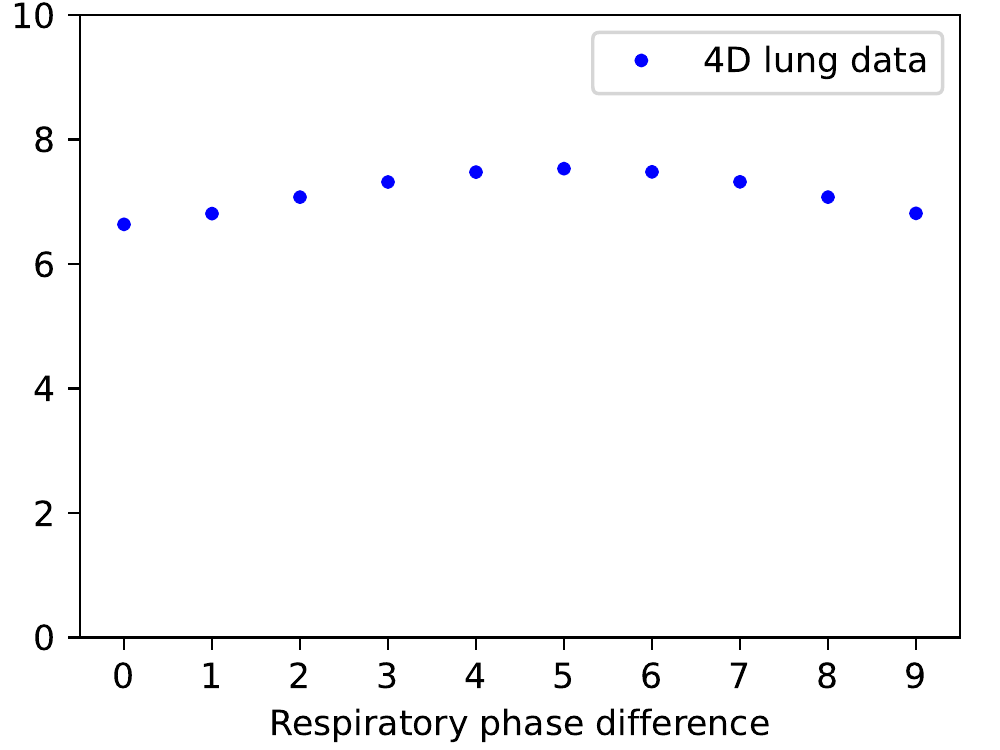}
\label{subfig:ErrorOverPhaseTorso}
}
\end{minipage}

\caption{\modified{The distribution of model performance over system errors for Pix2pixGAN with $0^\circ$ and $180^\circ$ polar inputs. (a) The rotation error is the rotation inaccuracy of the second (complementary) view; (b) The Pix2pixGAN model is trained with $D_\text{si}=750$\,mm; (c) The shift is the absolute distance of principal points on two complementary views; (d) One breath cycle is divided into 10 phases.}}
\label{Fig:SystemErrors}
\end{figure}

The robustness of Pix2pixGAN to geometric inaccuracy using $0^\circ$ and $180^\circ$ polar inputs on the bead phantom and chest data is displayed in Fig.\,\ref{Fig:SystemErrors}. Here we focus on Pix2pixGAN due to its persistence of decent performance on multiple datasets described above.
In Fig.\,\ref{subfig:ErrorOverDeg}, the RMSE values over rotation errors (i.e. the amount of deviation to $180^\circ$ for the second view) are displayed. When the rotation error increases from $0^\circ$ to $4.0^\circ$, RMSE increases from 1.22 to 2.72 on the bead phantom data, while it increases from 4.98 to 8.20 on the chest data. Nevertheless, within the mechanically achievable calibration accuracy of $0.5^\circ$, RMSE increases from 1.22 to 1.26 (3.3\% more) on the bead phantom data and from 4.98 to 5.08 (2.0\% more) on the chest data.

In Fig.\,\ref{subfig:ErrorOverSID}, the RMSE values over source-to-isocenter distance $D_\text{si}$ errors are displayed, where the Pix2pixGAN model is trained with $D_\text{si} = 750$\,mm. On both datasets, when new test data is acquired in a CBCT system with shorter source-to-isocenter distances than 750\,mm, obvious RMSE increase is observed. In contrast, when new test data is acquired in a CBCT system with longer source-to-isocenter distances than 750\,mm, RMSE increases relatively slower. For example, when $D_\text{si} = 650$\,mm, RMSE increases from 1.22 to 3.40 for the bead phantom data and from 4.98 to 6.02 for the chest data; when $D_\text{si} = 850$\,mm with the same 100\,mm deviation, RMSE increases from 1.22 to 2.03 for the bead phantom data and from 4.98 to 5.13 for the chest data. For a deviation of 10\,mm which is within the achievable calibration accuracy range, RMSE increases from 1.22 to 1.28 (4.9\% more) for the bead phantom data and from 4.98 to 5.04 (1.2\%) for the chest data.

In Fig.\,\ref{subfig:ErrorOverPP}, the RMSE values over the relative shifts of the two principal points on complementary views are displayed. The detector principal point shift is caused by mechanical instability. It is illustrated in the supplementary material (Fig.\,2 and Fig.\,3), where the principal point of the $0^\circ$ view is rebinned to the virtual detector center and the shift distance is measured in the virtual detector. When the shift increases from 0\,mm to 20\,mm, RMSE increases from 1.22 to 3.73 on the bead phantom data, while it increases from 4.98 to 7.15 on the chest data. In practice, such shift can be controlled in the range of 3.5\,mm. For example, in the real data used in this work, the majority of principal point shifts are within 10 pixels (1.6\,mm in the virtual detector) and all the views are within 20 pixels (3.2\,mm). Within this range (3.2\,mm), RMSE increases from 1.22 to 1.31 (6.9\% more) on the bead phantom data and from 4.98 to 5.05 (1.4\% more) on the chest data.

To investigate the influence of respiratory motion in the Pix2pixGAN model, a 4D lung dataset \cite{balik2013evaluation} is used like the chest data. In this dataset, one breath cycle is divided into 10 phases. Different phase combinations from 20 patients, in total 2000 input pairs are used for evaluation. Without phase difference (no motion influence), the mean RMSE is 6.64. Note that all the parameters to generate the 4D lung test projection data are the same as those for the chest data. The main difference is that the original volumes in the 4D lung dataset have a coarse slice distance of 3\,mm, while those in the previous chest data mainly have a slice distance of 1\,mm. Because of this resolution difference, the baseline RMSE changes from 4.98 to 6.64. In Fig.\,\ref{subfig:ErrorOverPhaseTorso}, when the phase differs from 1 to 5, the largest anatomical deformation caused by respiratory motion is reached. Correspondingly, RMSE increases from 6.81 to 7.53. When the phase differs from 5 to 9, RMSE decreases to 6.81 again. With additional respiratory motion tracking surrogates and gating techniques as in \cite{balik2013evaluation}, respiratory motion can be controlled within one phase for two CBCT views. In this case, RMSE increases from 6.64 to 6.81, which is only 2.5\% more.

\subsection{Real Knee Data with Metal Implants}
\begin{figure}
\centering
\begin{minipage}[t]{0.32\linewidth}
\subfigure[]{
\includegraphics[width=\linewidth]{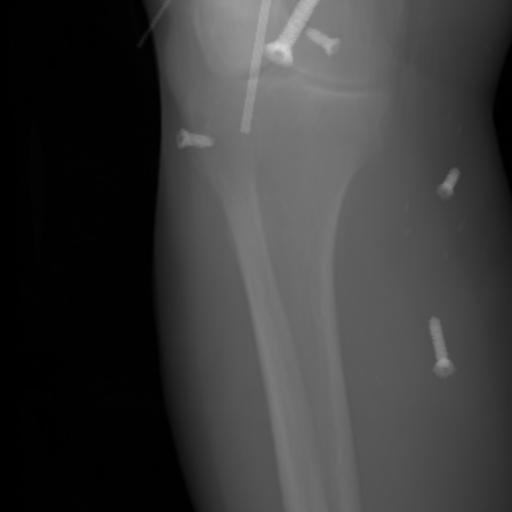}
\label{subfig:legTrain0}
}
\end{minipage}
\begin{minipage}[t]{0.32\linewidth}
\subfigure[]{
\includegraphics[width=\linewidth]{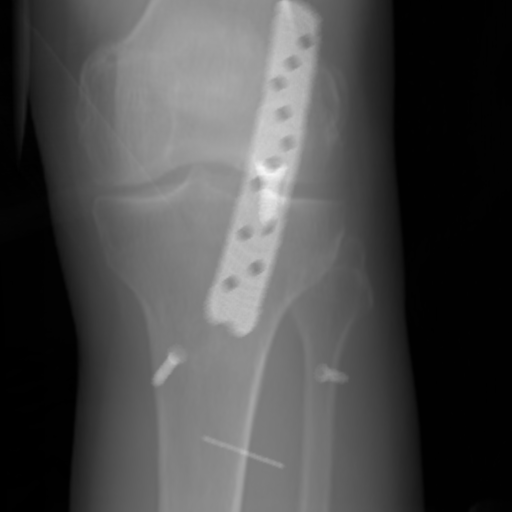}
\label{subfig:legTrain2}
}
\end{minipage}
\begin{minipage}[t]{0.32\linewidth}
\subfigure[]{
\includegraphics[width=\linewidth]{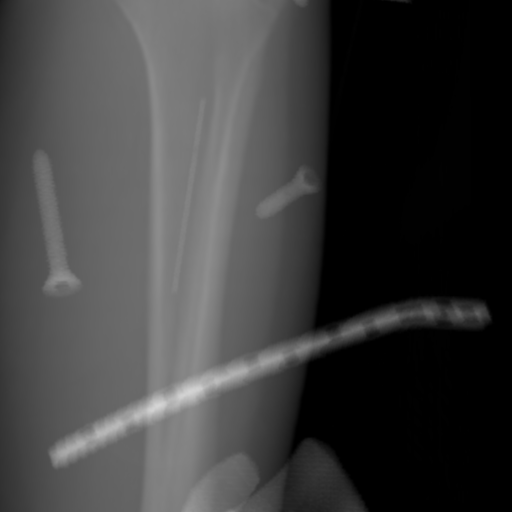}
\label{subfig:legTrain3}
}
\end{minipage}
\caption{\modified{Three examples of synthetic perspective projection images for training, intensity window: [0, 4]. The appearance (e.g., image contrast and metal image resolution) of such DRR training images is different from that of real projection images in Fig.\,\ref{Fig:ResultsReal}.}}
\label{Fig:cadaverTrainingExamples}
\end{figure}

\begin{figure}[h]
\centering
\begin{minipage}[t]{0.06\linewidth}
\centering
{\rotatebox{90}{\footnotesize{$0^\circ$ perspective}}}
\end{minipage}
\begin{minipage}[b]{0.3\linewidth}
\subfigure[]{
\includegraphics[width=\linewidth]{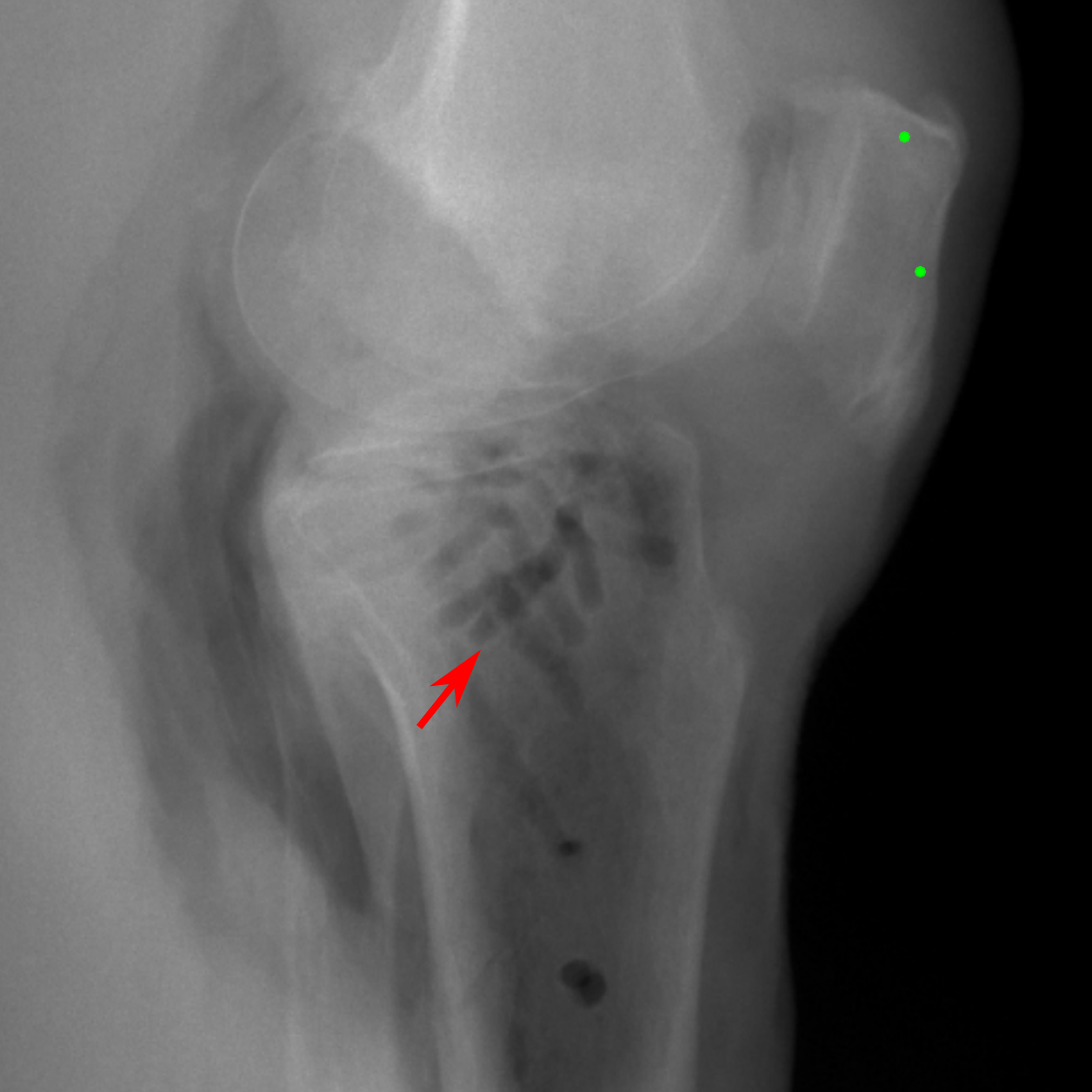}
\label{subfig:leg15120View0}
}
\end{minipage}
\begin{minipage}[b]{0.3\linewidth}
\subfigure[32.56, 109.85]{
\includegraphics[width=\linewidth]{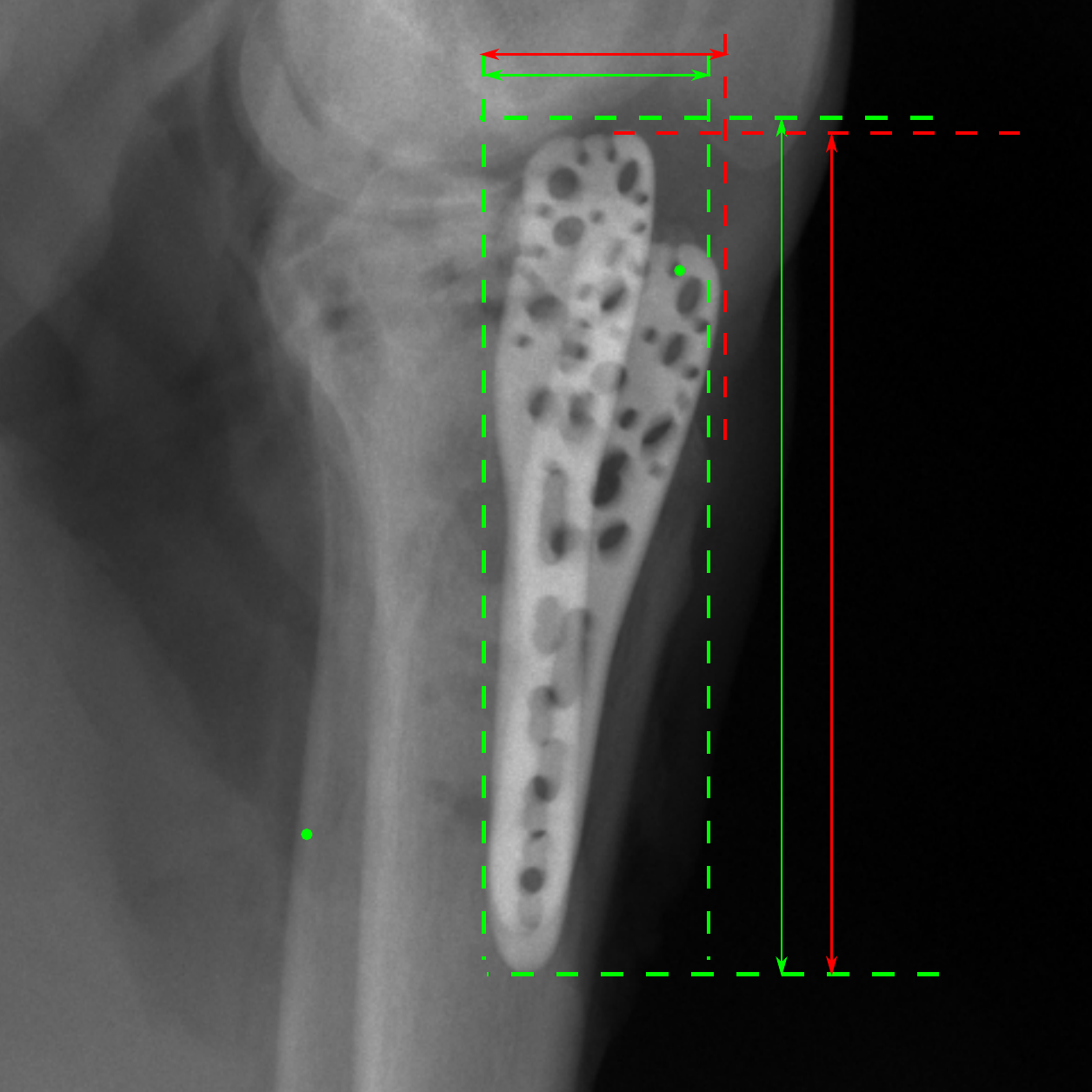}
\label{subfig:leg1250View0}
}
\end{minipage}
\begin{minipage}[b]{0.3\linewidth}
\subfigure[18.91, 19.54]{
\includegraphics[width=\linewidth]{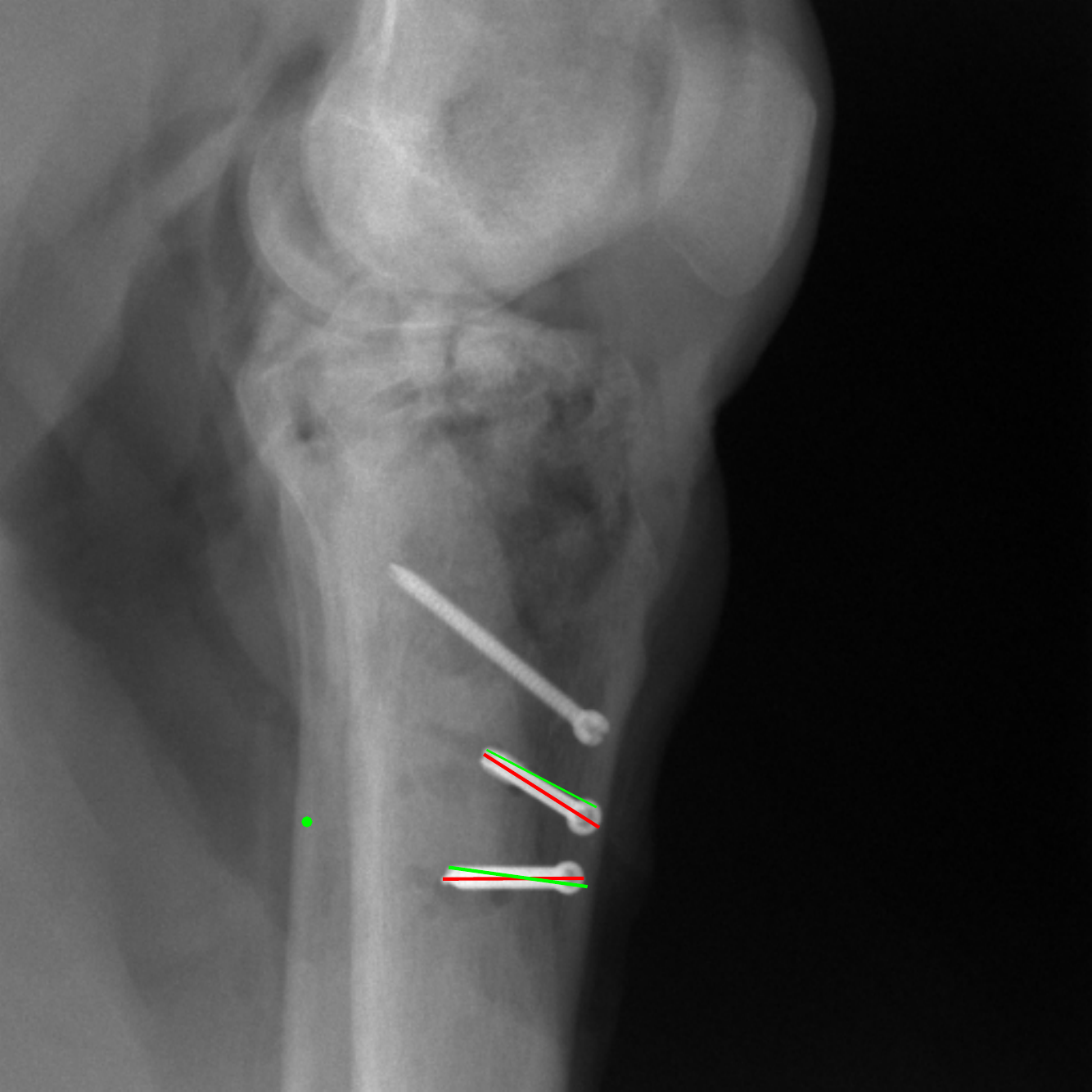}
\label{subfig:leg2270View0}
}
\end{minipage}

\begin{minipage}[t]{0.06\linewidth}
\centering
{\rotatebox{90}{\footnotesize{$180^\circ$ perspective}}}
\end{minipage}
\begin{minipage}[b]{0.3\linewidth}
\subfigure[]{
\includegraphics[width=\linewidth]{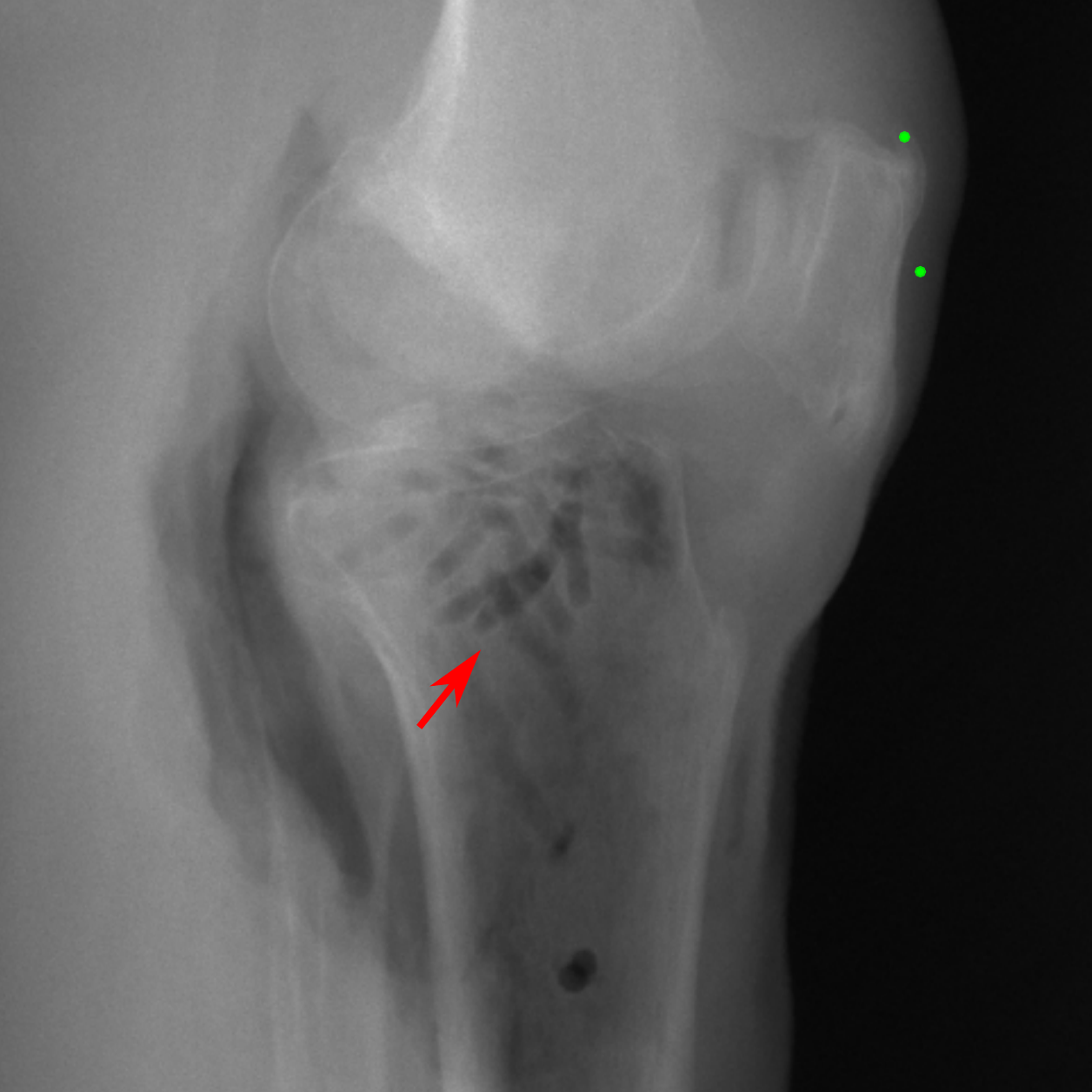}
\label{subfig:leg15120View180}
}
\end{minipage}
\begin{minipage}[b]{0.3\linewidth}
\subfigure[28.11, 112.16]{
\includegraphics[width=\linewidth]{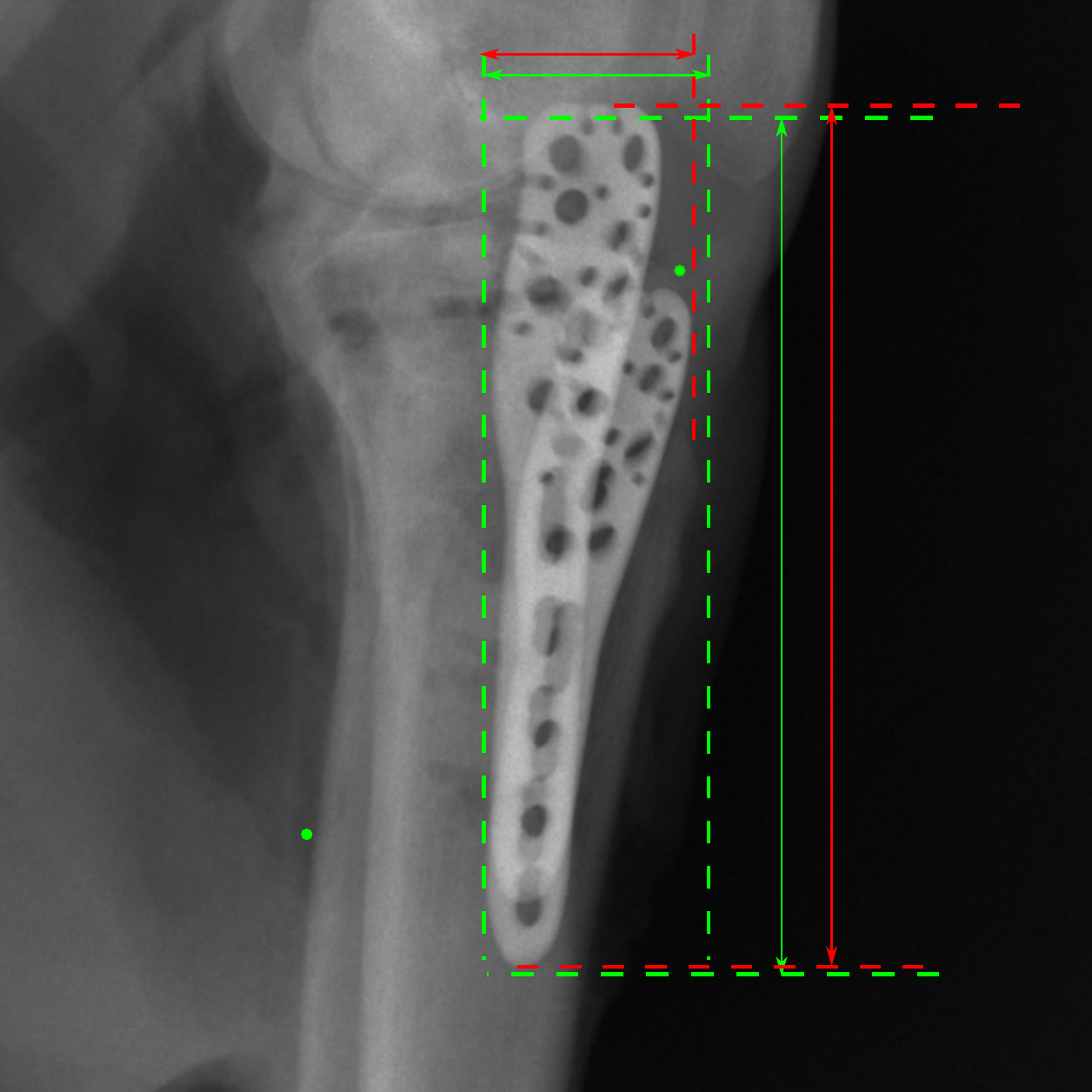}
\label{subfig:leg1250View180}
}
\end{minipage}
\begin{minipage}[b]{0.3\linewidth}
\subfigure[15.80, 19.27]{
\includegraphics[width=\linewidth]{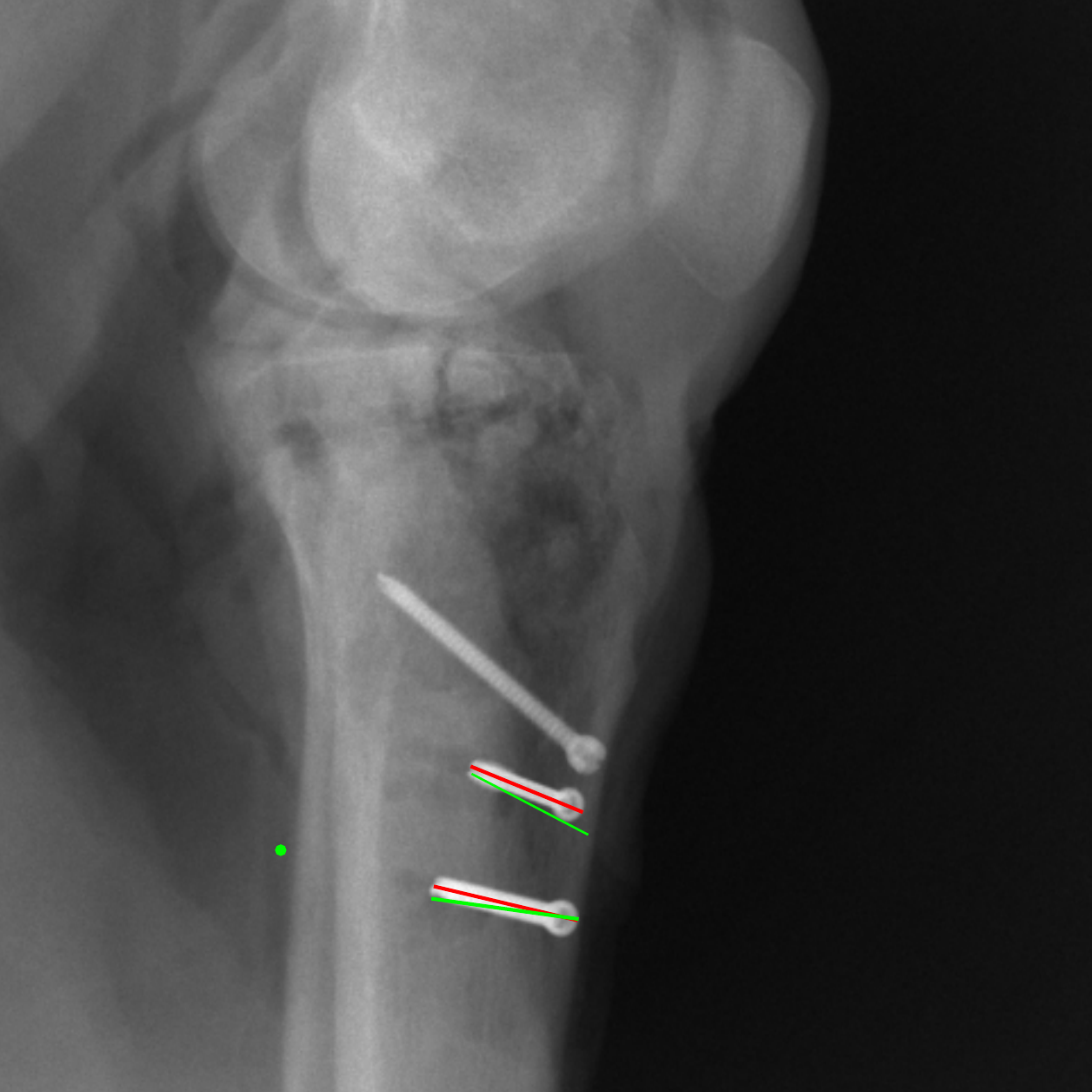}
\label{subfig:leg2270View180}
}
\end{minipage}

\begin{minipage}[t]{0.06\linewidth}
\centering
{\rotatebox{90}{\footnotesize{View difference}}}
\end{minipage}
\begin{minipage}[b]{0.3\linewidth}
\subfigure[]{
\includegraphics[width=\linewidth]{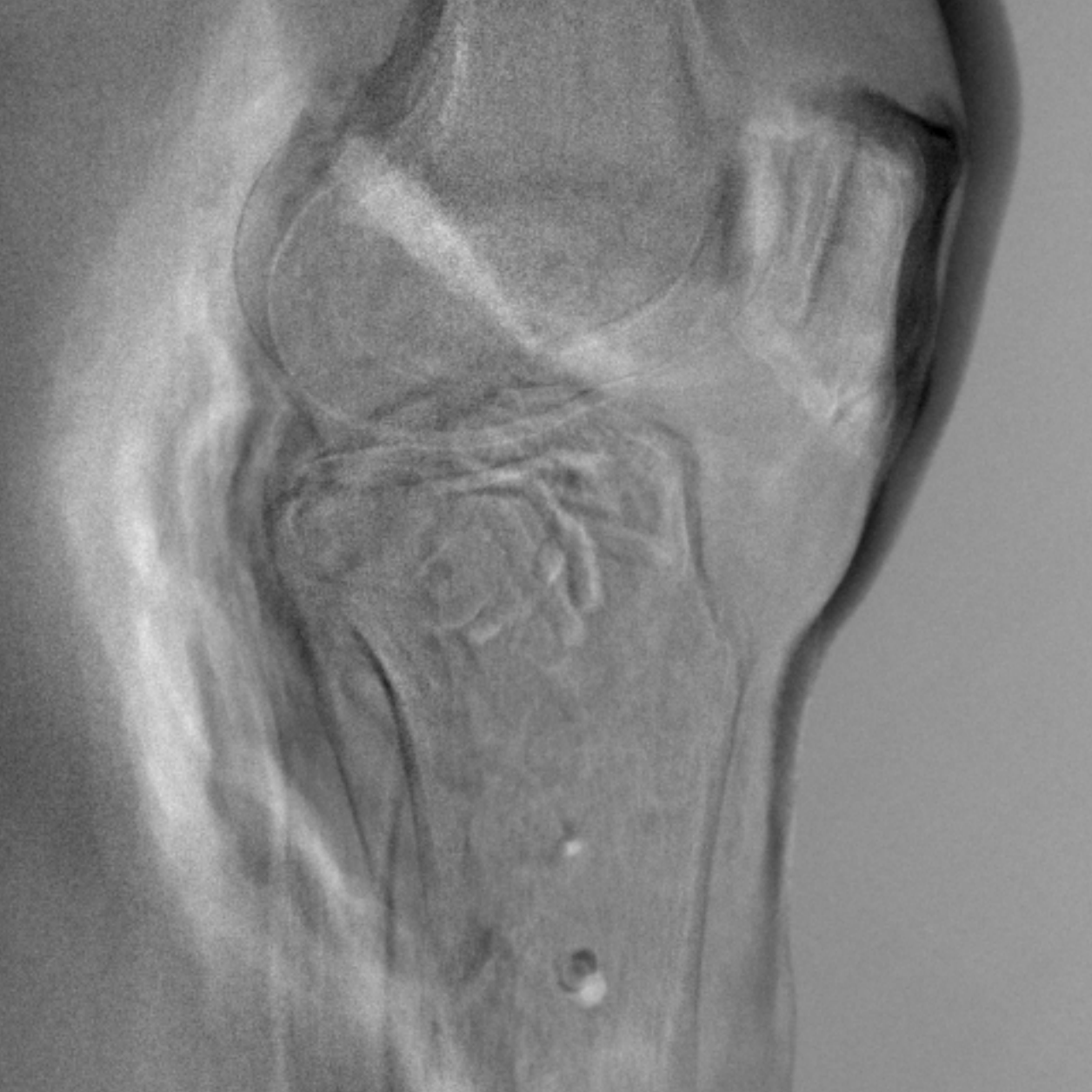}
\label{subfig:leg15120ViewDifference}
}
\end{minipage}
\begin{minipage}[b]{0.3\linewidth}
\subfigure[]{
\includegraphics[width=\linewidth]{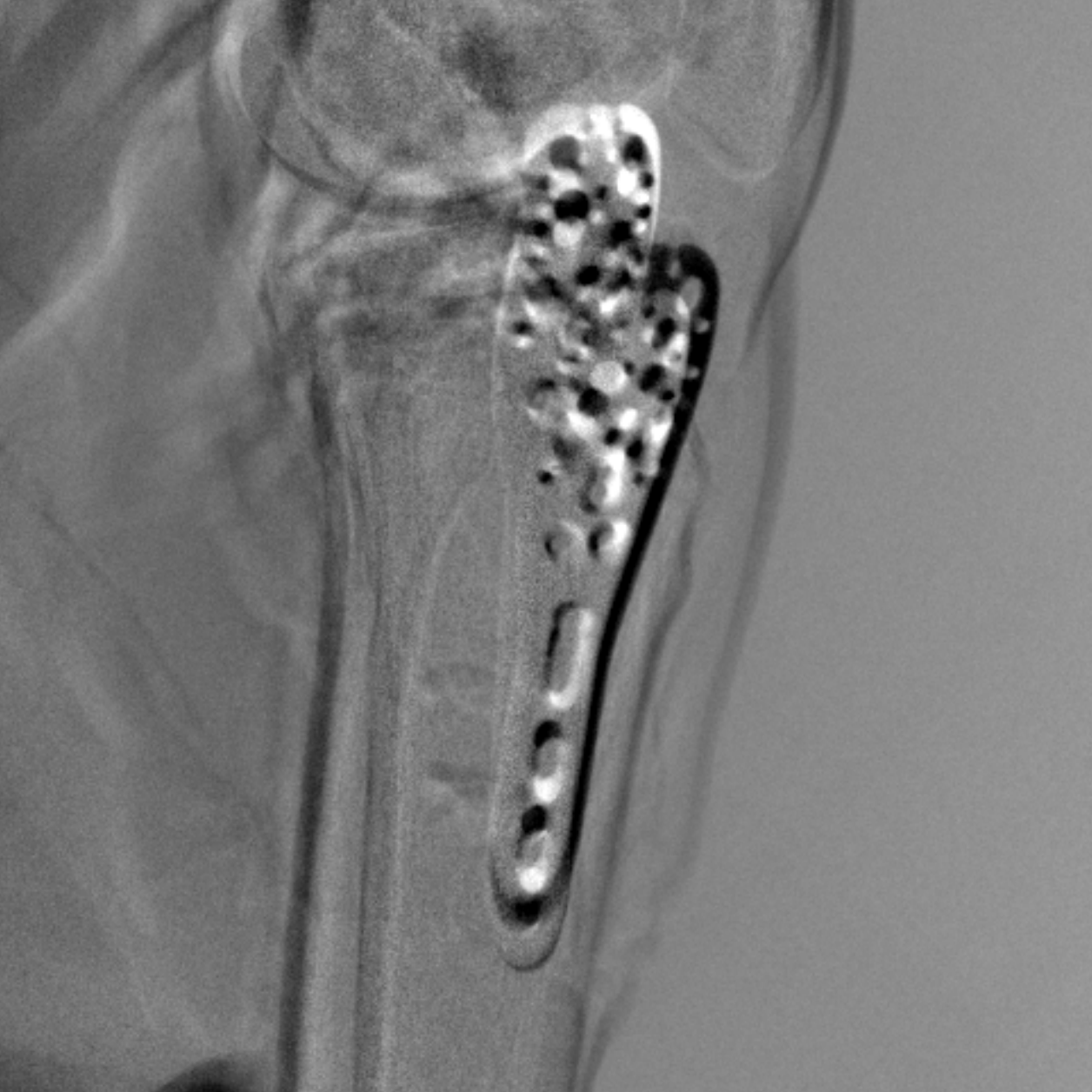}
\label{subfig:leg1250ViewDifference}
}
\end{minipage}
\begin{minipage}[b]{0.3\linewidth}
\subfigure[]{
\includegraphics[width=\linewidth]{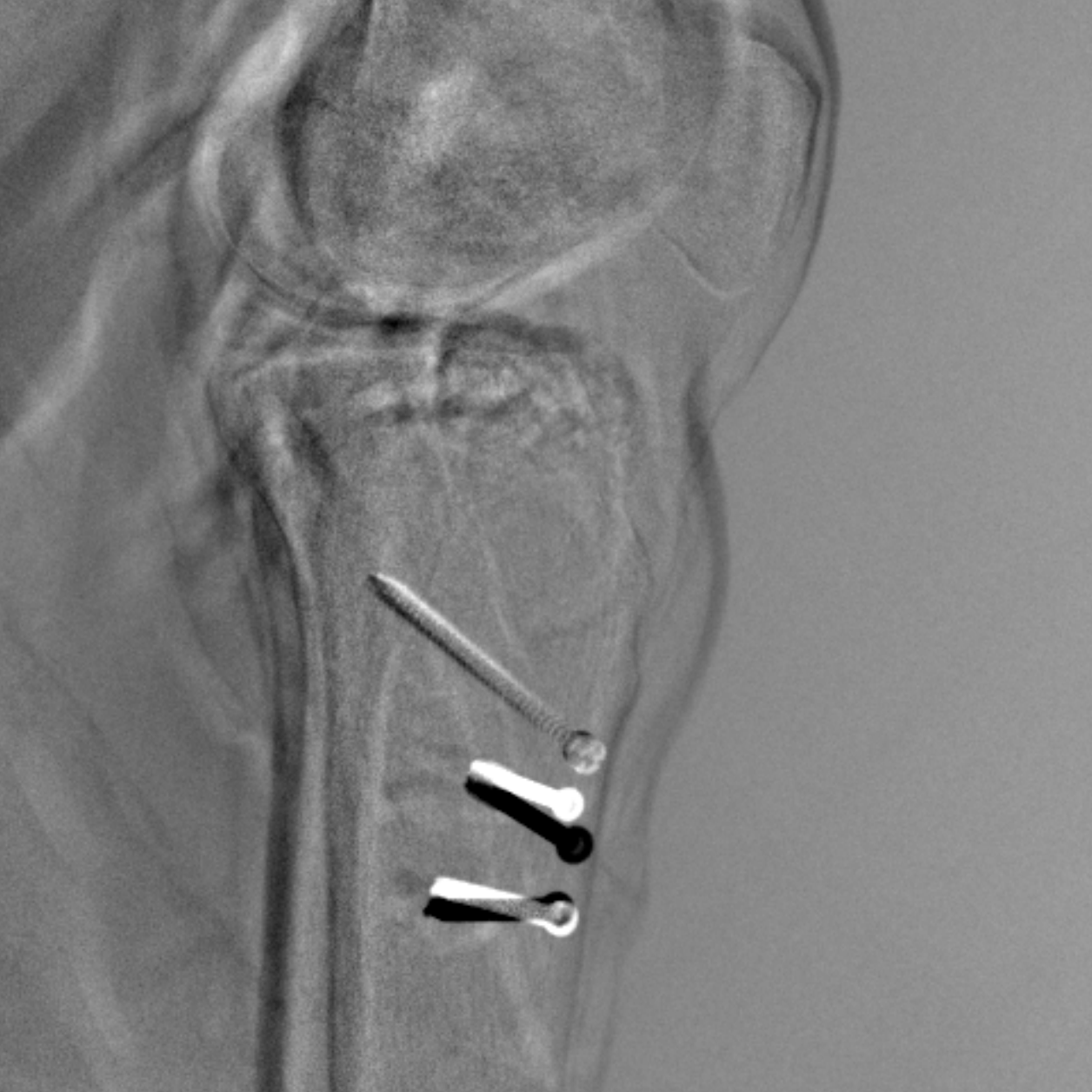}
\label{subfig:leg1250ViewDifference}
}
\end{minipage}

\begin{minipage}[t]{0.06\linewidth}
\centering
{\rotatebox{90}{\footnotesize{$0^\circ$ and $180^\circ$ RGB}}}
\end{minipage}
\begin{minipage}[b]{0.3\linewidth}
\subfigure[]{
\includegraphics[width=\linewidth]{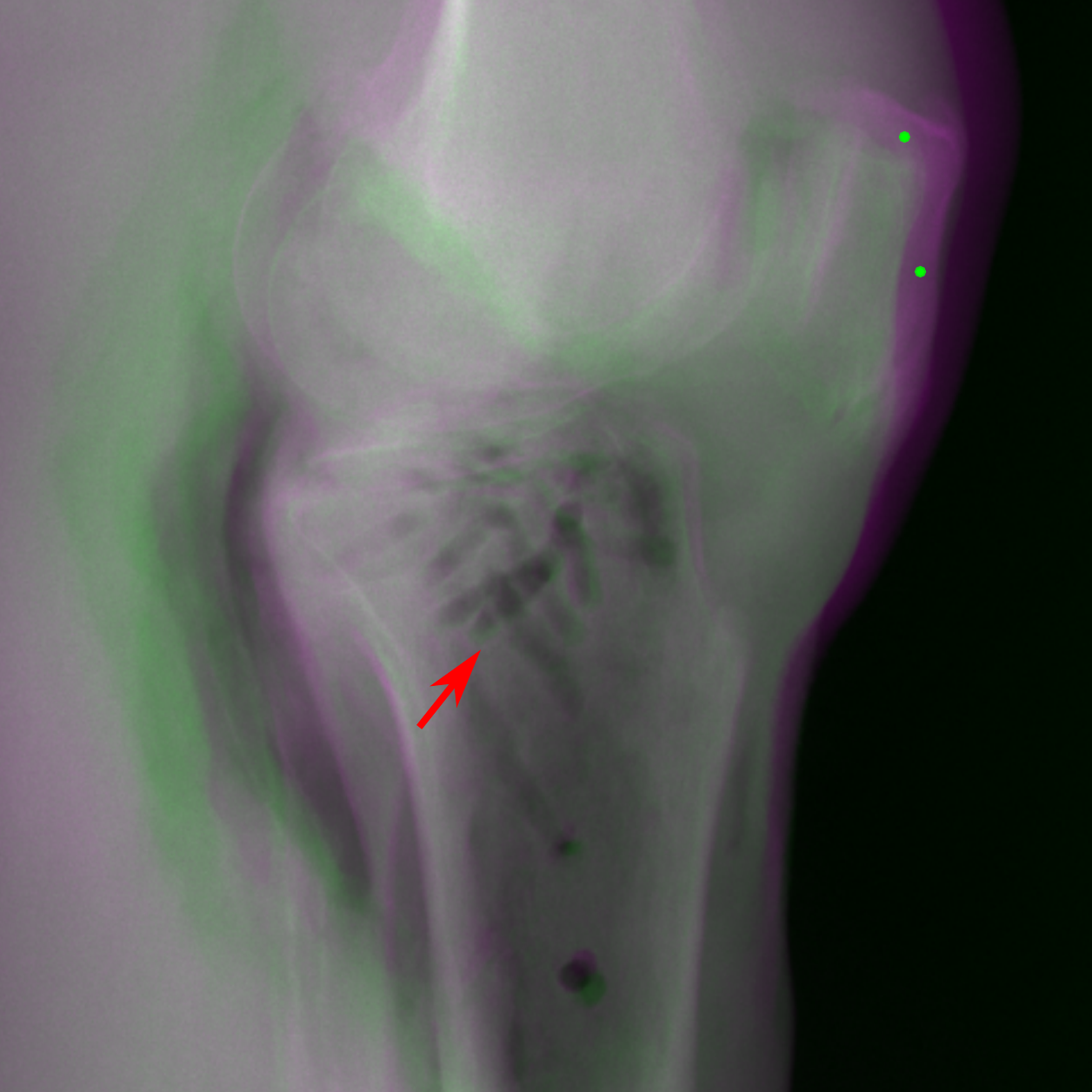}
\label{subfig:leg15120ViewRGB}
}
\end{minipage}
\begin{minipage}[b]{0.3\linewidth}
\subfigure[]{
\includegraphics[width=\linewidth]{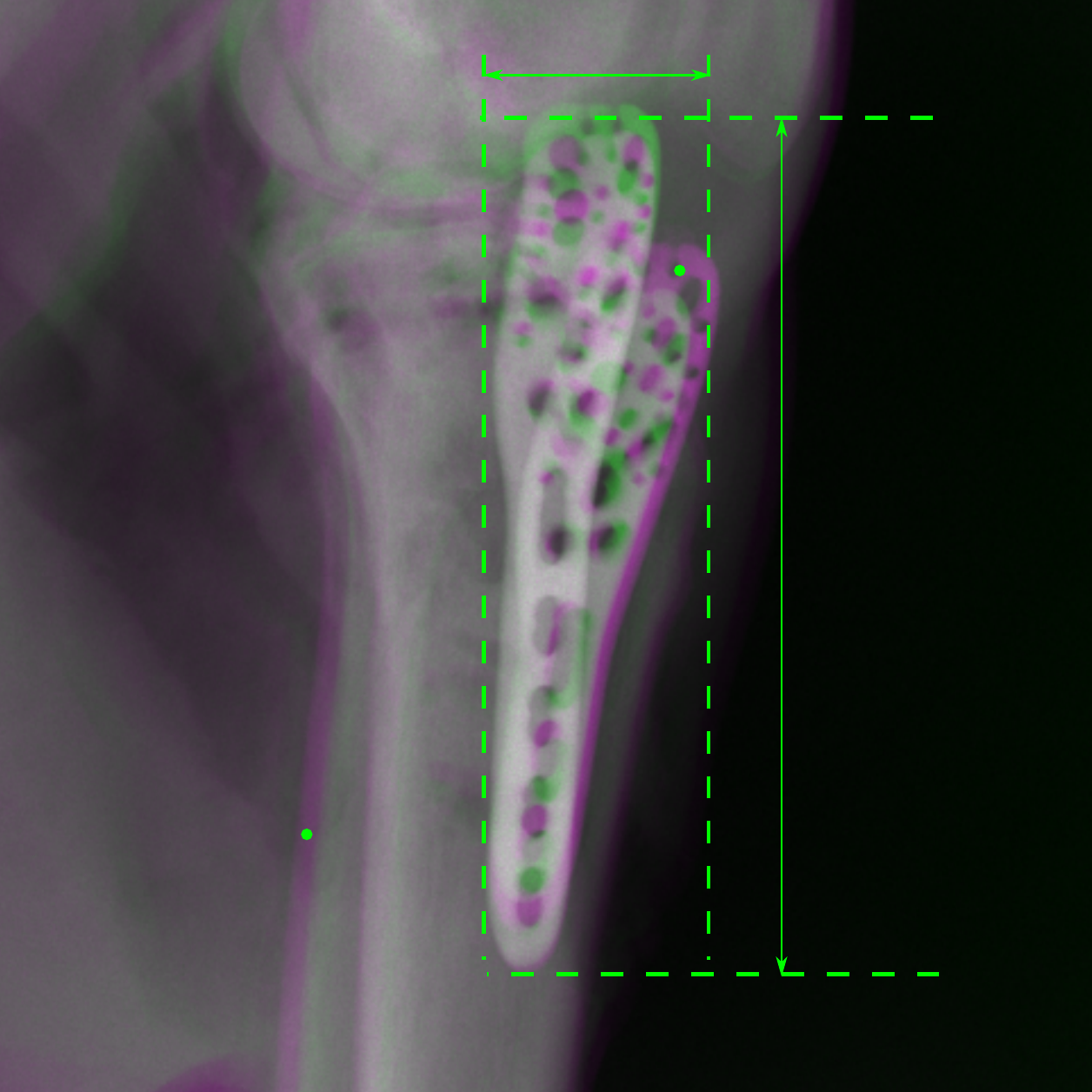}
\label{subfig:leg1250ViewRGB}
}
\end{minipage}
\begin{minipage}[b]{0.3\linewidth}
\subfigure[]{
\includegraphics[width=\linewidth]{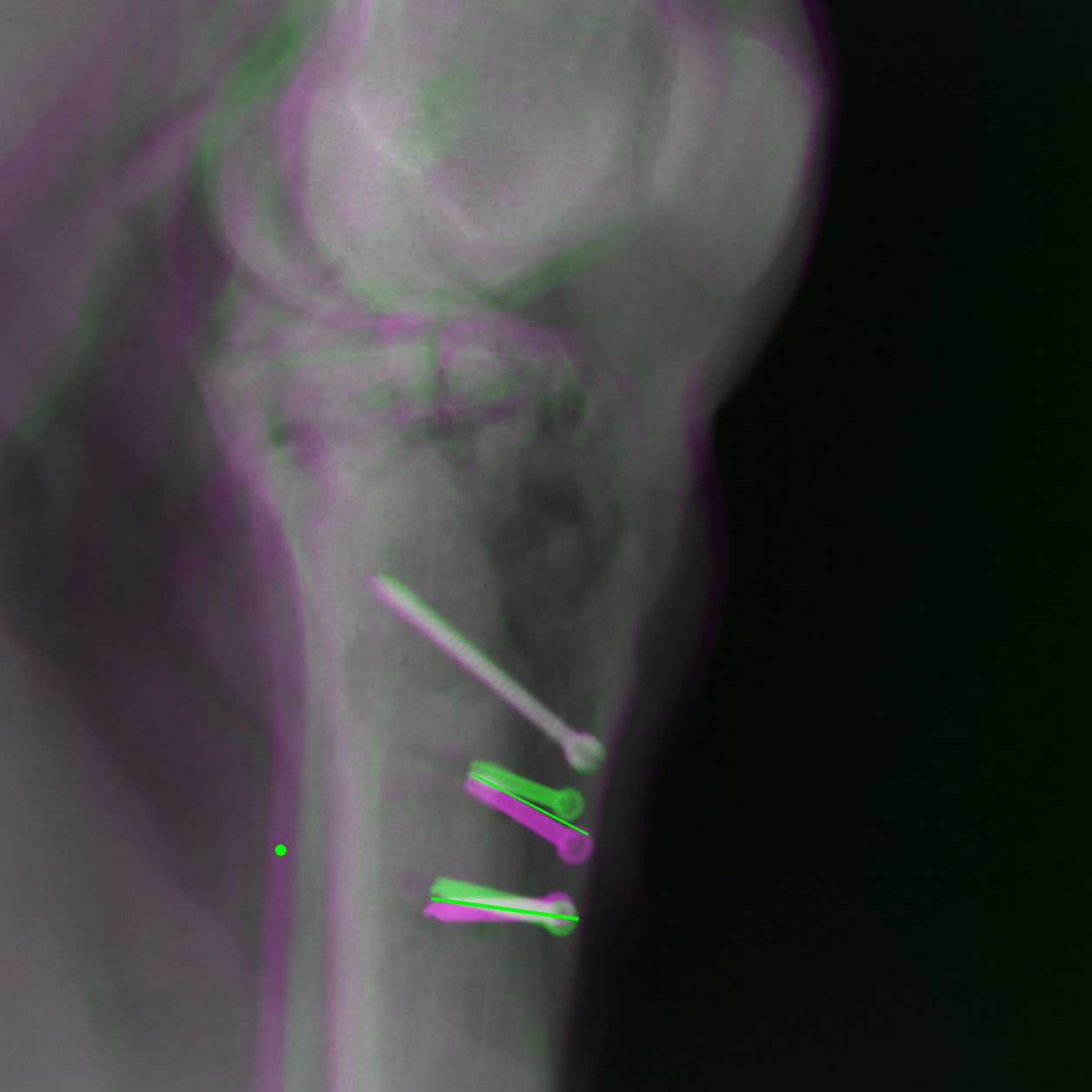}
\label{subfig:leg2270ViewRGB}
}
\end{minipage}

\begin{minipage}[t]{0.06\linewidth}
\centering
{\rotatebox{90}{\footnotesize{DRR reference}}}
\end{minipage}
\begin{minipage}[b]{0.3\linewidth}
\subfigure[]{
\includegraphics[width=\linewidth]{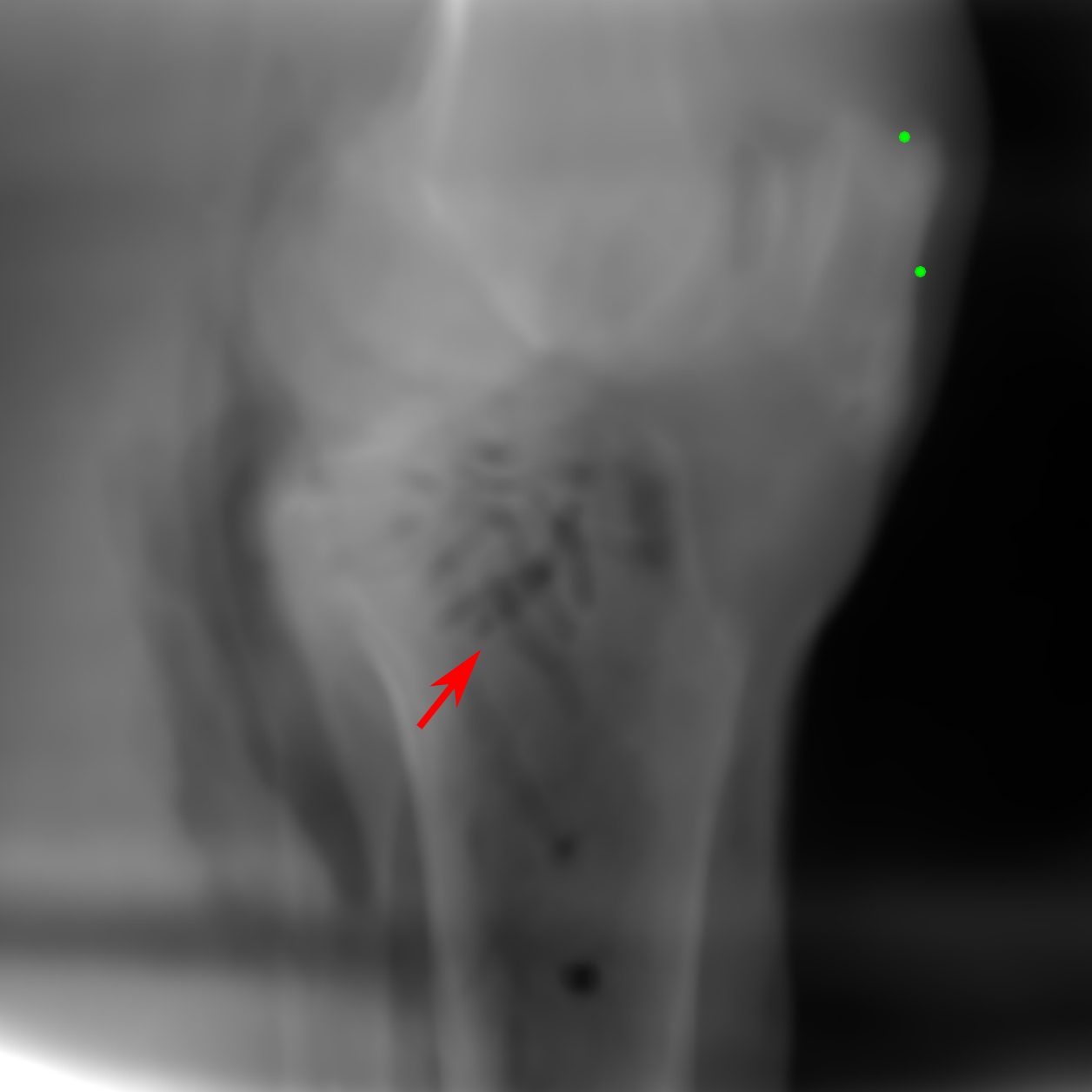}
\label{subfig:leg15120ViewReference}
}
\end{minipage}
\begin{minipage}[b]{0.3\linewidth}
\subfigure[29.72, 111.80]{
\includegraphics[width=\linewidth]{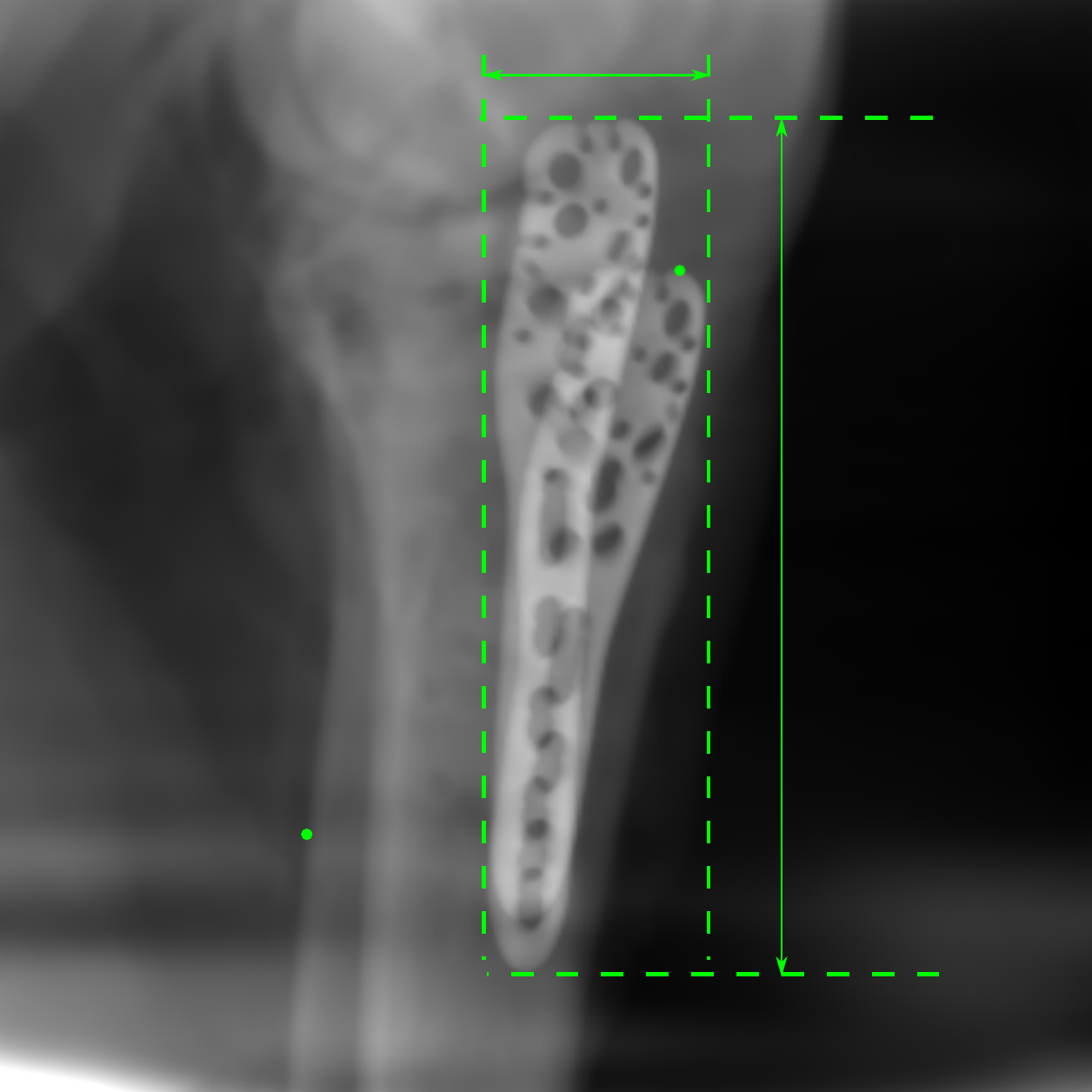}
\label{subfig:leg1250ViewReference}
}
\end{minipage}
\begin{minipage}[b]{0.3\linewidth}
\subfigure[17.14, 19.38]{
\includegraphics[width=\linewidth]{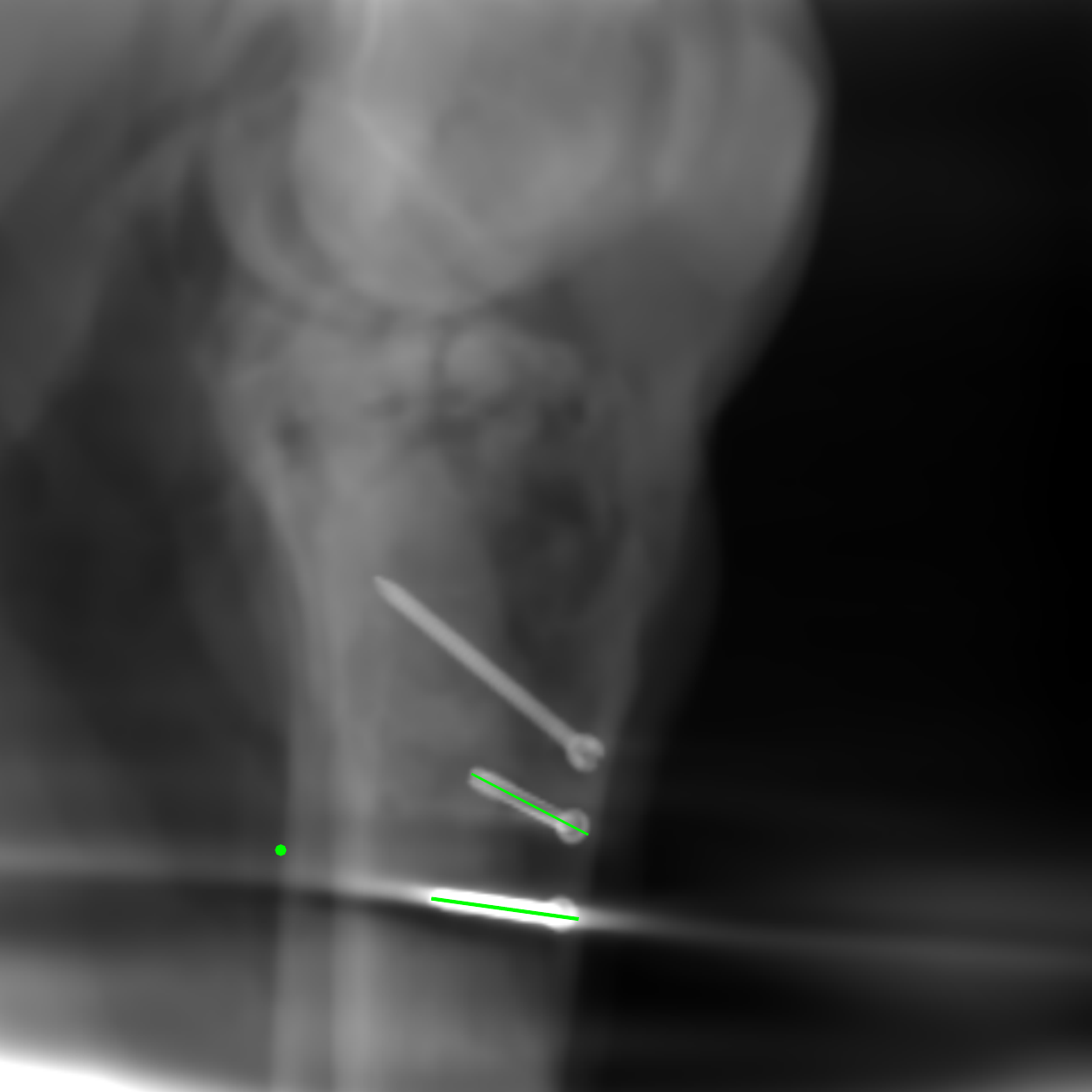}
\label{subfig:leg2270ViewReference}
}
\end{minipage}

\begin{minipage}[t]{0.06\linewidth}
\centering
{\rotatebox{90}{\footnotesize{Pix2pixGAN}}}
\end{minipage}
\begin{minipage}[b]{0.3\linewidth}
\subfigure[]{
\includegraphics[width=\linewidth]{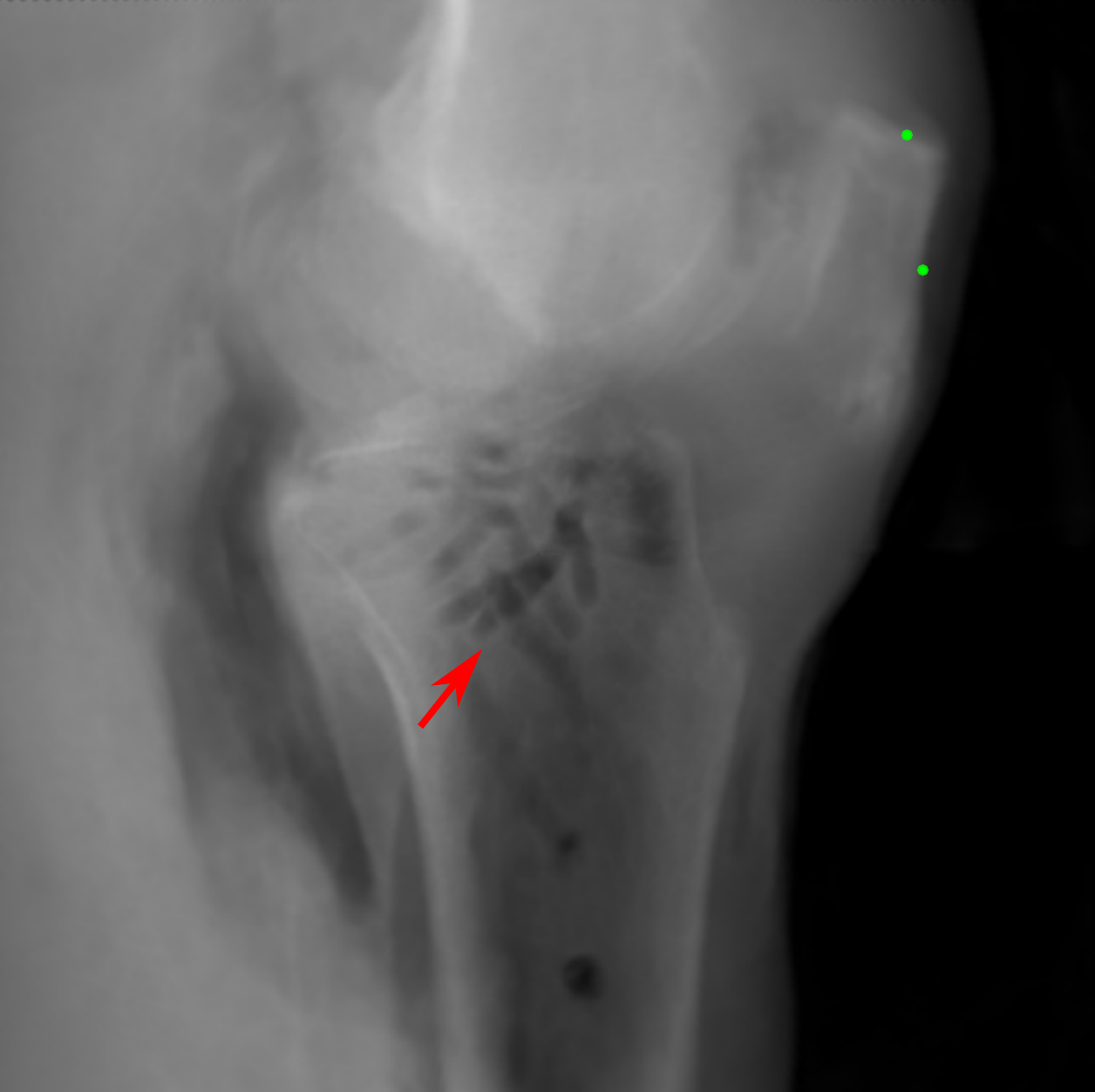}
\label{subfig:leg15120ViewPix2pixGAN}
}
\end{minipage}
\begin{minipage}[b]{0.3\linewidth}
\subfigure[29.72, 111.80]{
\includegraphics[width=\linewidth]{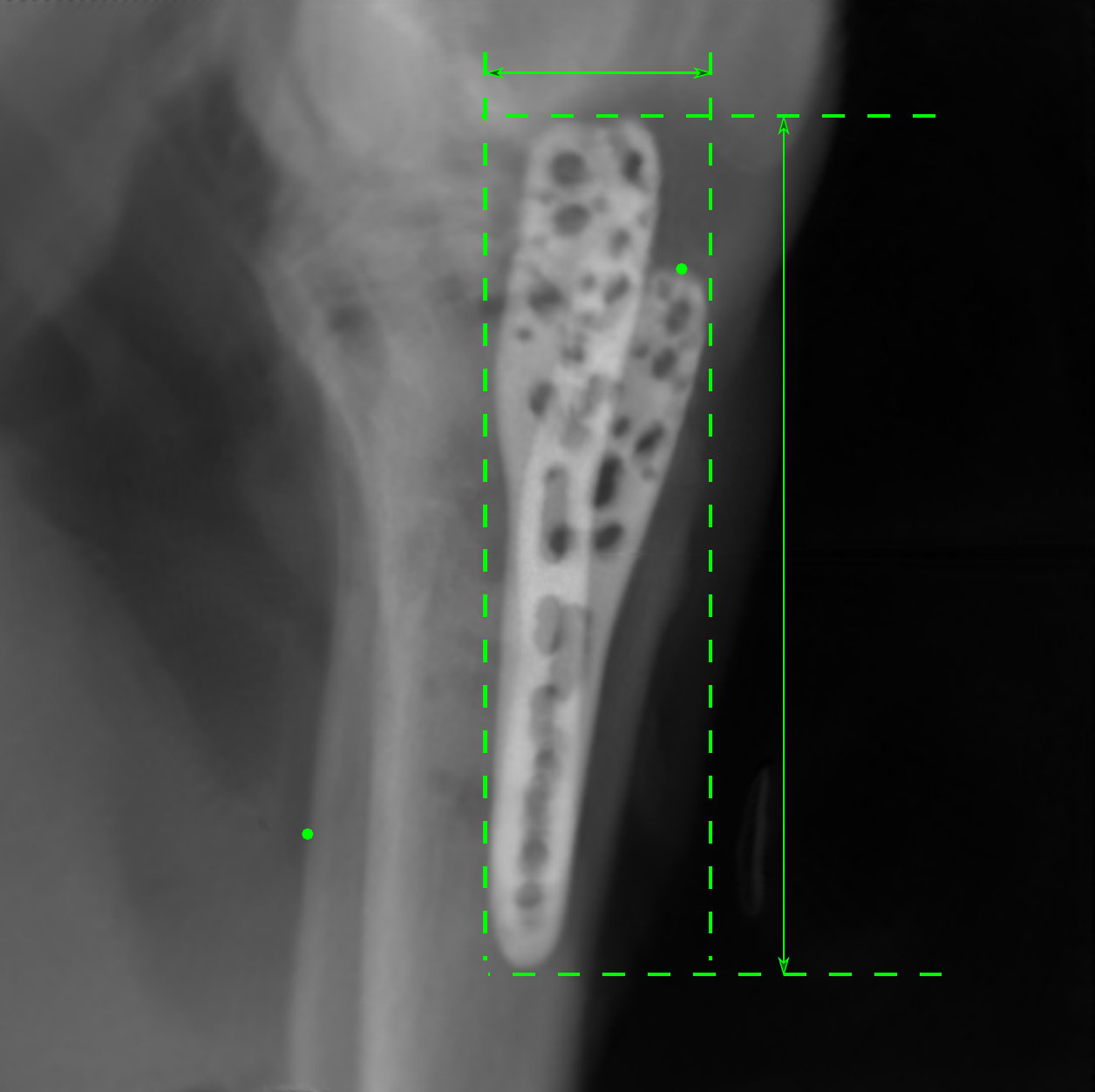}
\label{subfig:leg1250ViewPix2pixGAN}
}
\end{minipage}
\begin{minipage}[b]{0.3\linewidth}
\subfigure[17.50, 19.70]{
\includegraphics[width=\linewidth]{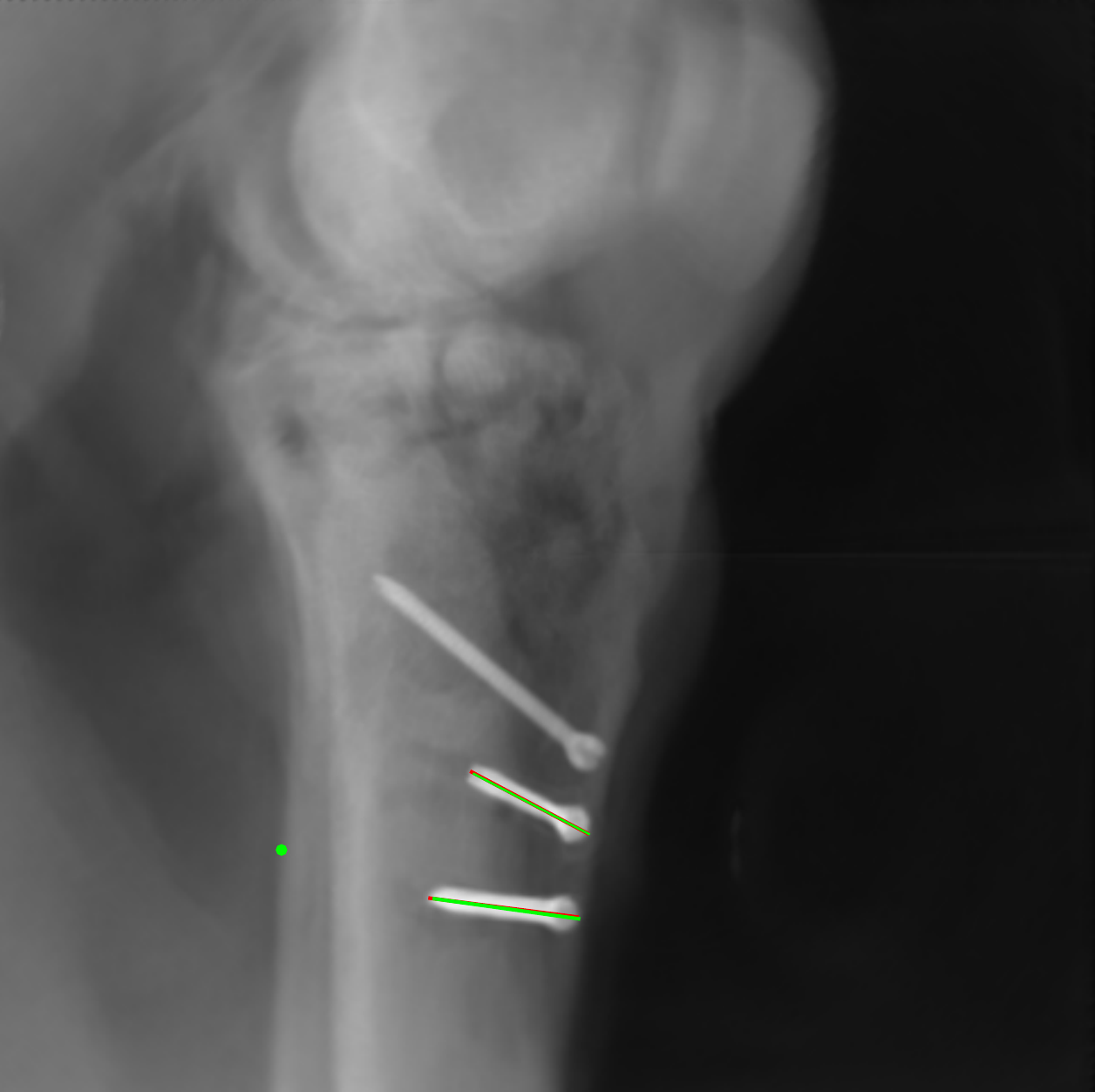}
\label{subfig:leg2270ViewPix2pixGAN}
}
\end{minipage}
(Fig.\,\ref{Fig:ResultsReal} continues in the next column.)
\end{figure}

\begin{figure}
\begin{minipage}[t]{0.06\linewidth}
\centering
{\rotatebox{90}{\footnotesize{TransU-Net}}}
\end{minipage}
\begin{minipage}[b]{0.3\linewidth}
\subfigure[]{
\includegraphics[width=\linewidth]{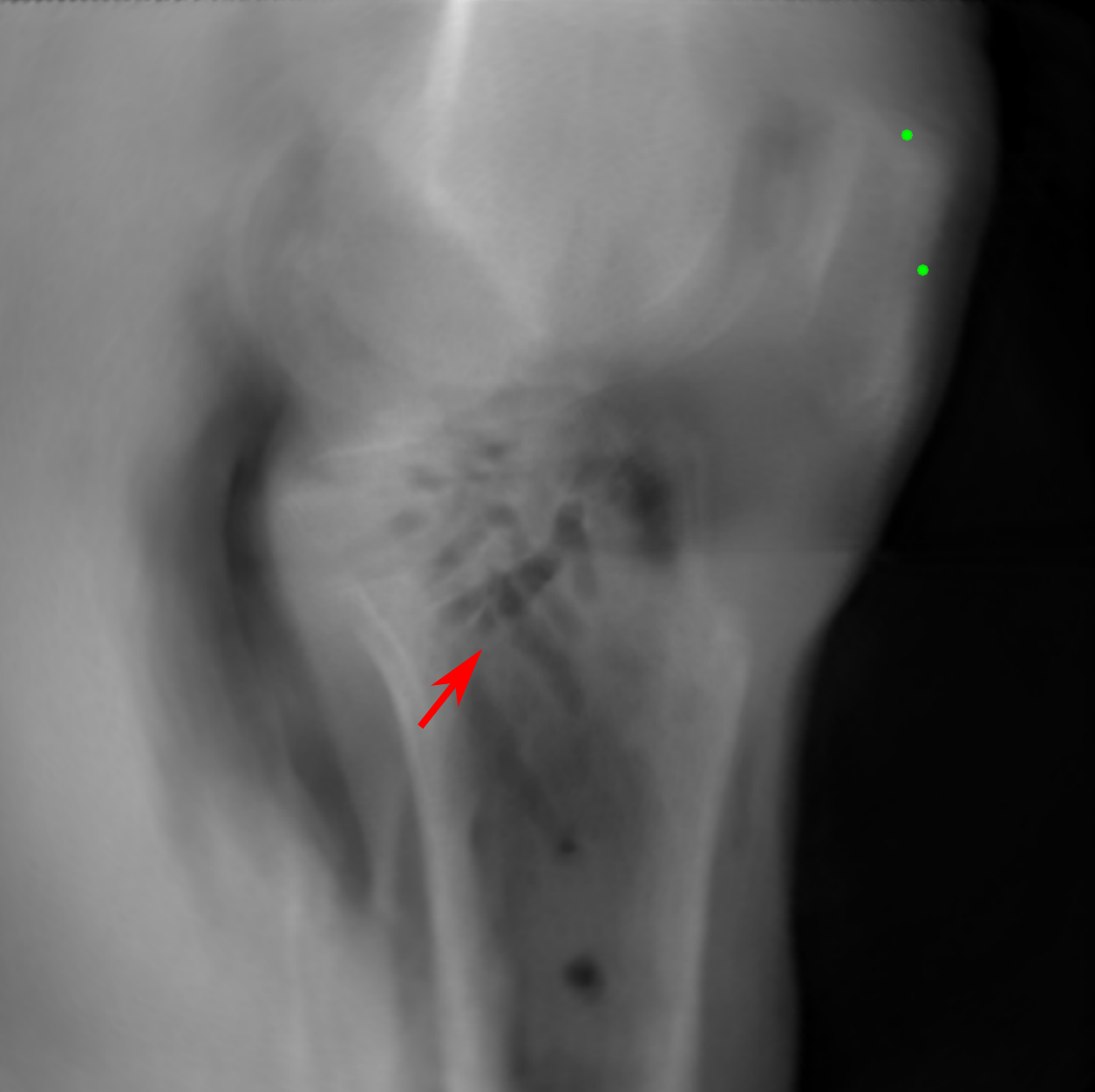}
\label{subfig:leg15120ViewTransUNet}
}
\end{minipage}
\begin{minipage}[b]{0.3\linewidth}
\subfigure[29.72, 111.80]{
\includegraphics[width=\linewidth]{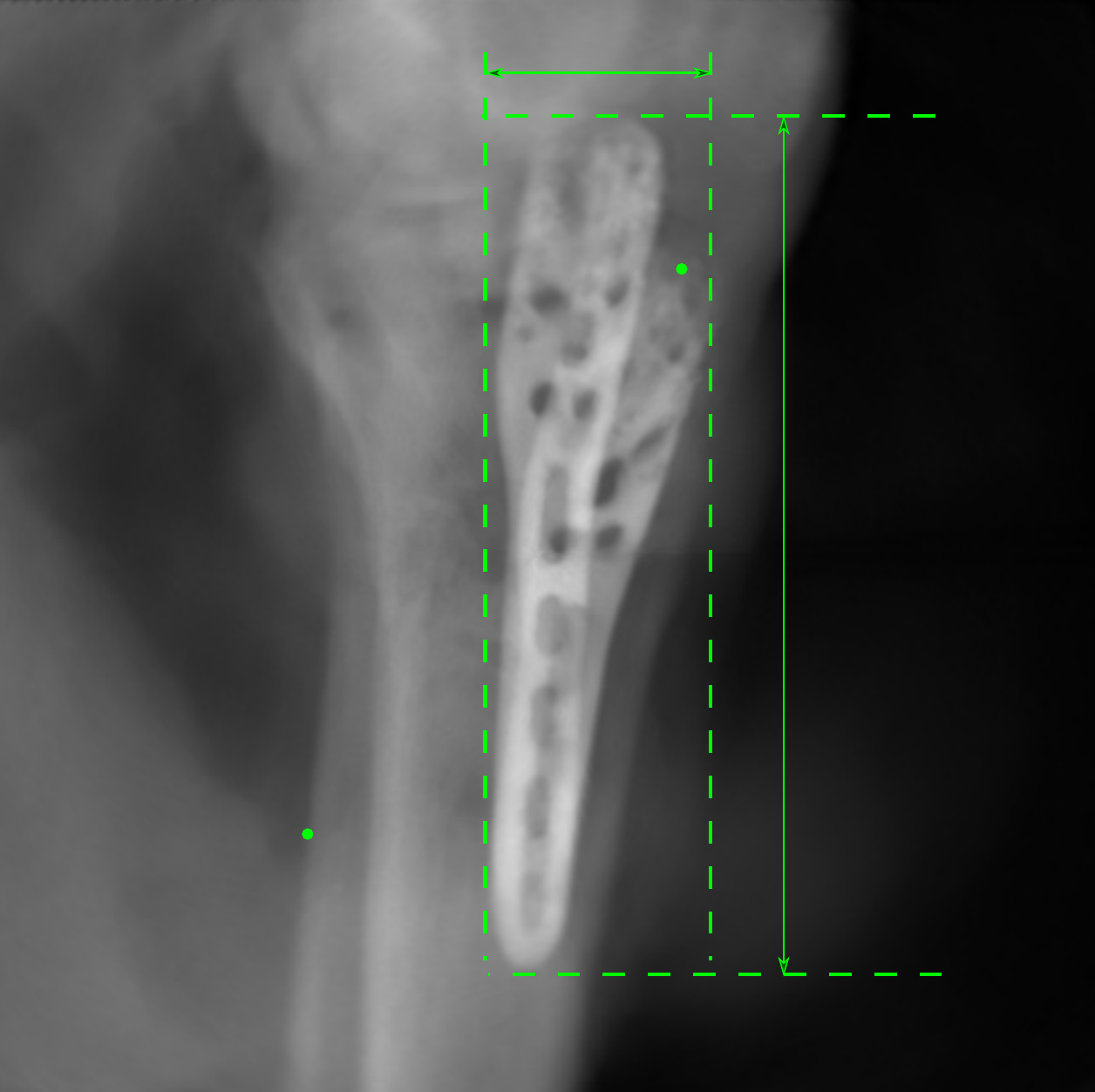}
\label{subfig:leg1250ViewTransUNet}
}
\end{minipage}
\begin{minipage}[b]{0.3\linewidth}
\subfigure[16.23, 19.25]{
\includegraphics[width=\linewidth]{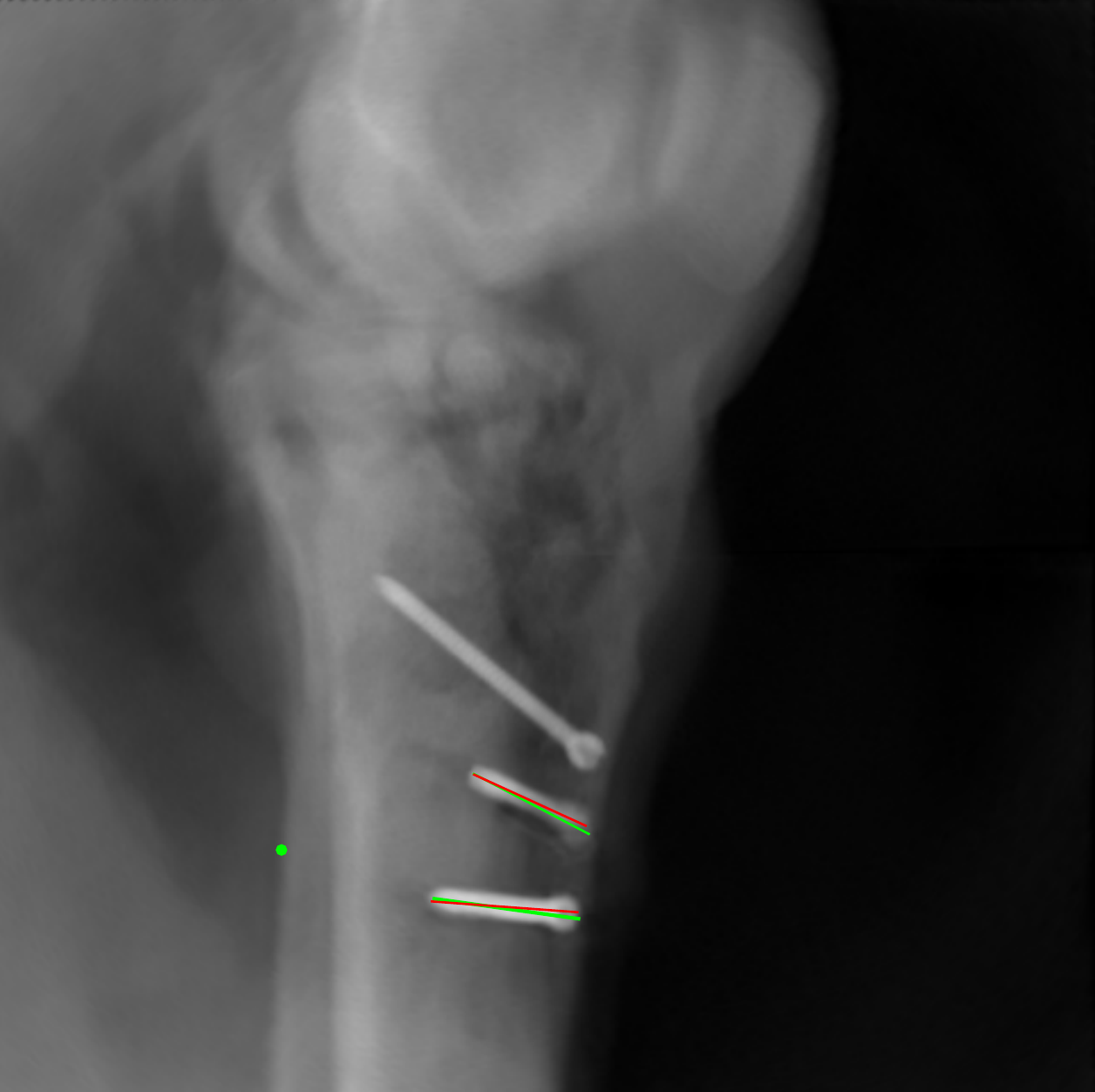}
\label{subfig:leg2270ViewTransUNet}
}
\end{minipage}
\caption{\modified{The results for the real cadaver leg data. The green dots mark the positions of landmarks in the corresponding reference images. In the first column, the arrows indicate the bone cavity/fracture structures. In the second column, the solid lines mark the widths and heights of the metals. The green ones are measured from the DRR reference, while the red ones are measured in the corresponding perspective projection image. In the last column, the lines mark the centerlines of the metal screws, where the green lines are those from the reference image. The lengths of the lines in mm are displayed in their corresponding subcaptions. Intensity window: [0, 4]. Please zoom in for better visualization.}}
\label{Fig:ResultsReal}
\end{figure}

The complementary view setting for learning perspective deformation is also evaluated on real CBCT projection data. In this evaluation, real CBCT projection data from a dataset of knees with metal implants is used for testing, while DRRs created from volumetric CT datasets with inserted metals is used for training. Three exemplary DRR perspective projection images for training are displayed in Fig.\,\ref{Fig:cadaverTrainingExamples}, in which synthetic metal implants are inserted \cite{fan2022simulation}. The appearance, e.g., image contrast and metal image resolution, of such DRR training images is different from that of real projection images in Fig.\,\ref{Fig:ResultsReal}. In Fig.\,\ref{Fig:ResultsReal}, the results for three knees, with and without metal implants, are displayed. The first and second rows are the $0^\circ$ and $180^\circ$ perspective projections, respectively, rebinned to the virtual detector with geometric calibration based on their respective principal points and projections of the world origin. The third row displays their difference images, where the magnitude of deviation increases from the center towards the outside like it does in DRRs with an ideal scan trajectory (e.g., Fig.\,\ref{subfig:diffPers0Sub180Example}), although real projection data suffer from various physical effects like beam hardening and Poisson noise. The fourth row displays the RGB stacks of $0^\circ$ and $180^\circ$ perspective projection images. The magenta and green regions indicate structures with considerable perspective deformation, for example, the knee patella in Fig.\,\ref{subfig:leg15120ViewRGB}, the top parts of the two metals in Fig.\,\ref{subfig:leg1250ViewRGB}, and the bottom two screws in Fig.\,\ref{subfig:leg2270ViewRGB}. The fifth row displays reference images, which are orthogonal projections of iterative reconstruction volumes from measured CBCT projection data. In the reference images, a total of five landmarks are selected, with the positions being marked by the green dots: In Fig.\,\ref{subfig:leg15120ViewReference}, two positions at the edges of the knee patella are marked; In Fig.\,\ref{subfig:leg1250ViewReference} and Fig.\,\ref{subfig:leg2270ViewReference}, one position at the left edge of the fibula is marked for each image. In addition, a rectangular frame for the two metals is marked by the green dashed lines, while its width and height are indicated by the green solid lines, which are 29.71\,mm and 111.99\,mm, respectively. In Fig.\,\ref{subfig:leg2270ViewReference}, the centerlines of the bottom two screws are sketched by the green lines, which have the lengths of 17.14\,mm (middle screw) and 19.38\,mm (bottom screw). The corresponding rectangular frame for the two metals and the screw centerlines in the perspective projection images are marked as well, but in red color. In Fig.\,\ref{subfig:leg1250View0}, the width and height of the metals are 32.56\,mm and 109.85\,mm, which have deviations of 2.84\,mm and -1.95\,mm to the reference ones, respectively. In Fig.\,\ref{subfig:leg2270View0}, the centerline lengths are 18.91\,mm and 19.54\,mm, which have deviations of 1.77\,mm and 0.16\,mm, respectively. Although the bottom screw has little length deviation to the reference, the orientations of both screws are obviously deviated. The sixth row shows the results of Pix2pixGAN using $0^\circ$ and $180^\circ$ polar inputs. For all of the landmarks, the green reference dots are all located accurately in the Pix2pixGAN images. The rectangular reference frame also accurately covers the metals in Fig.\,\ref{subfig:leg1250ViewPix2pixGAN}. In Fig.\,\ref{subfig:leg2270ViewPix2pixGAN}, although the two red centerlines do not exactly overlap with the green one, they are very close in lengths and orientations. Please zoom in for better visualization.

\begin{figure}
\centering
\begin{minipage}[t]{0.32\linewidth}
\subfigure[Real $0^\circ$ perspective]{
\includegraphics[width=\linewidth]{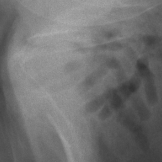}
\label{subfig:legROIperspective0}
}
\end{minipage}
\begin{minipage}[t]{0.32\linewidth}
\subfigure[DRR reference]{
\includegraphics[width=\linewidth]{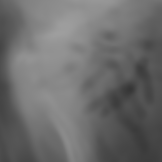}
\label{subfig:legROIreference}
}
\end{minipage}
\begin{minipage}[t]{0.32\linewidth}
\subfigure[Pix2pixGAN]{
\includegraphics[width=\linewidth]{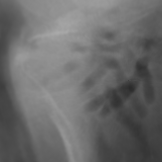}
\label{subfig:legROIpix2pixGAN}
}
\end{minipage}
\caption{The enlarged ROIs containing the cavity structures to demonstrate differences in image resolution.}
\label{Fig:LegROIs}
\end{figure}

The region-of-interest covering the cavity/fracture structures indicated by the red arrow in Fig.\,\ref{Fig:ResultsReal} has been enlarged in Fig.\,\ref{Fig:LegROIs} for better visualization of image resolution. In the real $0^\circ$ perspective projection ROI (Fig.\,\ref{subfig:legROIperspective0}), the cavities and bone edges appear sharp. The presence of Poisson noise can also be visualized to some degree. In the DRR reference ROI (Fig.\,\ref{subfig:legROIreference}), the cavities and bone edges appear blurry. This is likely caused by the partial volume effect in the intermediate 3D reconstruction volumes. In the Pix2pixGAN output (Fig.\,\ref{subfig:legROIpix2pixGAN}), there is a slight smoothing effect. For example, the fine edge indicated by the arrow is blurred and the Poisson noise is reduced. But in general, image resolution is preserved for most anatomical structures, e.g., the cavities.

\section{Limitations}

\begin{figure}
\centering
\begin{minipage}[b]{0.3\linewidth}
\subfigure[Reference]{
\includegraphics[width=\linewidth]{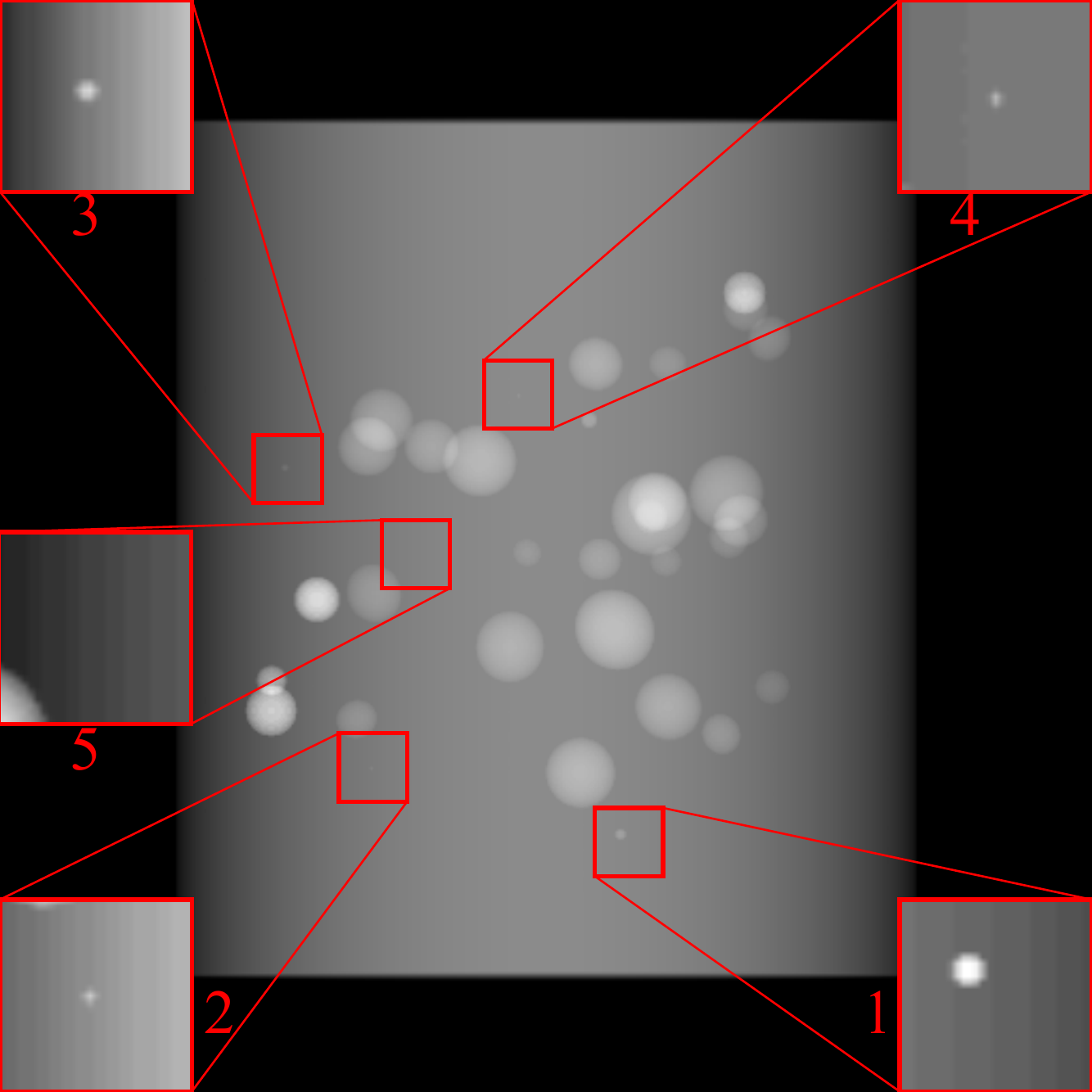}
\label{subfig:10TargetArrow}
}
\end{minipage}
\begin{minipage}[b]{0.3\linewidth}
\subfigure[Pix2pixGAN]{
\includegraphics[width=\linewidth]{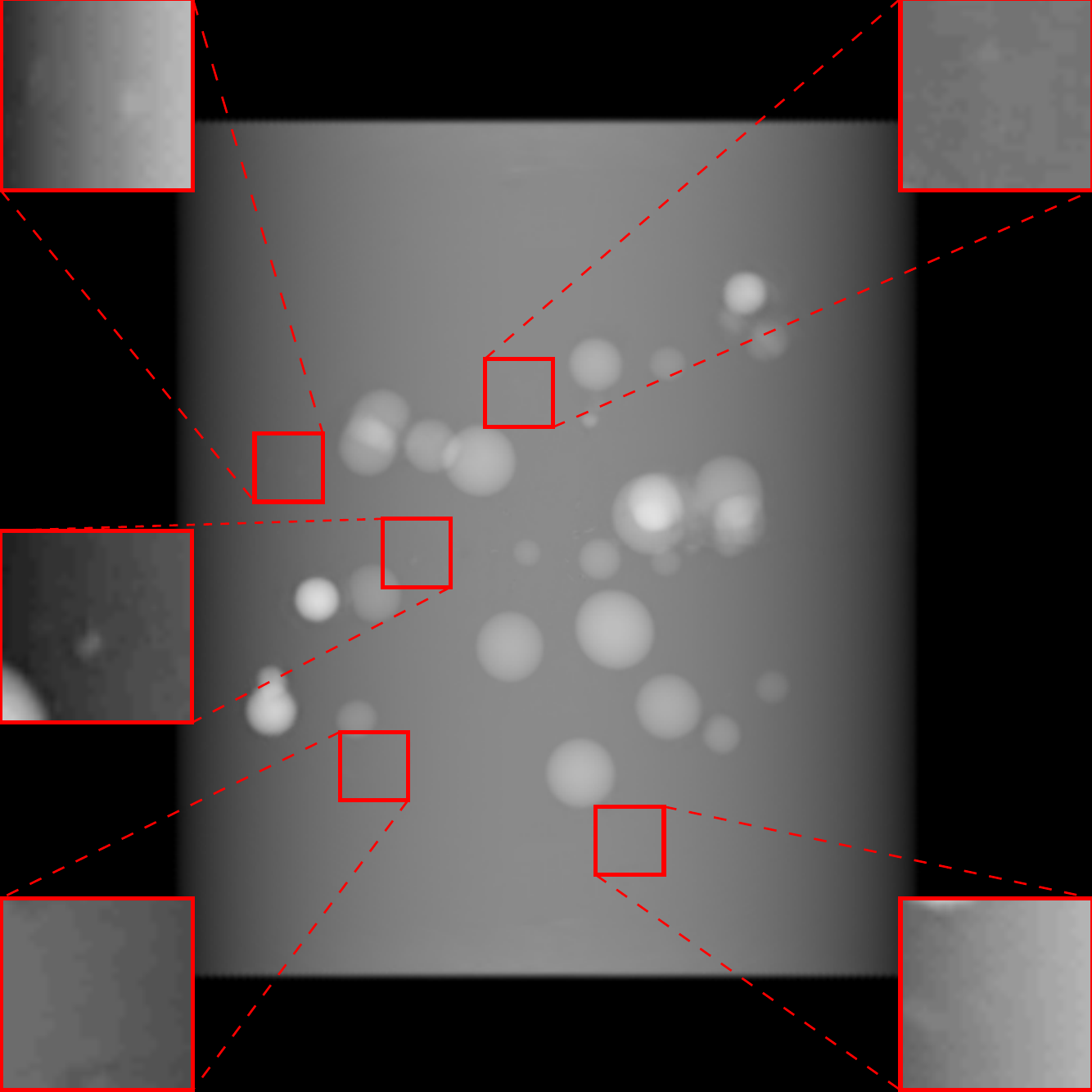}
\label{subfig:10PredictionArrow}
}
\end{minipage}
\begin{minipage}[b]{0.3\linewidth}
\subfigure[(b)-(a)]{
\includegraphics[width=\linewidth]{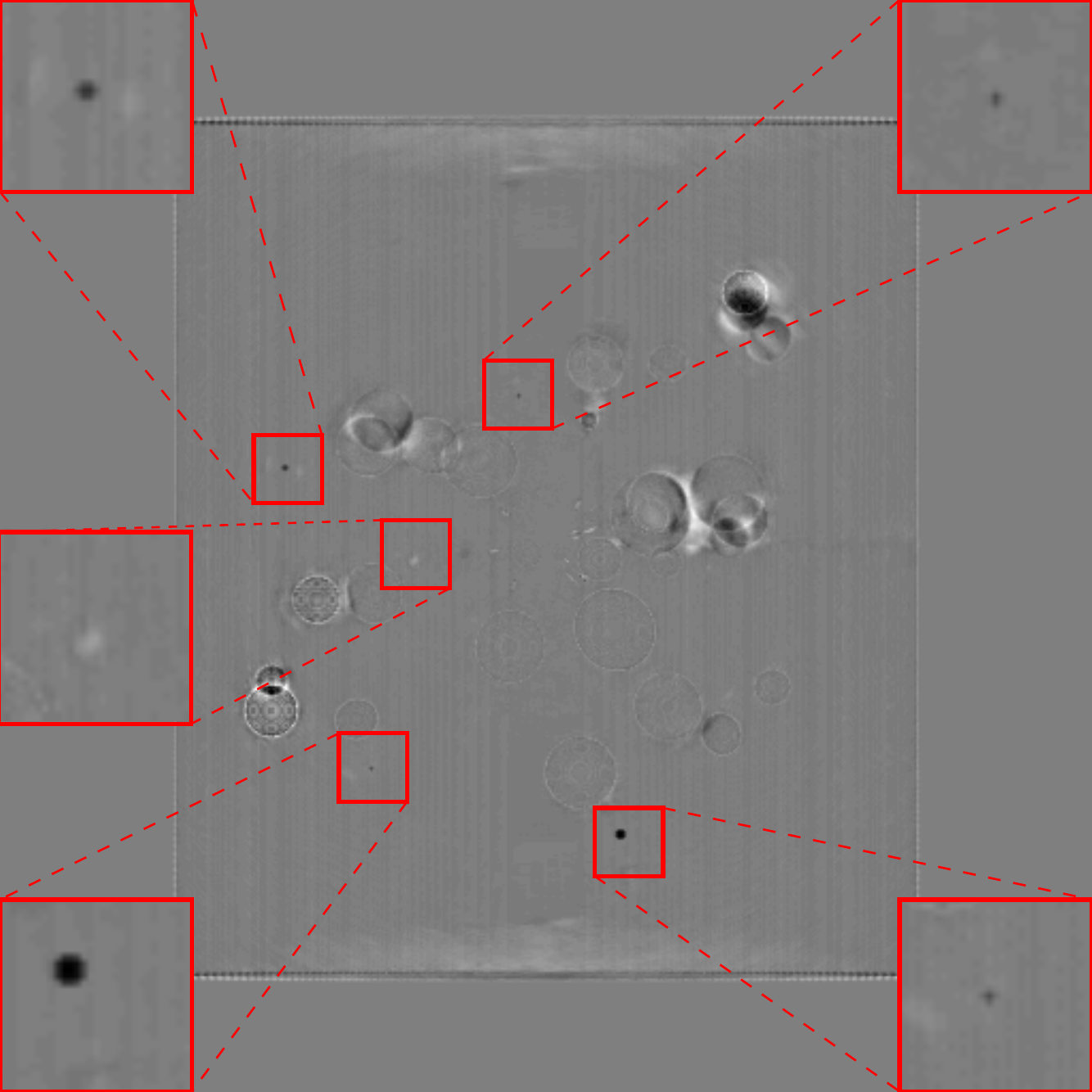}
\label{subfig:10PredictionDiffArrow}
}
\end{minipage}
\caption{An exemplary phantom contains tiny beads, where false positive and false negative tiny beads are observed. (a) is the reference image, where five zoomed-in ROIs (No.\,1-5) are displayed. (b) is the Pix2pixGAN prediction image (in Cartesian form) using two complementary views in polar coordinates, where the tiny beads in ROIs No.\,1-4 are hardly visible (completely missing in ROIs No.\,1 and 2, while blurred in the wrong locations in ROIs No.\,3 and No.\,4). The tiny bead in ROI\,5 of (b) does not exist in the reference image (a). (c) is the difference image. 
}
\label{Fig:FailureCases}
\end{figure}

In the experiments on numerical bead phantom data, false positive and false negative beads are observed, especially for tiny beads. The results of an exemplary phantom containing tiny beads are displayed in Fig.\,\ref{Fig:FailureCases}. Fig.\,\ref{subfig:10TargetArrow} is the reference image where five zoomed-in regions-of-interest (ROIs) (No.\,1-5) are displayed. Fig.\,\ref{subfig:10PredictionArrow} is the prediction image (in Cartesian form) using two complementary views in polar coordinates. The tiny beads in ROIs No.\,1-4 are hardly visible. The tiny beads in ROIs No.\,1 and 2 are completely missing, which are false negative cases. The tiny beads in ROIs No.\,3 and No.\,4 are blurred with very low contrast, and they are in wrong locations. The tiny bead in ROI\,5 of Fig.\,\ref{subfig:10PredictionArrow} does not exist in the reference image, which is a false positive case. 
\modified{After checking the corresponding polar image, we observe that the tiny beads are visible in the polar image despite of low contrast. Hence, the neural network is the main reason for the missing of tiny beads instead of resampling. 
}

\begin{figure}
\centering
\begin{minipage}[b]{0.32\linewidth}
\centering
\small{Reference}
\subfigure[]{
\includegraphics[width=\linewidth]{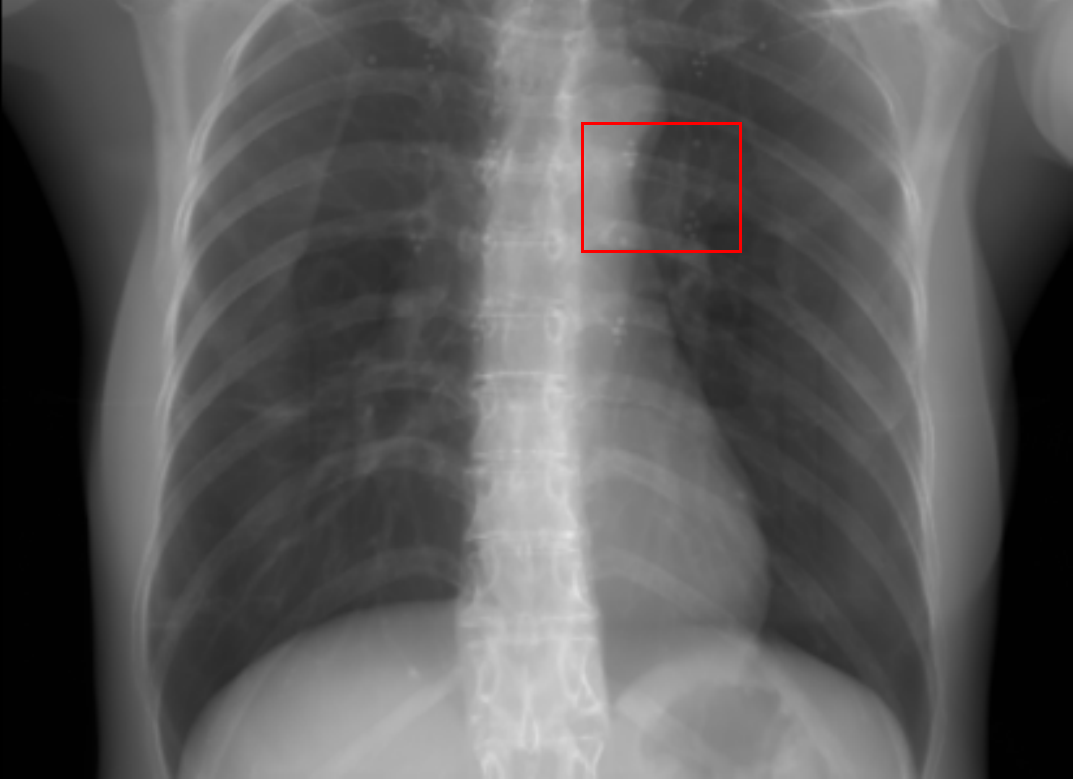}
\label{subfig:torsoTarget1}
}
\end{minipage}
\begin{minipage}[b]{0.32\linewidth}
\centering
\small{$0^\circ$ perspective}
\subfigure[18.16 0.6412]{
\includegraphics[width=\linewidth]{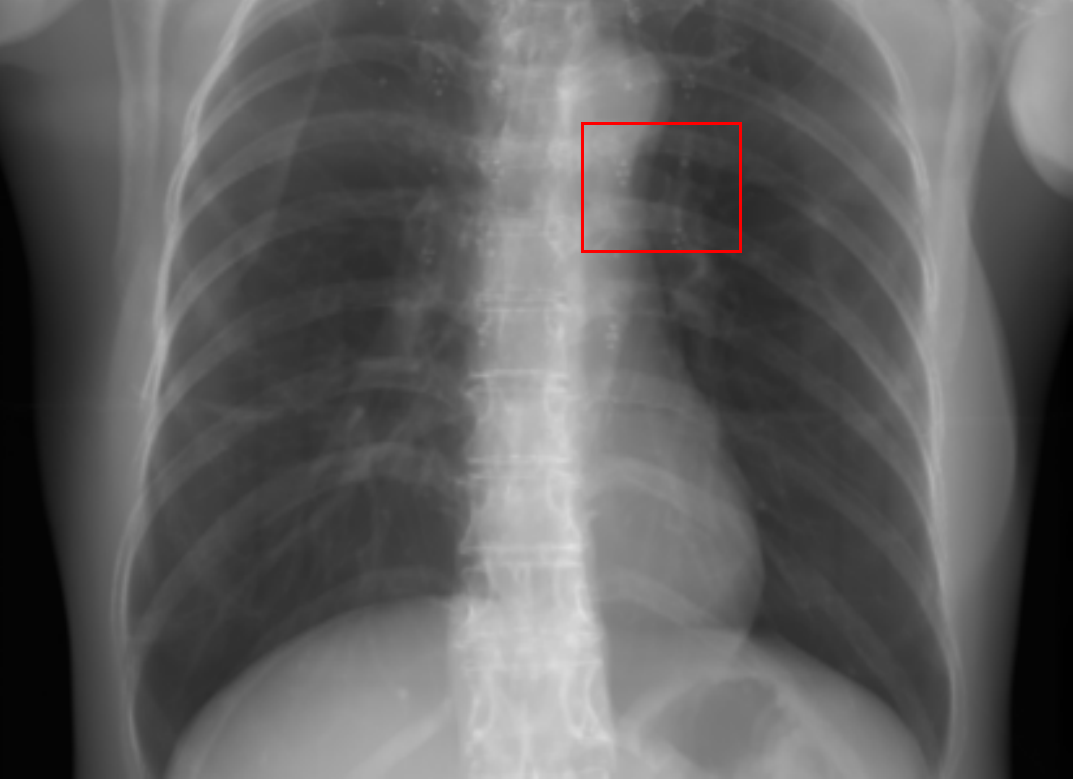}
\label{subfig:torsoPerspective1}
}
\end{minipage}
\begin{minipage}[b]{0.32\linewidth}
\centering
\small{Pix2pixGAN}
\subfigure[3.53 0.9595]{
\includegraphics[width=\linewidth]{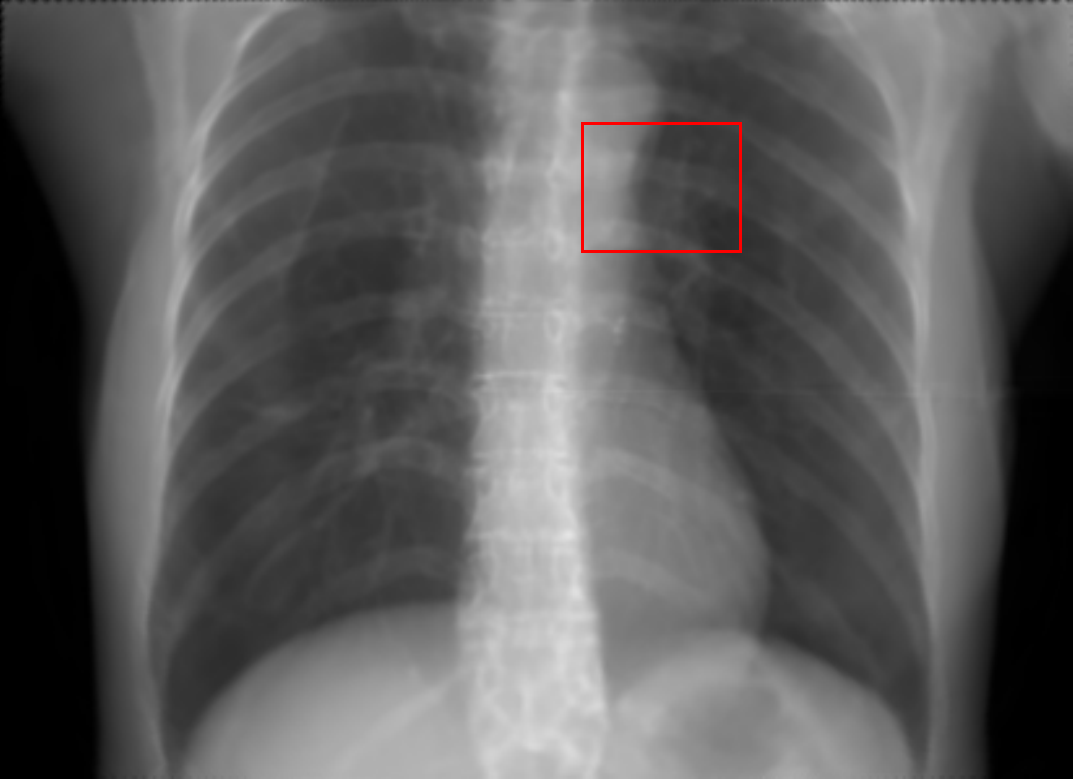}
\label{subfig:torsoDualPolarOutput1}
}
\end{minipage}

\begin{minipage}[t]{0.32\linewidth}
\subfigure[]{
\includegraphics[width=\linewidth]{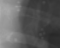}
\label{subfig:torsoTarget1ROI}
}
\end{minipage}
\begin{minipage}[t]{0.32\linewidth}
\subfigure[]{
\includegraphics[width=\linewidth]{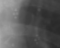}
\label{subfig:torsoPerspective1ROI}
}
\end{minipage}
\begin{minipage}[t]{0.32\linewidth}
\subfigure[]{
\includegraphics[width=\linewidth]{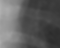}
\label{subfig:torsoDualPolarOutput1ROI}
}
\end{minipage}
\caption{An example of perspective deformation learning from the chest data. The ROIs in (a)-(c) are displayed in (d)-(f) respectively. The tiny metal implants in (d) and (e) are missing in (f).}
\label{Fig:moreTorsoExamples}
\end{figure}

For experiments on simulated anatomical data, similar to the results in the numerical bead phantom experiments (Fig.\,\ref{Fig:FailureCases}), tiny structures, which are around 1\,mm in radius, cannot be reconstructed reliably, especially when such structures are not present in the training data. For example, the tiny metal implants in Fig.\,\ref{subfig:torsoTarget1ROI} and Fig.\,\ref{subfig:torsoPerspective1ROI}, probably vessel stents, are \modified{hardly visible in Fig.\,\ref{subfig:torsoDualPolarOutput1ROI}}.

\begin{figure}
\centering
\begin{minipage}[t]{0.32\linewidth}
\subfigure[$0^\circ$ perspective]{
\includegraphics[width=\linewidth]{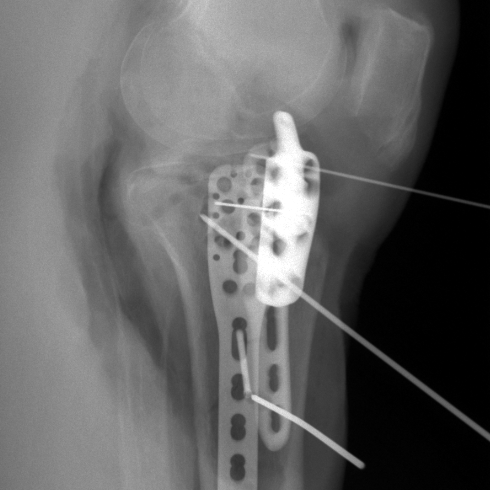}
\label{subfig:cadaver6260PerspectiveView0}
}
\end{minipage}
\begin{minipage}[t]{0.32\linewidth}
\subfigure[$0^\circ$ and $180$ RGB]{
\includegraphics[width=\linewidth]{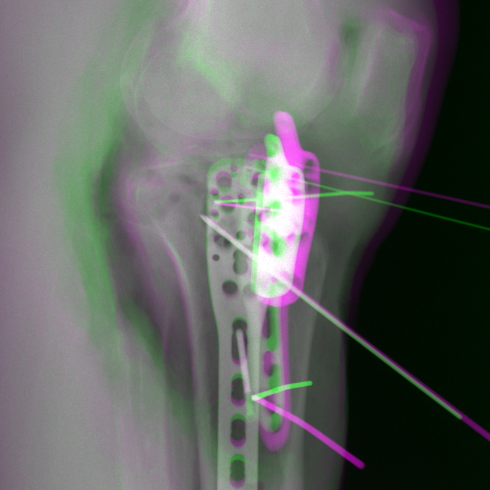}
\label{subfig:cadaver6260RGB}
}
\end{minipage}
\begin{minipage}[t]{0.32\linewidth}
\subfigure[Pix2pixGAN]{
\includegraphics[width=\linewidth]{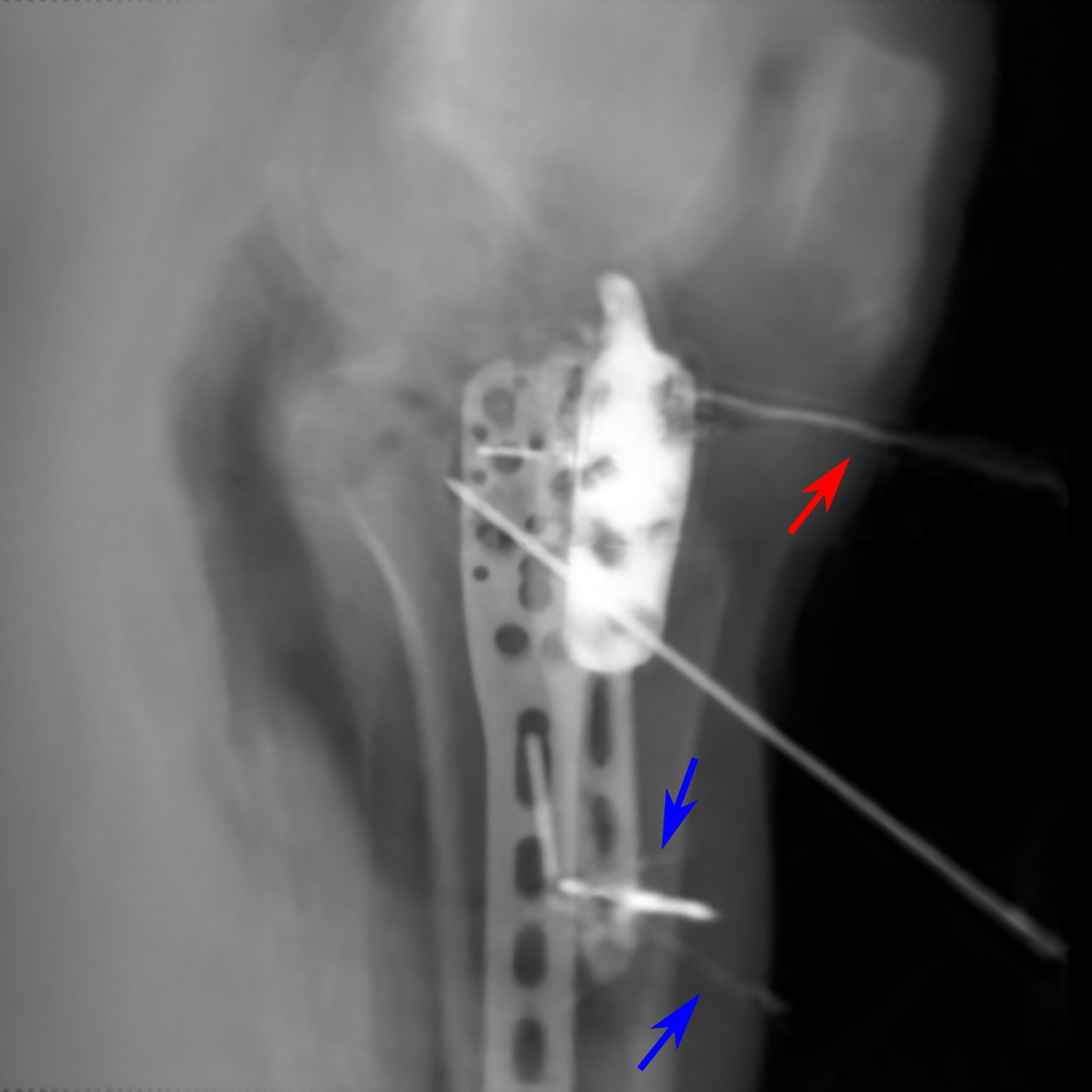}
\label{subfig:cadaver6260Pix2pixGAN}
}
\end{minipage}
\caption{\modified{An example from the real cadaver data, where the Pix2pixGAN predicted metal rod is distorted (indicated by the red arrow) and certain K-wire shadows from perspective projection images remain (indicated by the blue arrows).} }
\label{Fig:limitationKWire}
\end{figure}

In the real cadaver data, certain metal implants are distorted in Pix2pixGAN predictions. For example, the long metal rod indicated by the red arrow is no longer straight in Fig.\,\ref{subfig:cadaver6260Pix2pixGAN}, although its position is between the corresponding magenta and green rods in Fig.\,\ref{subfig:cadaver6260RGB}. Another limitation is that certain structures like thin K-wires from $0^\circ$ and $180^\circ$ perspective projection images will remain as shadows in the Pix2pixGAN prediction, for example, those indicated by the blue arrows in Fig.\,\ref{subfig:cadaver6260Pix2pixGAN}.

\modified{In practice, the perspective deformation corrected image can be compared with the input perspective images to check whether false positive and false negative structures occur.}
\section{Discussion}
\modified{The RGB stack of two complementary views provides a practical way to identify which structures suffer from perspective deformation, as displayed in Fig.\,\ref{subfig:perspectiveRGB0and180degExample} and Figs.\,\ref{Fig:ResultsReal}(j)-(l). Note that this is a sufficient but not necessary condition. The colorful structures, which deviate between two complementary views, must suffer from perspective deformation, given sufficient geometric calibration. However, certain structures which appear grey may also suffer from perspective deformation, due to the symmetry with respect to the virtual detector. For example, the grey cylinder background in the bead phantoms (Fig.\,\ref{subfig:perspectiveRGB0and180degExample}) also suffers from perspective deformation. Nevertheless, the networks are still able to correct these regions, as displayed in Fig.\,\ref{Fig:DataIPredictions}. This indicates that the networks exploit not only geometric features provided by the complementary view but also object/task-specific features for learning perspective deformation. Human anatomical structures are approximately symmetric with respect to the body's mid-sagittal plane. However, the mid-sagittal plane typically does not overlap with the virtual detector plane in practice.  Therefore, structures with perspective deformation are still indicated by colors (See Fig.\,7 in supplementary material). Such geometric information can still be exploited by neural networks, as demonstrated by the superior performance of the complementary view setting on the chest and head data (Fig.\,\ref{Fig:TorsoDataResults} and Fig.\,\ref{Fig:HeadDataResults}).}

Since gantry rotation is not needed for perspective deformation learning from dual orthogonal views in biplanar X-ray systems, it provides more convenience than learning from dual complementary views in practice. Therefore, there is still value for learning perspective deformation from dual orthogonal views. Learning perspective deformation using two orthogonal views did not have satisfactory performance in this work (Tab.\,\ref{Tab:OverallComparisonBead}). \modified{This is due to the large uncertainty for neural networks in learning point-to-point correspondences, as shown in Fig.\,\ref{Fig:uncertaintyAnalysis}.} When the number of beads is not large, it is easy for human visual systems to determine such correspondences based on the size and intensity information. However, learning feature correspondences, e.g. point-to-point correspondences, from two views is still a challenging task for neural networks. Currently, many efforts are being devoted to this topic \cite{yi2018learning,zhang2019learning,xie2021learning}. Nevertheless, it is still open for solutions. An exploration of correspondence learning and semantic reasoning in transmission images is a potential direction to improve perspective deformation learning from dual orthogonal views. Another potential direction is to seek an intermediate 3D reconstruction from two orthogonal views \cite{ying2019x2ct,montoya2022reconstruction}. However, the current state-of-art algorithms have insufficient accuracy to generate orthogonal projection images from such 3D volumes yet. For example, apparent distortions in the heart anatomical structures are observed in Fig.\,3 of \cite{montoya2022reconstruction}. Further research to improve such few-view reconstruction image quality is necessary for our application.

A gap between simulated and real X-ray images typically exists. In our work, appearance difference in image contrast and image resolution is observed in Fig.\,\ref{Fig:cadaverTrainingExamples} and Fig.\,\ref{Fig:ResultsReal}.  Depending on the specific applications, domain adaptation techniques \cite{tremblay2018training,zhang2018task} are necessary for deep learning models trained on synthetic data to generalize well to real test data. However, for various applications, deep learning models trained on synthetic data exclusively have been reported effective on real data directly \cite{chen2021synthetic,de2021next,mill2021synthetic}. The Pix2pixGAN and TransU-Net models trained on synthetic data (Fig.\,\ref{Fig:cadaverTrainingExamples}) have certain generalizability to real cadaver data in this work, as demonstrated by the results in Fig.\,\ref{Fig:ResultsReal} where the landmark positions like the knee patella are predicted accurately. Although real projection data suffer from various physical effects like beam hardening and Poisson noise, the magnitude of deviation between two complementary views in real data increases from the center towards the outside in a similar manner to DRRs with an ideal scan trajectory. This is the foundation for our model trained from DRRs to generalize to real data.

The influence of geometric inaccuracy such as source-to-isocenter distance error, rotation angle error, shift of detector principal point, and error caused by respiratory motion have been investigated in Fig.\,\ref{Fig:SystemErrors}. The results reveal that the Pix2pixGAN model using the complementary view setting has certain robustness to system errors within the geometric calibration accuracy and mechanical manufacturing accuracy. This indicates that one model trained for one CBCT scanner can potentially generalize to other CBCT scanners of the same type (with the same system configurations but slightly different configuration parameters). Because of this, the model trained from synthetic data in an ideal circular trajectory can still predict satisfying results for data acquired in a real trajectory in our work (Fig.\,\ref{Fig:ResultsReal}). To get better generalizability across different CBCT system setups, further techniques, e.g. using a conditional network \cite{mok2021conditional}, need to be explored.

The out-of-distribution (OOD) problem is one major obstacle to the practical use of deep learning algorithms. It is a common problem of all data-driven models, which is not specific to our application. The current practice to mitigate this problem is to train a model for a specific task with an as large as possible training dataset to cover various inference scenarios. With our limited access to real patient data, a comprehensive study on this issue is not possible at the current stage. Nevertheless, the real data experiments can provide some hints on the robustness and generalizability of our method in the presence of surgical tools, as metal implants like bulky metals, screws and K-wires are inserted. The real data experiments indicate that our proposed method is effective in the presence of bulky metal implants and surgical screws, as indicated by Fig.\,\ref{Fig:ResultsReal}. However, for thin and long K-wires, our method has limited performance, since some shadows from perspective projection images remain and a straight rod is no longer straight in the prediction, as indicated by Fig.\,\ref{Fig:limitationKWire}.

\section{Conclusion}
This work is a proof-of-concept study on learning perspective deformation in X-ray transmission imaging using a framework with two complementary views. The RGB stack of complementary views provides a practical way to identify which structures suffer from perspective deformation. The experiments on numerical bead phantom data demonstrate the advantage of complementary views over orthogonal views or a single view in learning perspective deformation. 
The study on spatial coordinate systems demonstrates that Pix2pixGAN as a FCN achieves better performance in polar space than Cartesian space, while TransU-Net as a transformer-based hybrid network achieves comparable performance (slightly worse) in Cartesian space to polar space. Further study shows that our method (Pix2pixGAN in particular) has certain tolerance to geometric inaccuracy such as source-to-isocenter distances, rotation angles, detector principal point shifts and respiratory motion within calibration accuracy. This indicates that one model trained from one CBCT scanner is applicable to other scanners of the same type.
The experiments on the chest and head data demonstrate that our method has the potential for accurate cardiothoracic ratio measurement and cephalometric imaging. The experiments on real data demonstrate the robustness in the presence of bulky metal implants and surgical screws. All in all, our experiments in this work indicate the promising aspects of learning perspective deformation with a complementary view setting to empower conventional CBCT systems with more future applications.

\appendices

\section{Experimental Setup}

\subsection{Numerical bead phantoms}
3D bead phantoms are generated to demonstrate the efficacy of our proposed algorithms for general perspective deformation learning. Each bead phantom is a cylinder containing spherical beads. The height and diameter of the cylinder are randomly generated, with values of $240 \pm 16$\,mm and $225 \pm 32$\,mm respectively. The background intensity value of the cylinder is a random value of $50\pm 35$\,HU.  Small and big beads have the sizes of $6.4 \pm 1.6$\,mm and $16 \pm 8$\,mm, respectively. Their intensities are either $3500 \pm 350$\,HU or $6000 \pm 1000$\,HU. The phantoms have a size of $512\,\times\,512\,\times\,512$ voxels with a voxel size of $0.625\,\times\,0.625\,\times\,0.625\,\text{mm}^3$. 200 bead phantoms are generated, with 185 phantoms for training, 5 phantoms for validation, and 10 phantoms for testing. Each phantom in the dataset contains approximately 50 spherical beads with random positions inside the cylinder. Note that the number of beads also varies randomly in the same data set. 

For data augmentation, each phantom is rotated by $15^\circ$, $30^\circ$, $\dots$, $75^\circ$. This is equivalent to acquire projection data from the view angles of $15^\circ$, $30^\circ$, $\dots$, $75^\circ$. Hence, for each original phantom, six orthogonal projection images with their corresponding $0^\circ$, $90^\circ$, and $180^\circ$ perspective projections images are acquired. This augmentation is applied to the training, validation, and test phantoms.

The projection images in Cartesian coordinates have an image size of $512 \times 512$ with a pixel size of $0.625\,\text{mm}\,\times\,0.625\,\text{mm}$. Images in polar coordinates have an image size of $512 \times 512$ with a pixel size of $0.375\,\text{mm}\,\times\,0.703^\circ$. Images in log-polar coordinates have an image size of $512 \times 512$ with an angular spacing of $0.703^\circ$ and an initial radial spacing of 0.0075\,mm. For display, the integral intensity range of [0, 6] in projection images is linearly converted to [0, 255] (the same for the following chest and head data).

For simulation studies on bead phantom data, a cone beam computed tomography (CBCT) system with a source-to-detector distance ($D_{\text{sd}}$) of 1200\,mm and a source-to-isocenter distance ($D_{\text{si}}$) of 750\,mm is used, which represents a common configuration for floor mounted C-arm CBCT systems. The detector has $1240\,\times\,960$ pixels with a pixel size of $0.308\,\times\,0.308\,\text{mm}^2$. Acquired projection images are rebinned to a virtual detector at the isocenter, which has $512\,\times\,512$ pixels with a pixel size of $0.625\,\times\,0.625\,\text{mm}^2$. In practice, higher pixel resolution for the virtual detector is achievable. Here a coarse resolution of $0.625\,\times\,0.625\,\text{mm}^2$ is used as a proof of concept. The volume centers of imaged objects are located at the origin of the world coordinates. Perspective projections are generated via forward projection of the volumes. The orthogonal projections are generated using a large virtual source-to-detector distance of 12000\,mm and a short isocenter-to-detector distance of 100\,mm.

\subsection{Chest CT data}
In chest X-ray imaging, a long source-to-detector distance is used to reduce perspective deformation. In order to acquire chest X-ray radiographs with regular CBCT systems, the proposed perspective deformation learning algorithms are evaluated on chest CT data. Three COVID-19 chest CT releases (MIDRC-RICORD-1a \cite{tsai2021rsna}, MIDRC-RICORD-1b \cite{tsai2021rsna}, and COVID-19 sequential data \cite{kassin2021generalized}) and a kidney  CT dataset \cite{heller2021state} from the cancer imaging archive (TCIA) \cite{clark2013cancer} are used. The volumes whose slice spacing is larger than 2.5\,mm (resolution in the Z-direction is too low) or the total slice number is smaller than 100 are removed (the volume cannot cover a complete chest). In total, 92 patients are used for training, 5 patients are used for validation, and 27 patients are used for testing. For data augmentation, each volume has a random anisotropic scaling transform along X, Y, and Z directions in spatial extents. The scaling factors are between 0.95 and 1.05. Each volume is augmented 5 times. In other words, 552 volumes are used for training, 30 volumes are used for validation, and 162 volumes are used for testing.

The same CBCT system, i.e., $D_{\text{sd}}$ = 1200\,mm and $D_{\text{si}}$ = 750\,mm, is used to generate perspective projections. The input perspective images are rebinned to a virtual detector of $472 \times 352$ pixels with a resolution of $0.8 \times 0.8$\,mm$^2$, where lateral and vertical data truncation occur since CBCT detectors typically cannot cover the complete chest. The target orthogonal projection images have $412 \times 300$ pixels with a resolution of $0.8 \times 0.8$\,mm$^2$. During training, both images are zero-padded to the size of $512 \times 512$, which is convenient for the down-sampling path of the U-Net. 

\begin{figure}
\centering
\begin{minipage}[t]{0.32\linewidth}
\subfigure[Phase 0]{
\includegraphics[width=\linewidth]{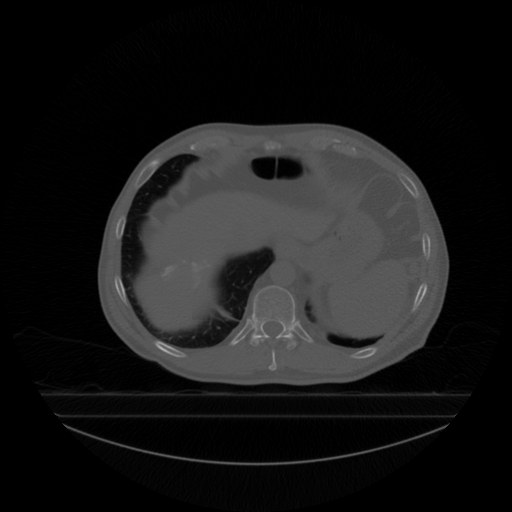}
\label{subfig:4DLungImageExample}
}
\end{minipage}
\begin{minipage}[t]{0.32\linewidth}
\subfigure[Phase 1 difference]{
\includegraphics[width=\linewidth]{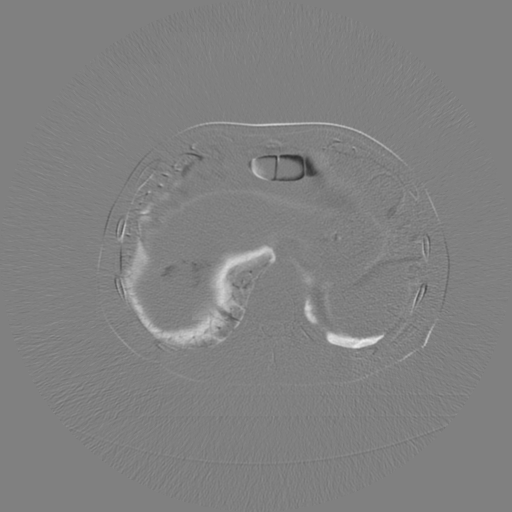}
\label{subfig:4DLungImageDiff1Example}
}
\end{minipage}
\begin{minipage}[t]{0.32\linewidth}
\subfigure[Phase 5 difference]{
\includegraphics[width=\linewidth]{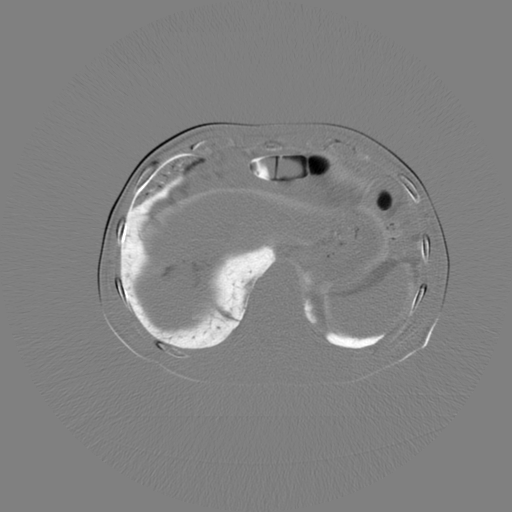}
\label{subfig:4DLungImageDiff5Example}
}
\end{minipage}
\caption{\modified{The influence of respiratory motion in anatomy change: (a) one slice example at phase 0, window [-1000, 2000]\,HU; (b) the difference image of phase 1 for the same slice, window [-1000, 1000]\,HU; (c) the difference image of phase 5 for the same slice, window [-1000, 1000]\,HU.}}
\label{Fig:4DLungDataExample}
\end{figure}

\color{black}
To investigate the influence of respiratory motion in learning perspective deformation, the 4D Lung dataset \cite{balik2013evaluation} is used. One breath cycle is divided into 10 phases in this dataset. Fig.\,\ref{Fig:4DLungDataExample} displays one exemplary slice with different respiratory phases. Between phase 0 and phase 1, considerable anatomical change can already be observed, as shown in Fig.\,\ref{subfig:4DLungImageDiff1Example}. The volume at phase 5 has the largest anatomical deviation to that at phase 0 and hence larger difference is observed in Fig.\,\ref{subfig:4DLungImageDiff5Example}. 20 patients from this dataset are selected with the criterion of more than 100 slices in one volume. The X and Y pixel spacing vary from 0.9766\,mm to 1.1172\,mm, but all the volumes have the same coarse Z spacing of 3\,mm. Perspective and orthogonal projection images are generated using the same parameters as described for the chest data above. For the complementary view setting, with different phase combinations, in total 2000 image pairs are obtained for evaluation.
\color{black}

\subsection{Head CT data}
 In dental imaging, orthogonal X-ray projections are preferred over perspective projections for cephalometric analysis. Hence, the proposed perspective deformation learning algorithms are also evaluated on head CT data. The CQ500 dataset \cite{chilamkurthy2018deep} as well as a public domain database for computational anatomy (PDDCA) \cite{raudaschl2017evaluation} and 10 complete human mandible data sets \cite{wallner2019computed} are used for this purpose. The CQ500 dataset consists of 491 scans, and the PDDCA consists of 48 complete patient head and neck CT images. Many volumes in the CQ500 dataset miss the lower head (neck) part. That is why the other two datasets are necessary for training. The training volumes are augmented by random anisotropic scalings. The volumes whose slice spacing is larger than 2.5\,mm or the total slice number is smaller than 100 are removed. In total, 960 volumes are used for training, 10 volumes are used for validation, and 30 volumes are used for testing.
 
For simulation studies on the head CT data, the CBCT system has a source-to-detector distance of 960\,mm and a source-to-isocenter distance of 600\,mm, since dental CBCT systems typically have a shorter source-to-detector distance than that of angiographic C-arm CBCT systems. The projection images in Cartesian coordinates have an image size of $512 \times 512$ with a pixel size of $0.5\,\text{mm}\,\times\,0.5\,\text{mm}$. Images in polar coordinates have an image size of $512 \times 512$ with a pixel size of $0.5\,\text{mm}\,\times\,0.703^\circ$. 

\color{black}
\subsection{Real data}
\begin{figure}
\centering
\begin{minipage}[t]{0.48\linewidth}
\subfigure[$0^\circ$ perspective]{
\includegraphics[width=\linewidth]{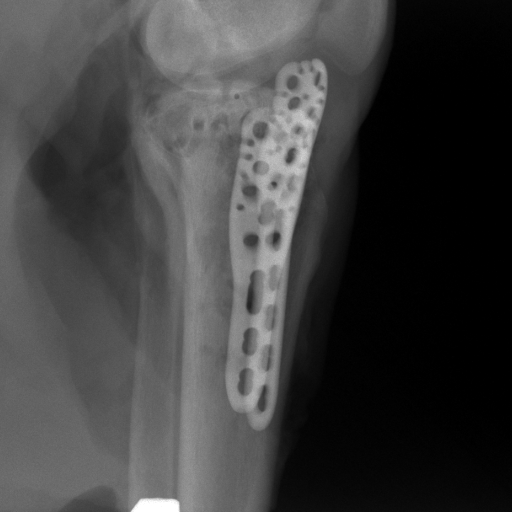}
\label{subfig:legView0}
}
\end{minipage}
\begin{minipage}[t]{0.48\linewidth}
\subfigure[$180^\circ$ perspective]{
\includegraphics[width=\linewidth]{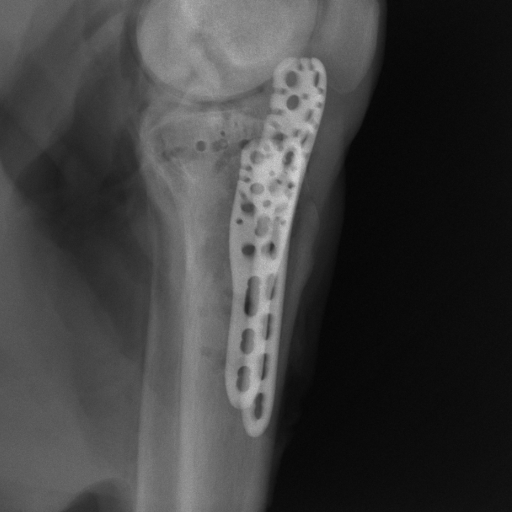}
\label{subfig:legView180}
}
\end{minipage}
\begin{minipage}[t]{0.48\linewidth}
\subfigure[Difference without calibration]{
\includegraphics[width=\linewidth]{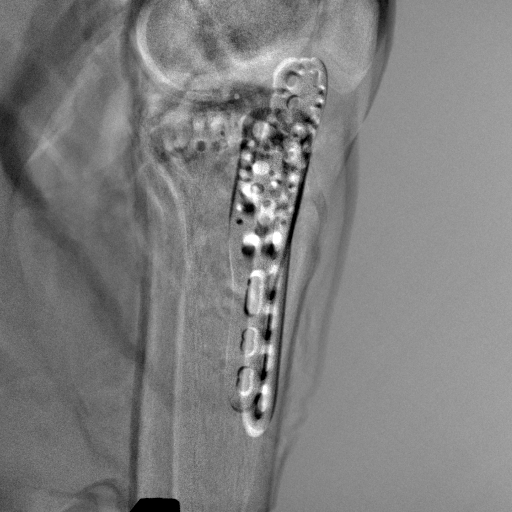}
\label{subfig:legViewDifferenceNaive}
}
\end{minipage}
\begin{minipage}[t]{0.48\linewidth}
\subfigure[Difference with calibration]{
\includegraphics[width=\linewidth]{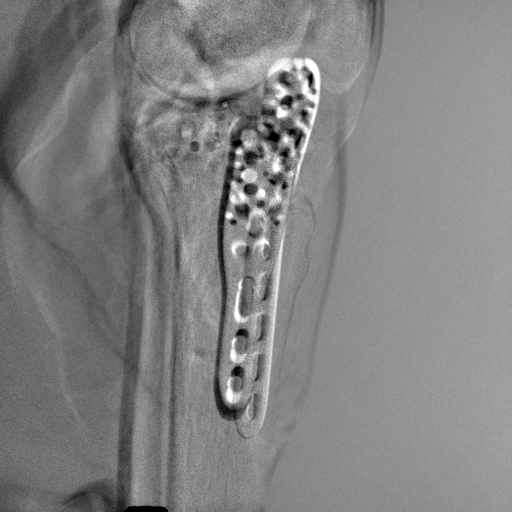}
\label{subfig:legViewDifferenceAligned}
}
\end{minipage}
\caption{\modified{The complementary view perspective projection images and their difference: (a) $0^\circ$ perspective projection image, window [0, 4]; (b) $180^\circ$ perspective projection image, window [0, 4]; (c) difference image between (a) and (b) without geometric calibration, window [-1, 1]; (d) difference image with geometric calibration, window [-1, 1].}}
\label{Fig:RealDataGeometricCalibration}
\end{figure}

In total, 20 real cadaver knee datasets are acquired from a mobile C-arm CBCT system, containing 10 knees without metal implants and 10 knees with metal implants. The CBCT system has a source-to-detector distance ($D_{\text{sd}}$) of approximately 1160\,mm and a source-to-isocenter distance ($D_{\text{si}}$) of around 621\,mm. Note that due to mechanical instability, these two distances vary slightly from view to view. The detector has $976\,\times\,976$ pixels with a pixel size of $0.304\,\times\,0.304\,\text{mm}^2$. Acquired projection images are rebinned to a virtual detector at the isocenter, which has $512\,\times\,512$ pixels with a pixel size of $0.295\,\times\,0.295\,\text{mm}^2$. 400 2D projection images for each scan are acquired with an equiangular increment of $0.4875^\circ$. The total angular range is $195^\circ$.  The scan trajectory is defined by projection matrices. For each scan, 31 projections ($0^\circ$ to $15^\circ$) have their corresponding complementary view projections ($180^\circ$ to $195^\circ$). One exemplary projection and its complementary view is displayed in Fig.\,\ref{Fig:RealDataGeometricCalibration}. 

\begin{figure}
\centering
\begin{minipage}[t]{0.48\linewidth}
\includegraphics[width=\linewidth]{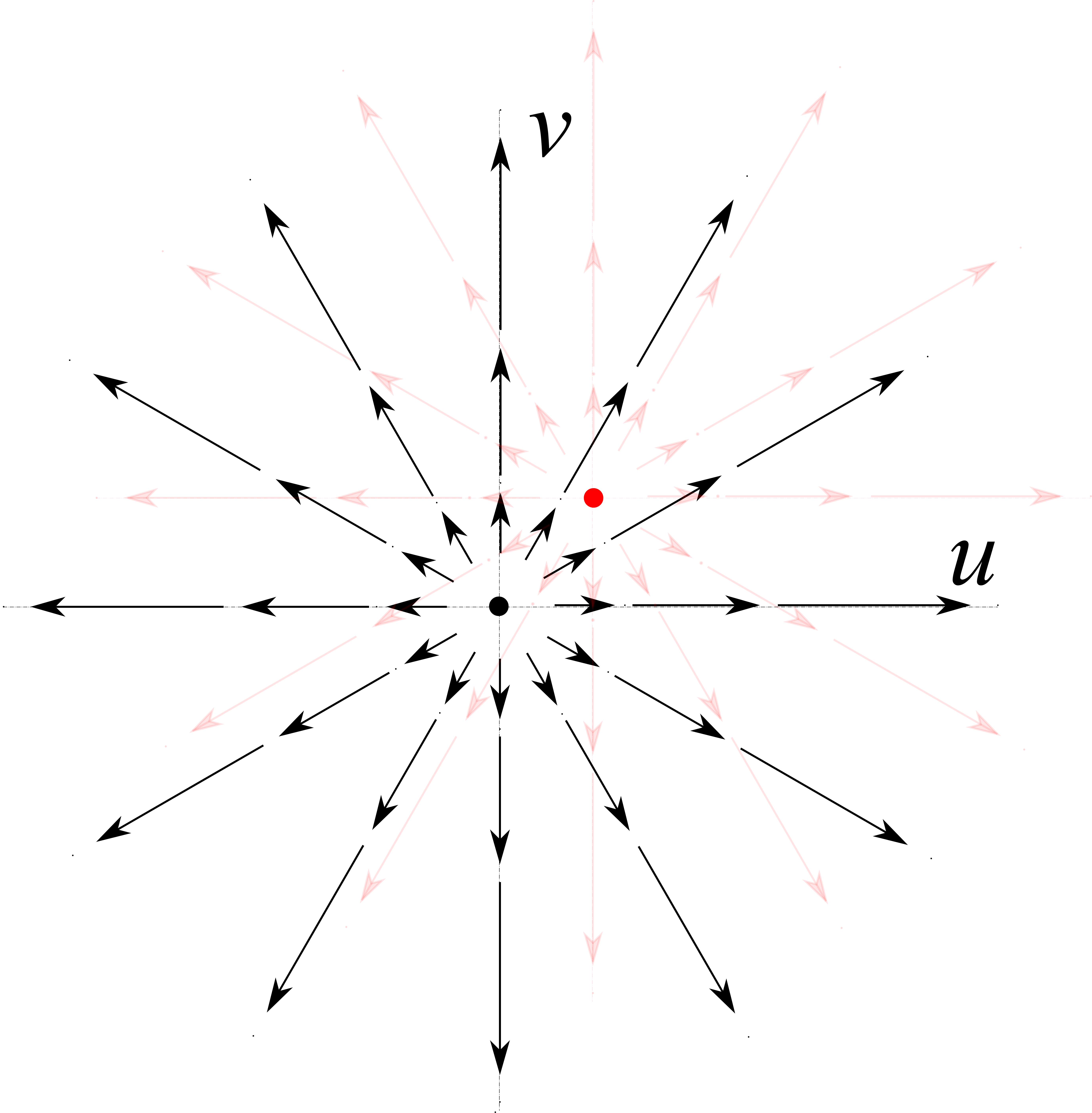}
\end{minipage}
\caption{\modified{An illustration of the relative position shift of the $180^\circ$ complementary view with respect to the $0^\circ$ view. The black point is the principal point of the $0^\circ$ view, which can be placed at the center after rebinning to the virtual detector. The red point is the principal point of the $180^\circ$ complementary view, which has a certain shift from the black one. }}
\label{Fig:principalPointShift}
\end{figure}

For real CBCT systems, the principal point of each view is not located perfectly at the detector center. 
 Hence, a simple horizontal flip of the complementary view will cause anatomy misalignment.  For example, in Fig.\,\ref{subfig:legViewDifferenceNaive} the difference between two complementary views is large at the central region, which violates the perspective deformation property (low deformation at the central region). When rebinning the $0^\circ$ view to the virtual detector, its principal point can be placed at the center, as displayed in Fig.\,\ref{Fig:principalPointShift}. For the $180^\circ$ complementary view, its structures need to be aligned to those in the $0^\circ$ view. In practice, the perspective projection of world origin, which is the last column of the perspective projection matrix, is located very close to the principal point. Therefore, shifting the second virtual detector based on its perspective projection of world origin will reduce such anatomy misalignment. Note that due to mechanical instability the principal point varies from view to view. Because of this, after structure alignment, the principal point of the complementary view still has slight shift from that of the $0^\circ$ view, as illustrated by the red point in Fig.\,\ref{Fig:principalPointShift}. In this real dataset, the majority of principal point shifts are within 10 pixels (1.6\,mm in the virtual detector) and all the views are with 20 pixels (3.2\,mm), which has limited influence on network performance. For example, after such calibration, the difference magnitude increases from the center towards the outside in Fig.\,\ref{subfig:legViewDifferenceAligned} like it does in DRRs with an ideal scan trajectory. 

To train deep learning models, 50 multi-slice CT volumes from the SIACS medical image repository \cite{kistler2013virtual} are used. Each volume contains a single knee. The volumes are resampled to a volume size of $300\times 600 \times 600$ with an isotropic voxel spacing of 0.5\,mm. Synthetic metal implants like K-wires, screws, and plates with holes are drawn by AutoCAD \cite{fan2022simulation}. These implants are randomly selected and placed in the CT volumes as ground truth volumes. These volumes are forward projected to generate projections by CONRAD \cite{maier2013conrad} with an ideal circular trajectory with $D_{\text{sd}}=1160$\,mm and $D_{\text{si}}=$ 621\,mm using the same detector configuration as the real mobile C-arm system. In total, 1400 images are used for training, 150 images are used for validation and 620 images from the real cadaver data are used for test.

\color{black}

\subsection{Neural network details}
\subsubsection{Pix2pixGAN and training parameters}
Learning perspective deformation is fundamentally an image-to-image translation problem, where GANs are the state-of-the-art approach. In this work, a pixel-to-pixel generative adversarial network (pix2pixGAN) is applied as an example of fully convolutional networks. Note that the CycleGAN is not chosen since the original CycleGAN is reported to have difficulty in tasks where geometric changes are involved. The Pix2pixGAN uses the U-Net, the most popular and successful neural network in biomedical imaging, as the generator $G$ and a 5-layer CNN as the discriminator $D$. $G$ learns to convert a perspective projection image to an orthogonal projection image. $D$ learns to distinguish the synthetic orthogonal projection image from the target orthogonal projection image. The objective of the conditional GAN is,
\begin{equation}
\begin{array}{l}
\mathcal{L}_{\text{cGAN}}(G,D) =  \\
\qquad \mathbb{E}_{\vx,\vy}\left[\log{D(\vx, \vy)}\right] + \mathbb{E}_{\vx}\left[\log{\left(1 - D(\vx, G(\vx)\right)}\right],
\end{array}
\end{equation}
where $\vx$ is the input, $\vy$ is the target, $G$ tries to minimize this objective against an adversarial $D$ that tries to maximize it, i.e., $G^{\ast}= \arg\min_{G}\max_{D}\mathcal{L}_{\text{cGAN}}(G,D)$. In addition, an $\ell_1$ loss function is applied to train the generator's output close to the target with less blurring compared to $\ell_2$ loss and a perceptual loss $\mathcal{L}_{\text{perc}}$ \cite{gholizadeh2020deep} using the VGG-16 model to further reduce blurring,
\begin{equation}
\mathcal{L}_{\ell_1}=\mathbb{E}_{\vx,\vy}\left[||\vy - G(\vx)||_1 \right].
\label{eqn:gradientweightloss}
\end{equation}
The overall objective function is 
\begin{equation}
G^\ast = \arg \min_G \max_D \mathcal{L}_{\text{cGAN}}(G,D) + \lambda_1 \mathcal{L}_{\ell_1} + \lambda_2 \mathcal{L}_{\text{perc}}.
\label{eqn:Pix2pixGANLoss}
\end{equation}

For training Pix2pixGAN, the Adam optimizer is used with an initial learning rate of 0.0002 and a momentum term of 0.5. $\lambda_1$ and $\lambda_2$ are set to 100 and 50  for the $\ell_1$ loss and the perceptual loss, respectively. Validation is performed during training to avoid over-fitting. In total, 150 epochs are used for training. 

\color{black}
\subsubsection{TransU-Net and training parameters}
TransU-Net is a hybrid CNN-Transformer network built upon the vision transformer (ViT), which leverages both high-resolution spatial information from CNN features and global context with long-range dependencies from transformers \cite{chen2021transunet}. In this work, the encoder of TransU-Net uses ResNet50 + ViT-B/16 configuration. The decoder is the expansive path of a regular U-Net. For the encoder, a CNN module is necessary to form a hybrid CNN-transformer encoder; otherwise an encoder with transformer only leads to coarse resolution. For the transformer, pretrained weights from the ViT work are used. Compared with the ViT-Base model, ViT-Large and ViT-Huge do not provide considerable improvement in our application. Therefore, the ViT-Base model is applied in our work. The overall objective function is
\begin{equation}
\mathcal{L} = \mathcal{L}_{\ell_1} + \lambda_3 \mathcal{L}_{\text{perc}}.
\end{equation}

To train TransU-Net, the momentum optimizer is used, with an initial learning rate of 0.01, a momentum term of 0.9 and a weight decay rate of 0.0001. The perceptual loss parameter $\lambda_3$ is set to 0.5. Note that setting $\lambda_3$ to a large value of 1.0 causes TransU-Net to predict the beads only without the cylinder background in our experiments. In total, 150 epochs are used for training.

\section{Supplementary Results}

\subsection{Influence of perceptual loss}

\begin{figure}
\centering

\begin{minipage}[b]{0.48\linewidth}
\subfigure[No perceptual loss]{
\includegraphics[width=\linewidth]{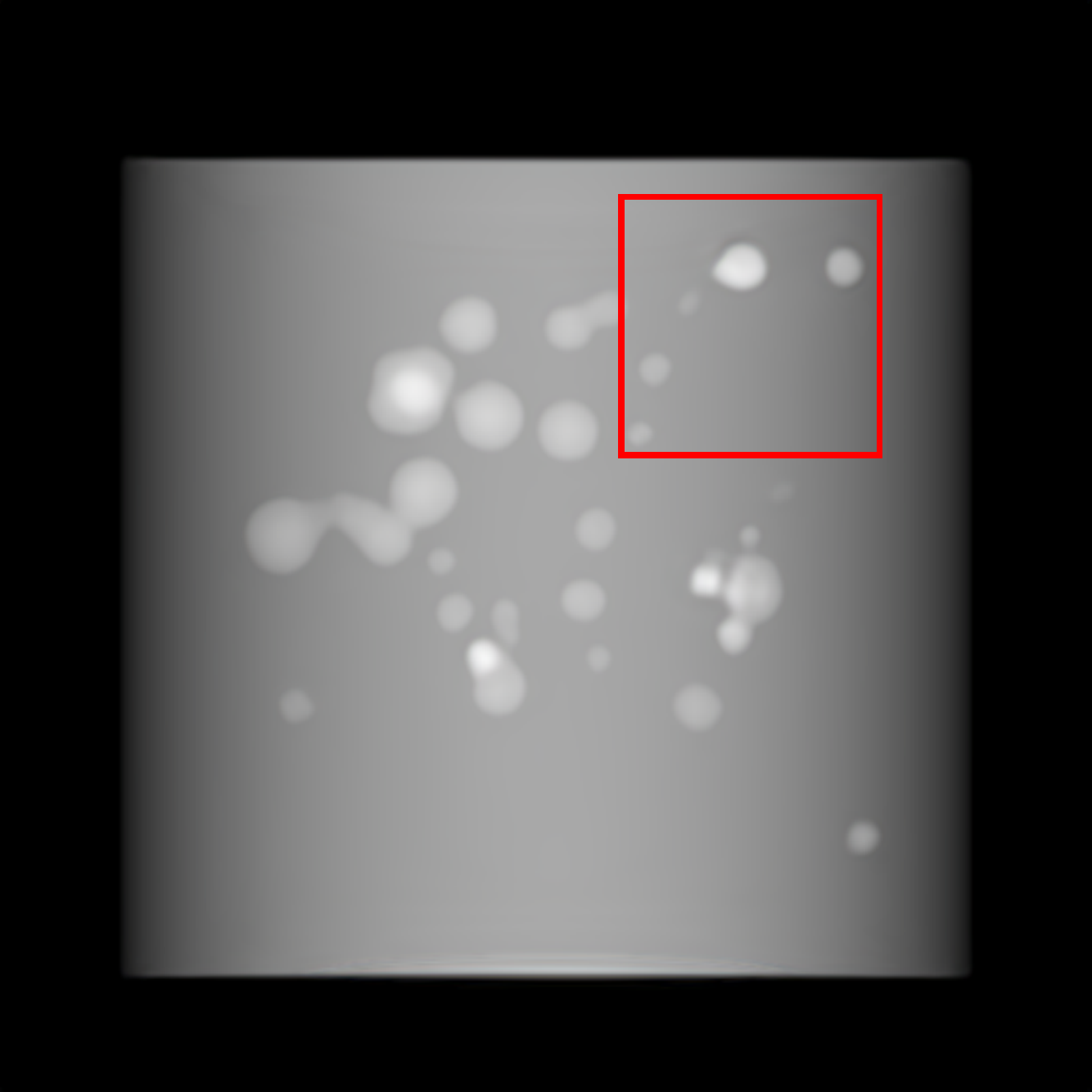}
\label{subfig:PLnoPL}
}
\end{minipage}
\begin{minipage}[b]{0.48\linewidth}
\subfigure[With perceptual loss]{
\includegraphics[width=\linewidth]{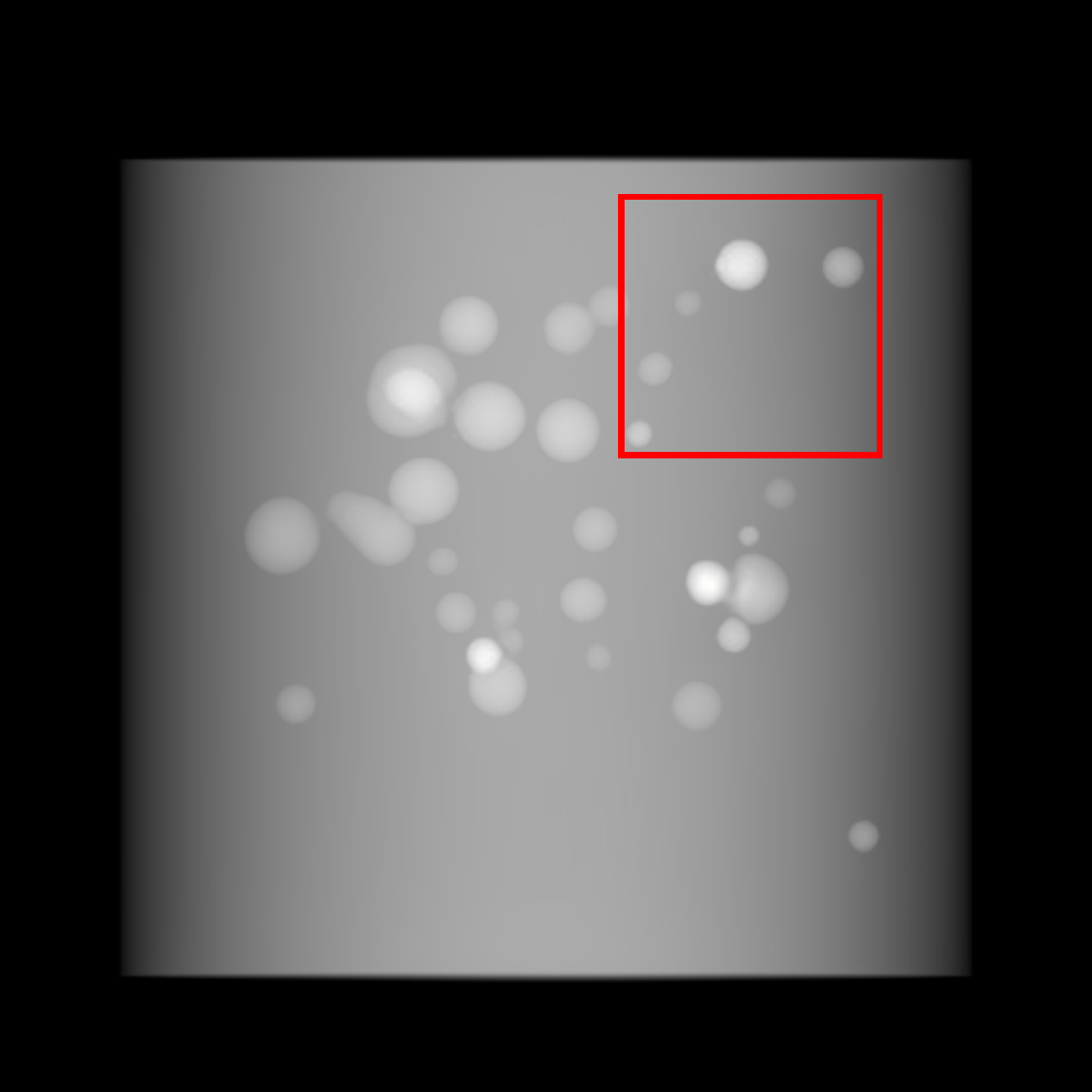}
\label{subfig:PLwithPL}
}
\end{minipage}

\begin{minipage}[b]{0.48\linewidth}
\subfigure[No perceptual loss]{
\includegraphics[width=\linewidth]{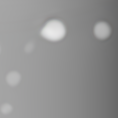}
\label{subfig:PLnoPLROI}
}
\end{minipage}
\begin{minipage}[b]{0.48\linewidth}
\subfigure[With perceptual loss]{
\includegraphics[width=\linewidth]{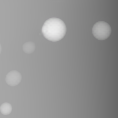}
\label{subfig:PLwithPLROI}
}
\end{minipage}
\caption{\modified{The influence of perceptual loss on image resolution for learning perspective deformation in polar space. (c) and (d) and zoomed-in ROIs (red boxes) from (a) and (b), respectively.}}
\label{Fig:PLinfluence}
\end{figure}

Perceptual loss has been demonstrated to be effective in improving image resolution in many applications. It can improve image resolution for learning perspective deformation as well as displayed in Fig.\,\ref{Fig:PLinfluence}. The depicted results are TransU-Net predictions using complementary views in the Cartesian space.

\subsection{Influence of periodic padding}
\begin{figure}
\centering
\begin{minipage}[b]{0.48\linewidth}
\subfigure[No periodic padding]{
\includegraphics[width=\linewidth]{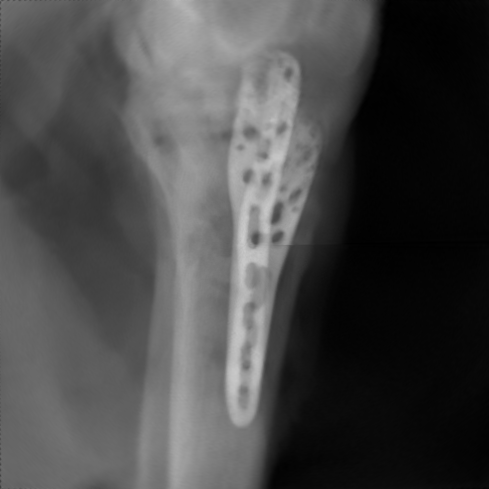}
\label{subfig:NoWrapingPadding}
}
\end{minipage}
\begin{minipage}[b]{0.48\linewidth}
\subfigure[With periodic padding]{
\includegraphics[width=\linewidth]{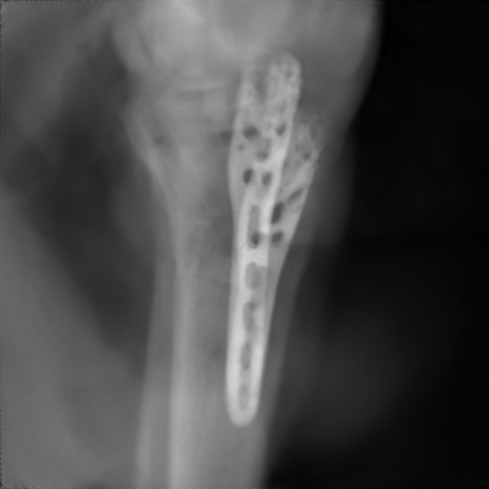}
\label{subfig:DecoderWrapingPadding}
}
\end{minipage}
\begin{minipage}[b]{0.48\linewidth}
\subfigure[No periodic padding]{
\includegraphics[width=\linewidth]{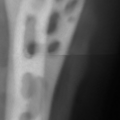}
\label{subfig:NoWrapingPaddingROI}
}
\end{minipage}
\begin{minipage}[b]{0.48\linewidth}
\subfigure[With periodic padding]{
\includegraphics[width=\linewidth]{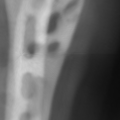}
\label{subfig:DecoderWrapingPaddingROI}
}
\end{minipage}
\caption{\modified{The influence of periodic padding for learning perspective deformation in polar space. (c) and (d) and zoomed-in ROIs from (a) and (b), respectively.}}
\label{Fig:periodicPadding}
\end{figure}

For perspective deformation learning in polar space, periodic padding in the angular dimension reduces stitching artifacts at the $0^\circ$ ($360^\circ$) radial direction. To demonstrate this effect, the TransU-Net prediction results of one knee dataset using complementary views are displayed in Fig.\,\ref{Fig:periodicPadding}. Fig.\,\ref{subfig:NoWrapingPadding} is the result without periodic padding, where the stitching line is slightly visible. In its zoomed-in ROI image in Fig.\,\ref{subfig:NoWrapingPaddingROI}, the stitching line is better visualized. With periodic padding, this stitching line is hardly noticeable even in the zoomed-in ROI in Fig.\,\ref{subfig:DecoderWrapingPaddingROI}.

\subsection{Additional TransU-Net results on bead phantom data}
\begin{figure}
\centering

\begin{minipage}[b]{0.06\linewidth}
\ 

\tiny{\ }
\end{minipage}
\begin{minipage}[t]{0.3\linewidth}
\centering
\footnotesize{Single view (0$^\circ$) $\qquad$}

\tiny{\ }
\end{minipage}
\begin{minipage}[b]{0.3\linewidth}
\centering
\footnotesize{Dual orthogonal views}

\tiny{\ }
\end{minipage}
\begin{minipage}[b]{0.3\linewidth}
\centering
\footnotesize{Dual complementary views}

\tiny{\ }
\end{minipage}

\begin{minipage}[b]{0.06\linewidth}
\centering
{\rotatebox{90}{\footnotesize{Cartesian}}}
\end{minipage}
\begin{minipage}[b]{0.3\linewidth}
\subfigure[4.26, 0.9740]{
\includegraphics[width=\linewidth]{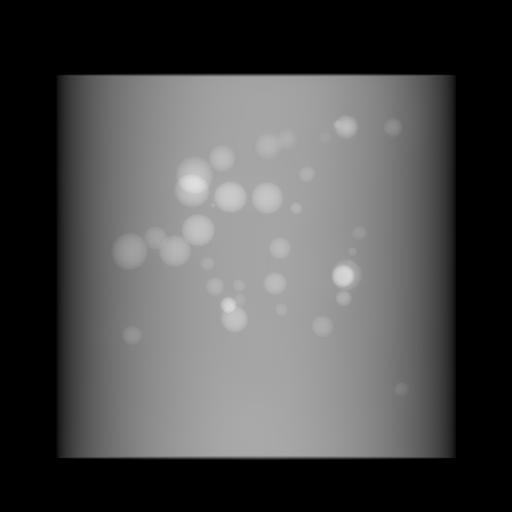}
\label{subfig:singularOutputTransUNet}
}
\end{minipage}
\begin{minipage}[b]{0.3\linewidth}
\subfigure[5.03, 0.9721]{
\includegraphics[width=\linewidth]{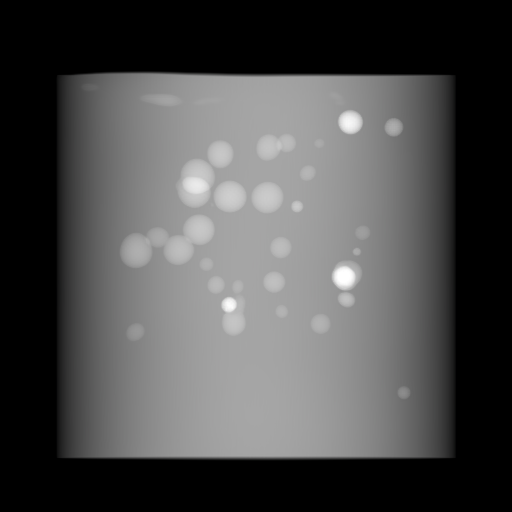}
\label{subfig:dual90OutputOPBPTransUNet}
}
\end{minipage}
\begin{minipage}[b]{0.3\linewidth}
\subfigure[2.52, 0.9947]{
\includegraphics[width=\linewidth]{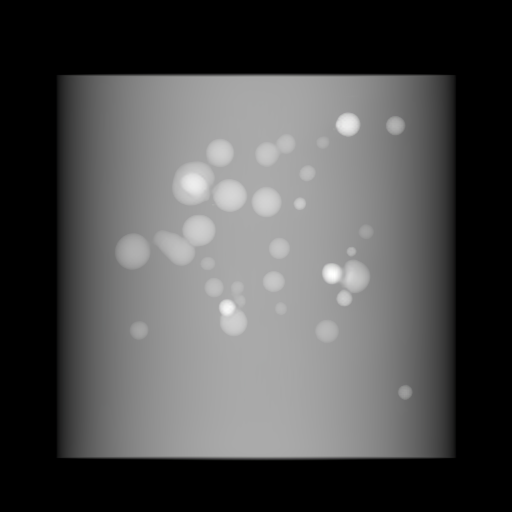}
\label{subfig:dual180OutputTransUNet}
}
\end{minipage}

\begin{minipage}[b]{0.06\linewidth}
\centering
{\rotatebox{90}{\footnotesize{Polar}}}
\end{minipage}
\begin{minipage}[b]{0.3\linewidth}
\subfigure[4.94, 0.9712]{
\includegraphics[width=\linewidth]{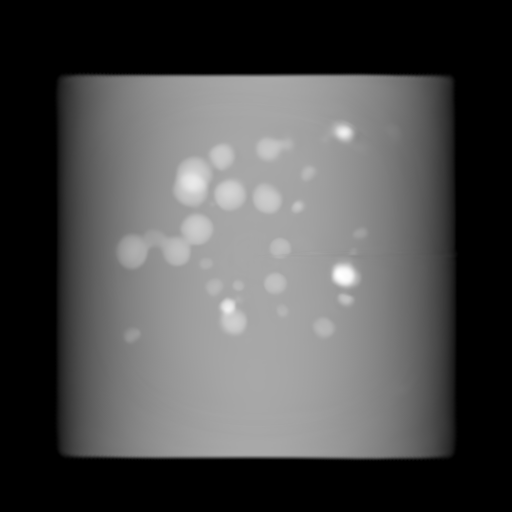}
\label{subfig:singularPolarOutputTransUNet}
}
\end{minipage}
\begin{minipage}[b]{0.3\linewidth}
\subfigure[4.34, 0.9736]{
\includegraphics[width=\linewidth]{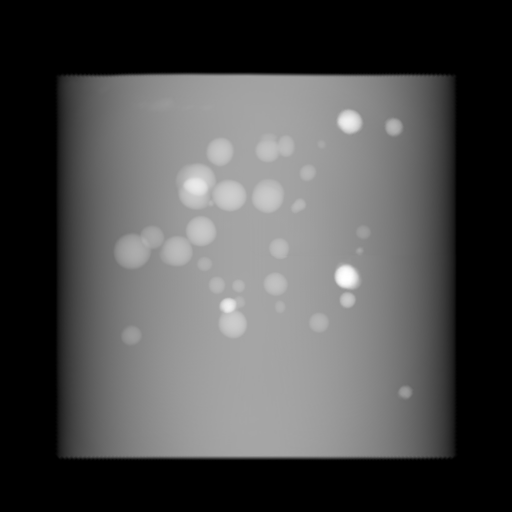}
\label{subfig:dual90PolarOutputOPBPTransUNet}
}
\end{minipage}
\begin{minipage}[b]{0.3\linewidth}
\subfigure[1.59, 0.9957]{
\includegraphics[width=\linewidth]{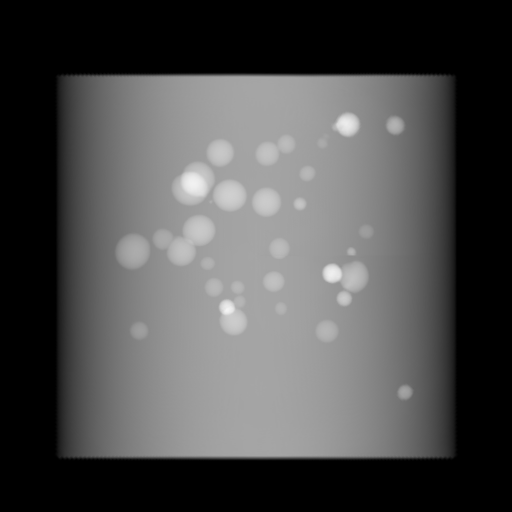}
\label{subfig:dual180PolarOutputTransUNet}
}
\end{minipage}

\begin{minipage}[b]{0.06\linewidth}
\centering
{\rotatebox{90}{\footnotesize{Cartesian error}}}
\end{minipage}
\begin{minipage}[b]{0.3\linewidth}
\subfigure[4.26, 0.9740]{
\includegraphics[width=\linewidth]{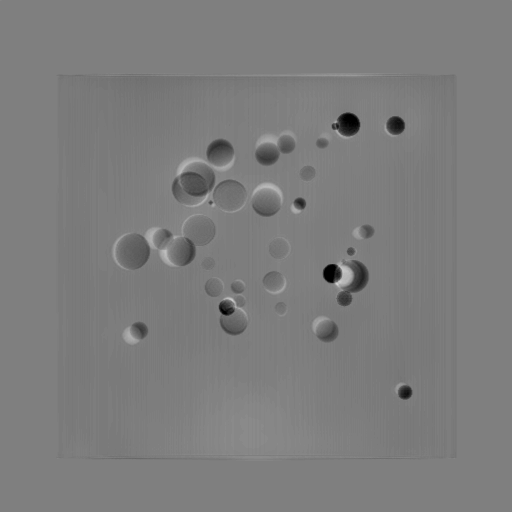}
\label{subfig:singularOutputTransUNetDiff}
}
\end{minipage}
\begin{minipage}[b]{0.3\linewidth}
\subfigure[5.03, 0.9721]{
\includegraphics[width=\linewidth]{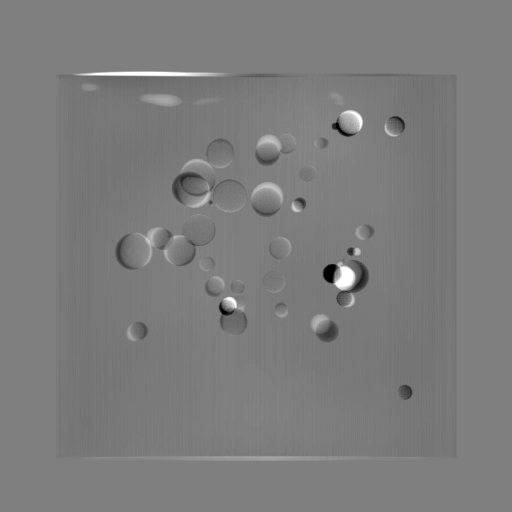}
\label{subfig:dual90OutputOPBPTransUNetDiff}
}
\end{minipage}
\begin{minipage}[b]{0.3\linewidth}
\subfigure[2.52, 0.9947]{
\includegraphics[width=\linewidth]{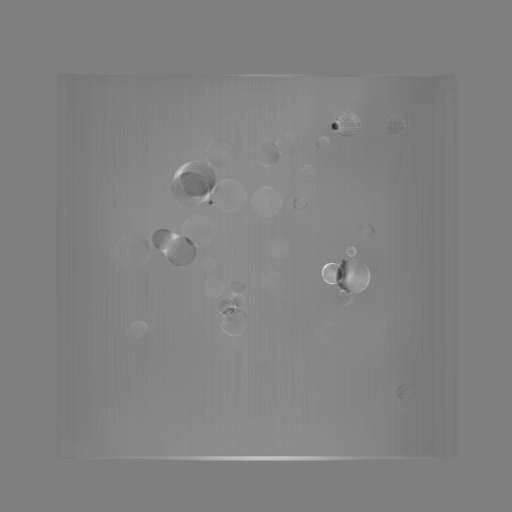}
\label{subfig:dual180OutputTransUNetDiff}
}
\end{minipage}

\begin{minipage}[b]{0.06\linewidth}
\centering
{\rotatebox{90}{\footnotesize{Polar error}}}
\end{minipage}
\begin{minipage}[b]{0.3\linewidth}
\subfigure[4.94, 0.9712]{
\includegraphics[width=\linewidth]{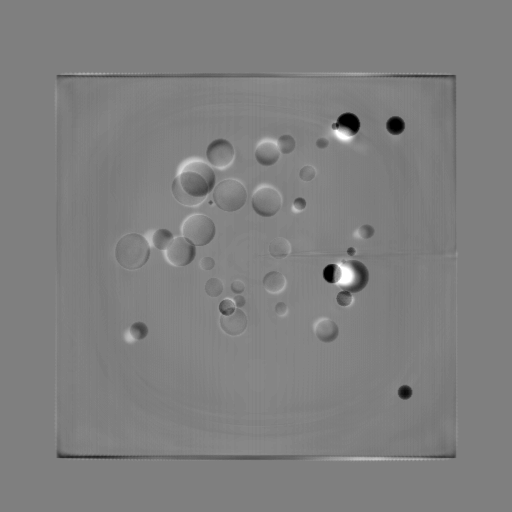}
\label{subfig:singularPolarOutputTransUNetDiff}
}
\end{minipage}
\begin{minipage}[b]{0.3\linewidth}
\subfigure[4.34, 0.9736]{
\includegraphics[width=\linewidth]{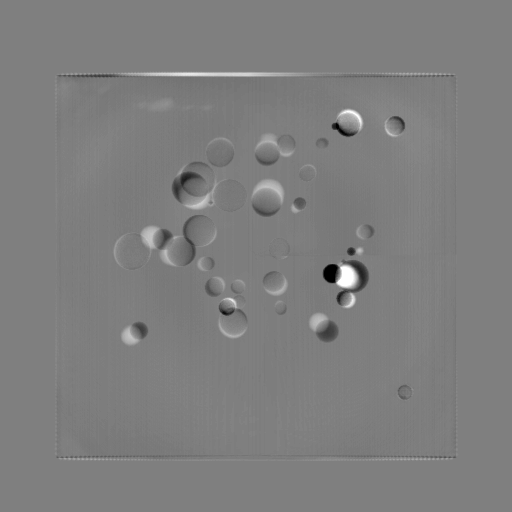}
\label{subfig:dual90PolarOutputOPBPTransUNetDiff}
}
\end{minipage}
\begin{minipage}[b]{0.3\linewidth}
\subfigure[1.59, 0.9957]{
\includegraphics[width=\linewidth]{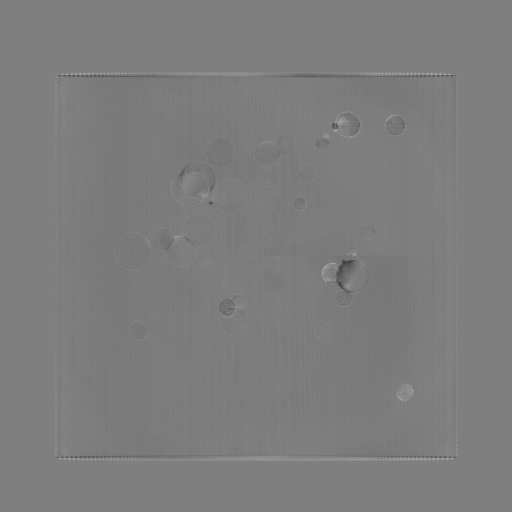}
\label{subfig:dual180PolarOutputTransUNetDiff}
}
\end{minipage}
\caption{\modified{Predictions of TransU-Net in different spaces with different views. Error images are displayed in a window of [-50, 50].}}
\label{Fig:TransUNetBeads}
\end{figure}

\modified{The TransU-Net prediction results of the same image in Fig.\,9 in the main text are displayed in Fig.\,\ref{Fig:TransUNetBeads}.}
\subsection{The selection of ``0$^\circ$ \& 180$^\circ$" \textit{vs.} ``0$^\circ$ \& 180$^\circ$ +" for view combination}

\begin{figure}
\centering
\begin{minipage}[b]{0.32\linewidth}
\subfigure[Input]{
\includegraphics[width=\linewidth]{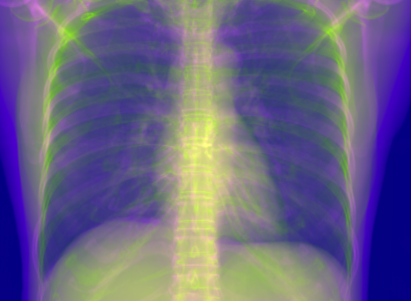}
\label{subfig:torsoDualPlusInputTransUNet}
}
\end{minipage}
\begin{minipage}[b]{0.32\linewidth}
\subfigure[Cartesian]{
\includegraphics[width=\linewidth]{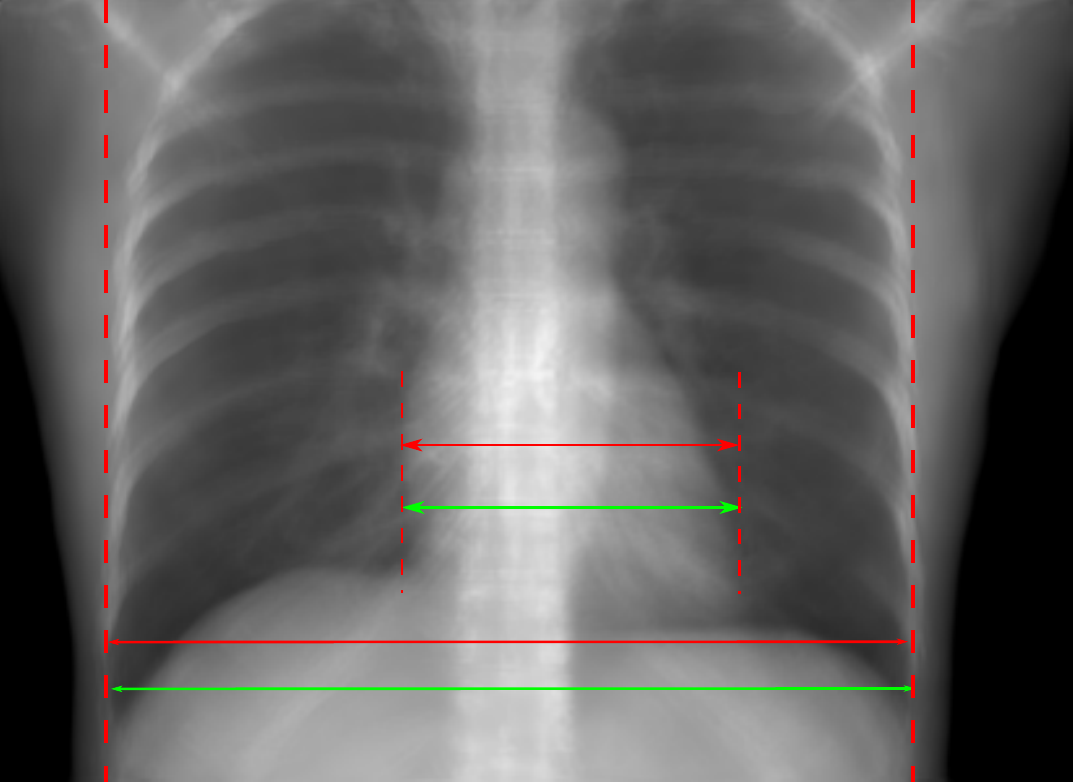}
\label{subfig:torsoDualPlusOutputTransUNet}
}
\end{minipage}
\begin{minipage}[b]{0.32\linewidth}
\subfigure[Polar]{
\includegraphics[width=\linewidth]{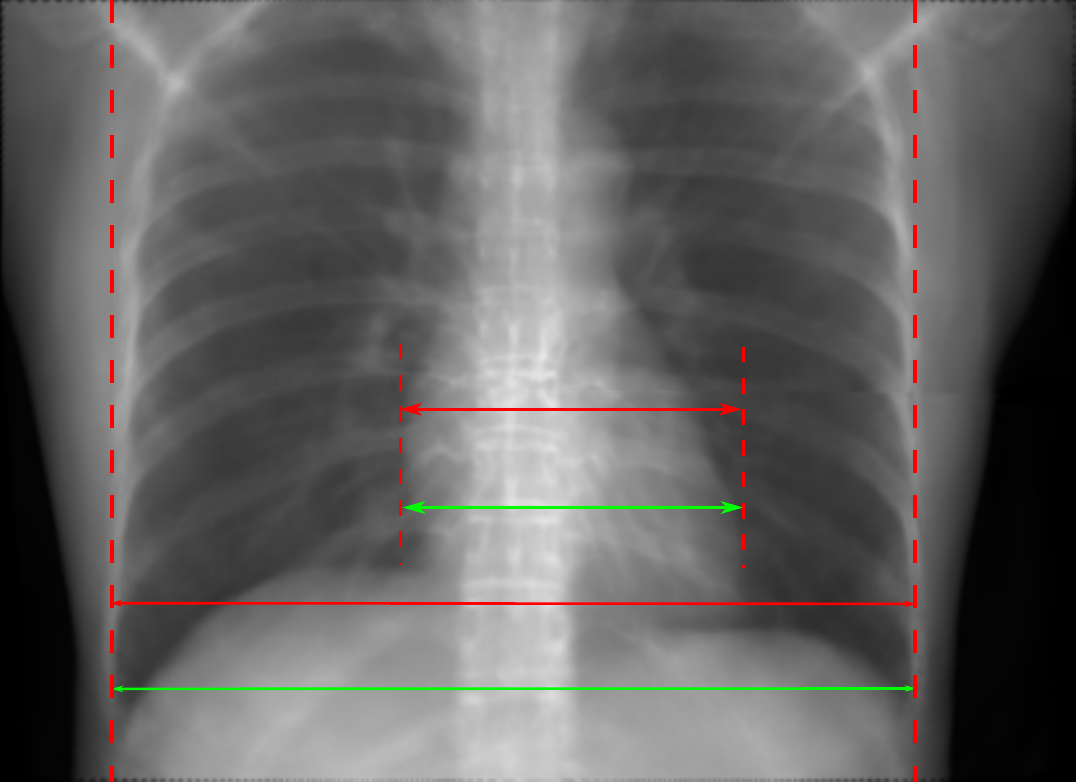}
\label{subfig:torsoDualPlusPolarOutputTransUNet}
}
\end{minipage}

\begin{minipage}[b]{0.32\linewidth}
\centering
\ 
\vspace{5pt}
\end{minipage}
\begin{minipage}[b]{0.32\linewidth}
\centering
\scriptsize{0.4239, 7.68, 0.8509}
\vspace{5pt}
\end{minipage}
\begin{minipage}[b]{0.32\linewidth}
\centering
\scriptsize{0.4217, 6.39, 0.8500}
\vspace{5pt}
\end{minipage}

\begin{minipage}[b]{0.32\linewidth}
\ 
\end{minipage}
\begin{minipage}[b]{0.32\linewidth}
\subfigure[Error of (b)]{
\includegraphics[width=\linewidth]{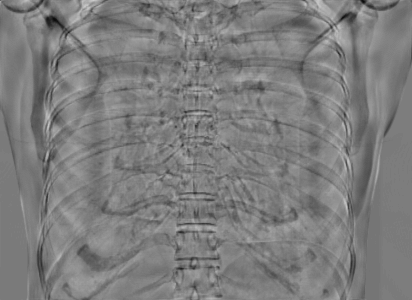}
\label{subfig:torsoDualPlusOutputDiffTransUNet}
}
\end{minipage}
\begin{minipage}[b]{0.32\linewidth}
\subfigure[Error of (c)]{
\includegraphics[width=\linewidth]{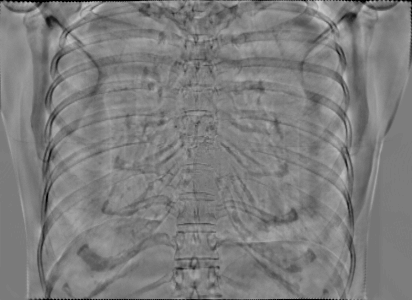}
\label{subfig:torsoDualPlusPolarOutputDiffTransUNet}
}
\end{minipage}
\caption{\modified{Perspective deformation learning on the exemplary patient data for chest X-ray imaging using the ``0$^\circ$ \& 180$^\circ$ +" view combination. The maximal horizontal cardiac diameter and the maximal horizontal thoracic diameter in (b) and (c) are indicated by the horizontal red lines, while those in the reference image are green lines. The cardiothoracic ratio, RMSE, and SSIM for each image is displayed in its corresponding subcaption.}}
\label{Fig:TorsoComplementaryPlus}
\end{figure}

With the complementary view setting, different view combination choices can be plugged into our proposed framework depicted in Fig.\,4 in the main manuscript. In this work, the two choices ``0$^\circ$ \& 180$^\circ$ +" and ``0$^\circ$ \& 180$^\circ$" are demonstrated. For ``0$^\circ$ \& 180$^\circ$ +", the difference image between the two complementary views serves as a straightforward guidance map for the network to learn perspective deformation. Such guidance information is beneficial for Pix2pixGAN in the Cartesian space. However, it brings no additional benefit in the polar space, since ideally the network is able to extract beneficial features automatically by the network itself. In contrast to Pix2pixGAN, such guidance information is beneficial for TransU-Net in both spaces, as suggested by the bead phantom experiments in Tab.\,1. However, ``0$^\circ$ \& 180$^\circ$ +" does not bring considerable visual improvement on all the datasets. For example, the results of ``0$^\circ$ \& 180$^\circ$ +" on the same patient for chest X-ray imaging is displayed in Fig.\,\ref{Fig:TorsoComplementaryPlus}, where the error images look similar to those of Fig.\,10 in the main manuscript. In addition, ``0$^\circ$ \& 180$^\circ$" provides a practical way to identify which structures are perspectively deformed. Therefore, ``0$^\circ$ \& 180$^\circ$" is our main view combination choice in this work.

\subsection{RGB stack examples of chest and head data}
\begin{figure}
\centering
\begin{minipage}[b]{0.563\linewidth}
\subfigure[Chest]{
\includegraphics[width=\linewidth]{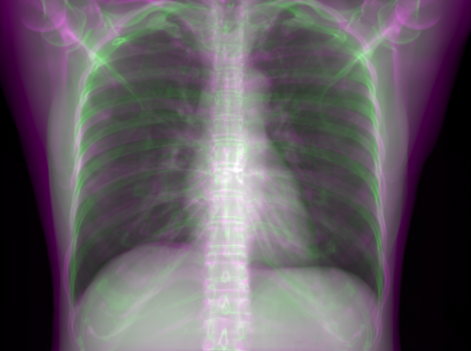}
\label{subfig:RGBChest}
}
\end{minipage}
\begin{minipage}[b]{0.42\linewidth}
\subfigure[Head]{
\includegraphics[width=\linewidth]{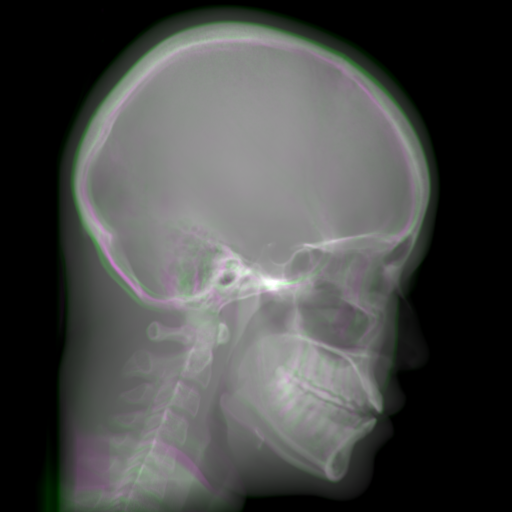}
\label{subfig:RGBHead}
}
\end{minipage}
\caption{\modified{The RGB stack examples of chest and head data for visualization of perspective deformation. (a) and (b) correspond to Fig.\,10 and Fig.\,11 in the main manuscript, respectively.}}
\label{Fig:RGBExamples}
\end{figure}

The RGB stack examples of chest and head data, consisting of $0^\circ$ and $180^\circ$ perspective projection images, are displayed in Fig.\,\ref{Fig:RGBExamples} for the visualization of perspective deformation. Note that for chest X-ray the body's mid-sagittal symmetry plane is approximately perpendicular to the virtual detector plane, while for cephalometric analysis the body's mid-sagittal symmetry plane is approximately but not exactly parallel to the virtual detector plane. Therefore, in Fig.\,\ref{subfig:RGBHead} fewer structures appear colorful than those in Fig.\,\ref{subfig:RGBChest}.

\color{black}


%




\ifCLASSOPTIONcaptionsoff
  \newpage
\fi

\end{document}